%% file: main.tex
\documentclass[acmlarge]{acmart}
\usepackage{epsfig,endnotes}
\usepackage{flushend}
\usepackage{times}
\usepackage{threeparttable}
\usepackage{multirow}
\usepackage{tabularray}
\usepackage{array}
\usepackage{siunitx}

\usepackage{epsfig,endnotes}
\usepackage[british]{babel}
\usepackage{times}
\usepackage{threeparttable}
\usepackage{amsthm}
\usepackage{ragged2e} 
\newcommand{\presec}{\vspace{-0.05in}}
\newcommand{\postsec}{\vspace{-0.05in}}

\usepackage{caption} 

\usepackage{amsmath,amssymb,amsfonts}

\usepackage{makecell}

\usepackage{graphicx}
\usepackage{textcomp}
\usepackage{xcolor}
\def\BibTeX{{\rm B\kern-.05em{\sc i\kern-.025em b}\kern-.08em
    T\kern-.1667em\lower.7ex\hbox{E}\kern-.125emX}}
\usepackage{amsmath}
\usepackage{amssymb}

\usepackage{bm}

\newcommand{\HIDNET}{HIDNet}
\newcommand{\sys}{AccLock}

\input{new-commands}

\usepackage{hyperref}
\usepackage{hyperxmp}
\usepackage{cleveref}  

\newcommand{\nosection}[1]{\vspace{3pt}\noindent\textbf{#1}}

\begin{document}
\title{AccLock: Unlocking Identity with Heartbeat Using In-Ear Accelerometers 
}

\author{Lei Wang}
\affiliation{%
  \institution{Soochow University}
  \country{China}}
  
\author{Jiangxuan Shen}
\affiliation{%
 \institution{ Soochow University}
  \country{China}}

\author{Xi Zhang}
\affiliation{%
 \institution{Macquarie University}
  \country{Australia}}

\author{Dalin Zhang}
\affiliation{%
 \institution{Aalborg University}
  \country{Denmark}}

  \author{Jingyu Li}
\affiliation{%
 \institution{ Peking University}
  \country{China}}

  \author{Haipeng Dai}
\affiliation{%
 \institution{ Nanjing University}
  \country{China}}

  \author{Chenren Xu}
\affiliation{%
 \institution{ Peking University}
  \country{China}}

  \author{Daqing Zhang}
\affiliation{%
 \institution{ Peking University}
  \country{China}}

  \author{He Huang}
\affiliation{%
 \institution{ Soochow University}
  \country{China}}


\input{latex/abstract.tex}
\maketitle 
\input{latex/intro}

\input{latex/pre.tex}
\input{latex/view.tex}

\input{latex/model.tex}
\input{latex/hidnet.tex}

\input{latex/related}

\input{latex/conclusion}

\bibliographystyle{ACM-Reference-Format}
\typeout{}
\bibliography{ref}
\end{document}

%% file: new-commands.tex
\usepackage{url}
\usepackage{graphicx}
\usepackage{latexsym}
\usepackage{amsmath}
\usepackage{amssymb}
\usepackage{amsfonts}
\usepackage{comment}
\usepackage{array}
\usepackage{xspace}
\usepackage{psfrag}
\usepackage{subfigure}
\usepackage{color}
\usepackage{algorithmic}
\usepackage[linesnumbered,ruled]{algorithm2e}
\usepackage{alltt}
\usepackage{indentfirst}
\usepackage{multirow}
\usepackage{amstext}
\usepackage{hyperref}
\usepackage{enumitem}
\usepackage{bbm,multirow}
\usepackage{subfigure,wrapfig,amsmath}
\usepackage{balance}
\usepackage{amsfonts}
\usepackage{float}
\usepackage{amsmath,bm}
\usepackage{siunitx}
\usepackage{gensymb}

\newcommand{\ie}{i.e.\xspace}
\newcommand{\eg}{e.g.\xspace}

\newcommand{\Comment}[1]{}

\newcommand{\secref}[1]{Sec.~\ref{#1}}
\newcommand{\figref}[1]{Fig.~\ref{#1}}
\newcommand{\tabref}[1]{Tab.~\ref{#1}}

%% file: latex/abstract.tex
\begin{abstract}
The widespread use of earphones has enabled various sensing applications, including activity recognition, health monitoring, and context-aware computing. 
Among these, earphone-based user authentication has become a key technique by leveraging unique biometric features.
However, existing earphone-based authentication systems face key limitations: they either require explicit user interaction or active speaker output, or suffer from poor accessibility and vulnerability to environmental noise, which hinders large-scale deployment.
In this paper, we propose a passive authentication system, called \sys{}, which leverages distinctive features extracted from in-ear BCG signals to enable secure and unobtrusive user verification. Our system offers several advantages over previous systems, including zero-involvement for both the device and the user, ubiquitous, and resilient to environmental noise. 
To realize this, we first design a two-stage denoising scheme to suppress both inherent and sporadic interference. To extract user-specific features, we then  propose a disentanglement-based deep learning model, \HIDNET{}, which explicitly separates user-specific features from shared nuisance components. 
Lastly, we  develop a scalable authentication framework based on a Siamese network that eliminates the need for per-user classifier training.
We conduct extensive experiments with 33 participants, achieving an average FAR of 3.13\% and FRR of 2.99\%, which demonstrates the practical feasibility of \sys{}.
\end{abstract}

%% file: latex/intro.tex
\presec 
\section{Introduction} \label{sec:intro}
\postsec

%

The ubiquity of earphones greatly facilitates human–technology interaction, driving the growing adoption of numerous earphone-based sensing applications such as activity recognition \cite{cao2023live}, health monitoring \cite{jin2022earhealth,hu2024breathpro}, and context-aware computing \cite{ma2021oesense,yang2022deepear}. Among these applications, earphone-based authentication emerges as a promising solution for continuous user authentication, because earphones remain closely coupled with the wearer during everyday use. By continuously verifying whether the current wearer remains the legitimate user, such a capability protects persistent access to paired devices and services without requiring  explicit user actions. 
This capability also supports practical applications such as device pairing, automated entry access, and seamless ticketing in everyday settings including homes, vehicles, and public transportation \cite{wang2024earslide, zou2024earprint,gao2019earecho}.
In the continuous authentication setting, the primary threat arises not at first access, but during subsequent use, when a device or service remains trusted and may be exploited by an attacker.
Existing phone-based approaches often either rely on biometrics that remain vulnerable to spoofing and replay attacks \cite{alegre2013spoofing,erdogmus2014spoofing,de2012evaluation,matsumoto2002impact}, or require explicit user actions that are difficult to sustain for implicit continuous authentication \cite{wang2018unlock}.
Smartwatch-based approaches also face limitations: many require active user participation \cite{lee2021usable,buriro2018airsign}, while some passive solutions depend on heart-rate sensors \cite{tang2025exploring,shang2019usable} that are available on only a limited number of commodity devices.
Taken together, these observations further make earphones a compelling platform for continuous authentication. 

\begin{figure}[t]
\begin{center}
\includegraphics [width = 0.6\linewidth]{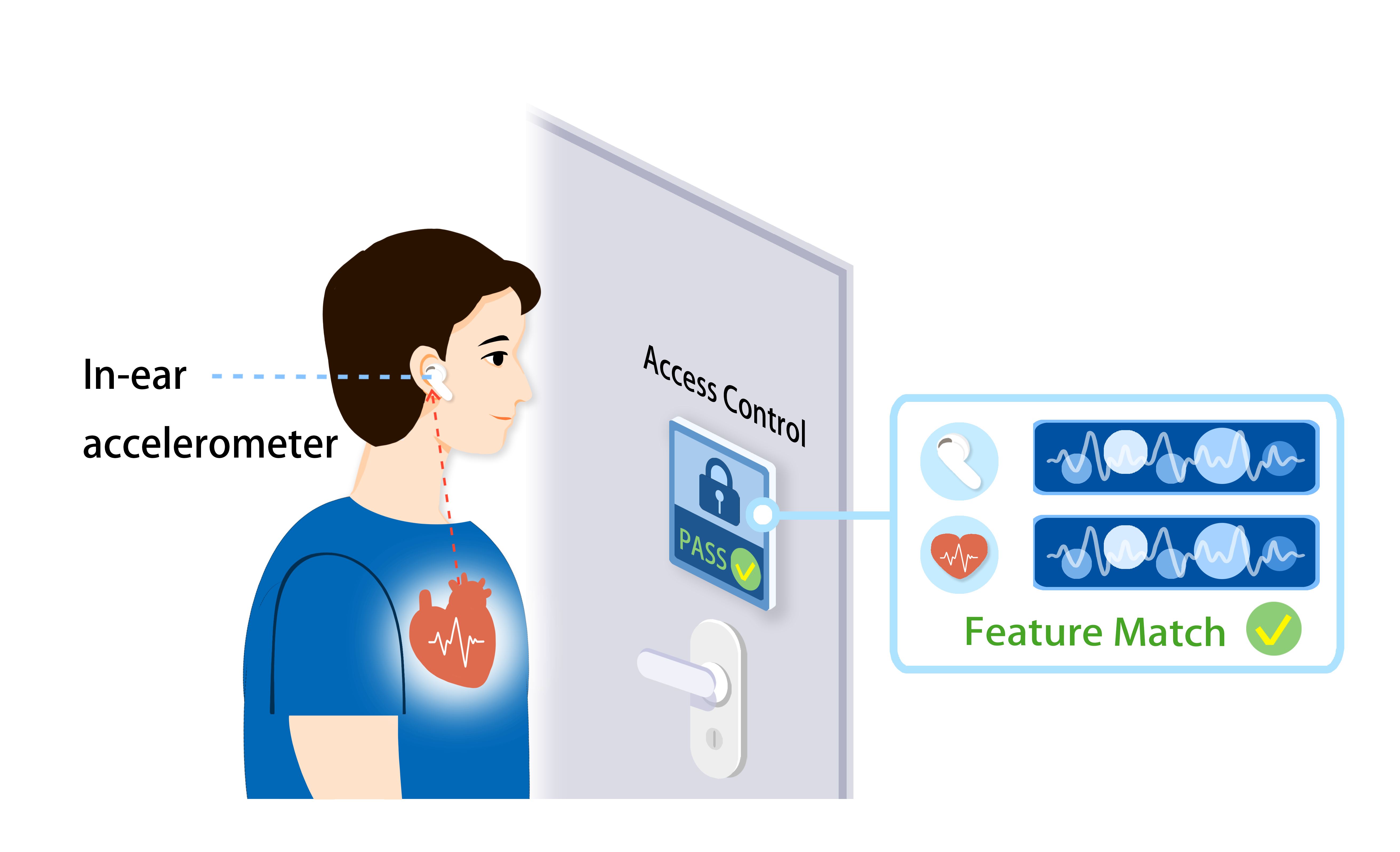}
\caption{Example of an application scenario for \sys{}.}
\label{fig:scenario}
\end{center}
\end{figure}

Existing earphone-based authentication systems typically leverage built-in components and fall into two categories. 
The first category relies on behavioral biometric features, which are typically extracted using microphones or inertial sensors based on user activities such as tooth clicking \cite{xie2022teethpass,wang2022toothsonic}, mandible vibration \cite{liu2021mandipass}, or walking patterns \cite{ferlini2021eargate}.
Clearly, they require explicit user engagement, potentially reducing convenience.
The second category focuses on physiological biometric features, which are inherent physical traits.
Most studies rely on user-specific ear canal biometric features captured using the earphone’s speaker and microphone, such as ear canal geometry \cite{gao2019earecho} and dynamic ear canal deformation \cite{wang2021eardynamic}, offering a user-friendly approach.
However, they require active speaker output, which interferes with normal earphone functionality \cite{cao2023heartprint}.
As is well known, the heartbeat also serves as a unique physiological biometric \cite{wang2018unlock}, and it has been demonstrated that in-ear microphones integrated into active noise cancellation (ANC) earphones can effectively capture heart sounds \cite{nirjon2012musicalheart}.
Building on this capability, recent study \cite{cao2023heartprint} has successfully leveraged these fine-grained and distinctive audio signals to enable passive user authentication, providing the key benefit of requiring no active involvement from either the user or the earphone.
However, this approach faces two key issues. 
First, the high cost of ANC earphones limits their accessibility and hinders large-scale deployment.
Second, the faint heartbeat sounds are passively affected by environmental noise. 
Although noise reduction techniques can suppress such interference to some extent, residual noise often remains, especially in scenarios with strong environmental disturbances.
Recent studies demonstrate that in-ear accelerometers can effectively capture subtle body micro-movements induced by cardiac mechanical activity \cite{lin2024ballistocardiogram,li2025optimizing,islam2025ballistobud}. 
These vibrations, transmitted through bones and tissues to the ear canal, manifest as measurable ballistocardiography (BCG) signals, which exhibit user-specific patterns and provide strong resistance to spoofing attacks due to their physiological origin.
%
%
In this paper, we present \sys{}, a passive system for continuous authentication using in-ear BCG signals. By leveraging distinctive biometric features embedded in these signals, \sys{} enables secure and unobtrusive verification of the current wearer.
\sys{} offers several advantages over prior solutions: it requires no explicit involvement from either the user or the device, builds on the ubiquitous accelerometer already available in modern earphones, and remains robust to environmental acoustic interference.
We focus on continuous authentication as the primary application scenario, in which the earphone passively captures in-ear BCG signals while being worn in everyday settings.
This capability naturally supports downstream applications such as seamless entry, where the authentication result is transmitted via Bluetooth or Wi-Fi to an access-control system, enabling frictionless access as shown in \figref{fig:scenario}.

\begin{figure}[t]
\begin{center}
\includegraphics [width = 0.5\linewidth]{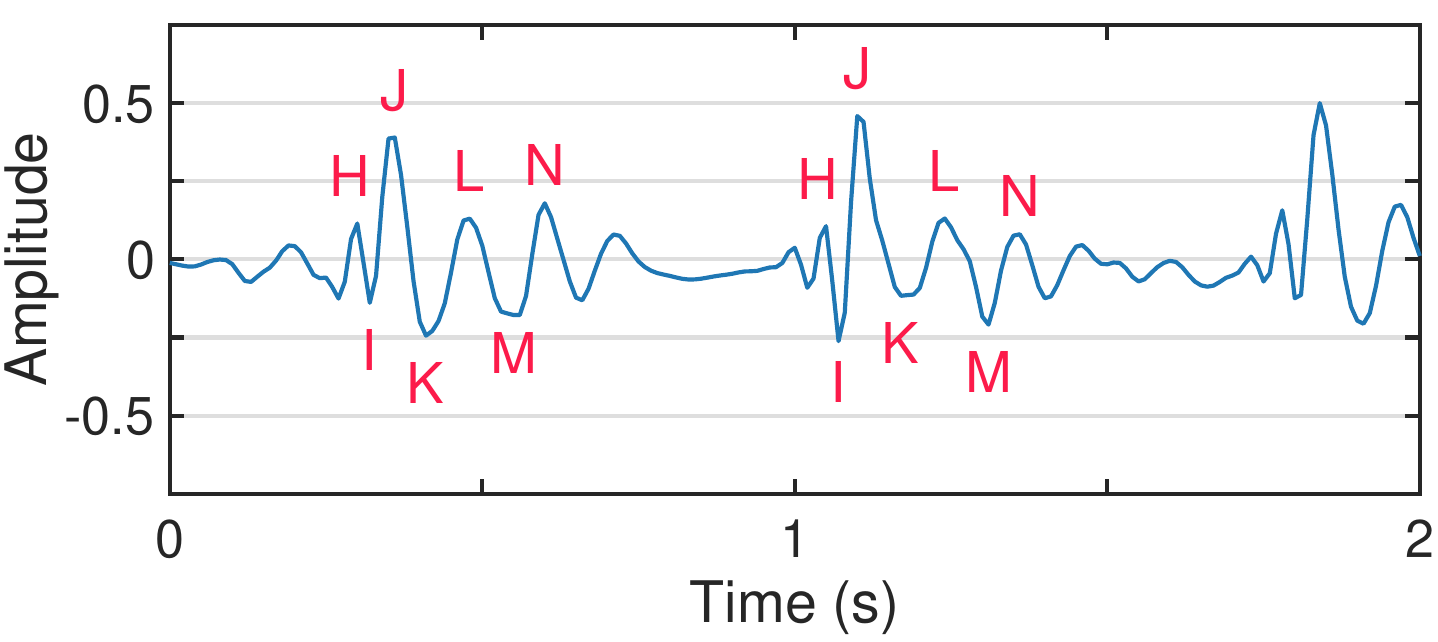}
\caption{In-ear BCG signal pattern.}
\label{fig:cycle}
\end{center}
\end{figure}

While promising, the concept faces several challenges before it can become a practical system.
The first challenge lies in mitigating the effects of noise interference.
During data collection, the in-ear accelerometer records both heartbeat signals and interference arising from diverse movements and inherent system imperfections.
To address this challenge, we divide the interference into two types: inherent interference, which refers to system noise and persistent, involuntary movements, and sporadic interference, which represents sporadic movements such as head shaking, and achieve noise reduction in two stages. 
First, for inherent interference, we propose a wavelet-based denoising scheme to remove the majority of noise, followed by a hyperbolic thresholding method to suppress minor residual components. 
Second, for sporadic interference, we initially exploit the periodic nature of heartbeats to identify sporadic events, followed by an RLS filtering framework with a dynamically adjusted forgetting factor to extract the underlying heartbeat pattern.

The second challenge lies in the entanglement of user-specific and nuisance features within the same BCG signals. User-specific features refer to subtle, individual-dependent physiological signatures, such as variations in heartbeat morphology, timing, and waveform dynamics, that remain stable across sessions and are critical for reliable identification. In contrast, nuisance features are not merely external noise but consist of intrinsic heartbeat patterns shared across all individuals. These common physiological structures tend to dominate BCG signals, often obscuring the fine-grained, identity-specific variations.
To address this issue, we propose a disentanglement-based deep learning model, \HIDNET{}, which explicitly separates user-specific features from shared nuisance components. \HIDNET{} employs gradient reversal adversarial training and orthogonal regularization to enforce disentanglement. This allows the model to isolate and amplify identity-relevant information while suppressing features that are ubiquitous across users, thereby enhancing the discriminative power and robustness of the learned representations.

The third challenge lies in the need to retrain the authentication model for every new user prior to deployment. Conventional methods typically frame authentication as a binary classification problem, where the model learns to distinguish between genuine and attacker samples for a specific user. While effective in controlled settings, this approach requires training or fine-tuning the model each time a new user is enrolled. Such a process is not user-friendly, as it incurs additional time, computational cost, and often demands labeled data from both the target user and attackers.
To overcome this limitation, we propose a user-agnostic authentication paradigm based on a Siamese network. Instead of training a separate classifier per user, the Siamese model learns a feature embedding space where samples from the same individual are pulled closer together, while those from different individuals are pushed farther apart in Euclidean space. During user registration, the system records the user’s BCG signals, extracts the corresponding feature representations, and computes a personalized decision threshold.
At authentication time, a new BCG is recorded, and its feature is extracted and compared to the registered template. If the Euclidean distance is below a predefined threshold, authentication succeeds; otherwise, it fails. This distance-based decision process eliminates the need for user-specific model training and enables fast, flexible, and scalable biometric authentication.

In summary, this paper makes the following contributions:

$\bullet$ To the best of our knowledge, this is the first passive authentication system based solely on in-ear accelerometers, which is zero-involvement for both user and device, widely deployable, and noise-resilient.

$\bullet$ We design a two-stage denoising scheme to suppress both inherent and sporadic interference. To extract user-specific features, we propose a disentanglement-based deep learning model, \HIDNET{}, which
explicitly separates user-specific features from shared nuisance
components. We then develop a scalable authentication framework based on a Siamese network that eliminates the need for per-user classifier training.

$\bullet$  We conduct extensive experiments with 33 participants. The excellent results demonstrate the promising practical usability of \sys{}.

%% file: latex/pre.tex
\section{Preliminary}

\subsection{In-Ear BCG Signal Characterization}
In-ear accelerometers have been shown to capture BCG signals \cite{islam2025ballistobud,zhou2025know}, which reflect subtle head movements induced by cardiac mechanical activity. 
These vibrations result from the inertial response of the body to blood ejection, particularly during ventricular contraction. 
As shown in \figref{fig:cycle}, a typical BCG cycle, recorded from the ear canal, exhibits a sequence of identifiable deflections labeled H through N. 
These include: the initial deflection H associated with atrial activity, a preceding dip I before ejection, the dominant J peak linked to the aortic ejection impulse, followed by K, L, M, and N, which represent subsequent mechanical responses during ventricular relaxation and refilling.
In our system, the vertical axis (\eg, X-axis) of the in-ear accelerometer, aligned with head–neck movement, is responsible for continuously capturing these heartbeat-induced dynamics.
Each cardiac cycle produces one such waveform, where the J peak serves as a reliable marker due to its pronounced amplitude. 
We collect 100 one-second samples from a user instructed to remain completely still and manually label the J peaks. 
For comparison, we also record 100 samples with the earphones placed stationary on a table. 
In our measurements, the average J-peak amplitude is 0.483 m/s², and the accelerometer’s background noise variance is 0.041 m/s², which provides a sufficient signal-to-noise ratio (SNR) for capturing detailed BCG signals.

\begin{figure*}[t]
    \centering
    \vspace{-0.2in}
    \subfigure[User 1]{\label{fig:user1}
    \includegraphics[width = 0.24\textwidth]{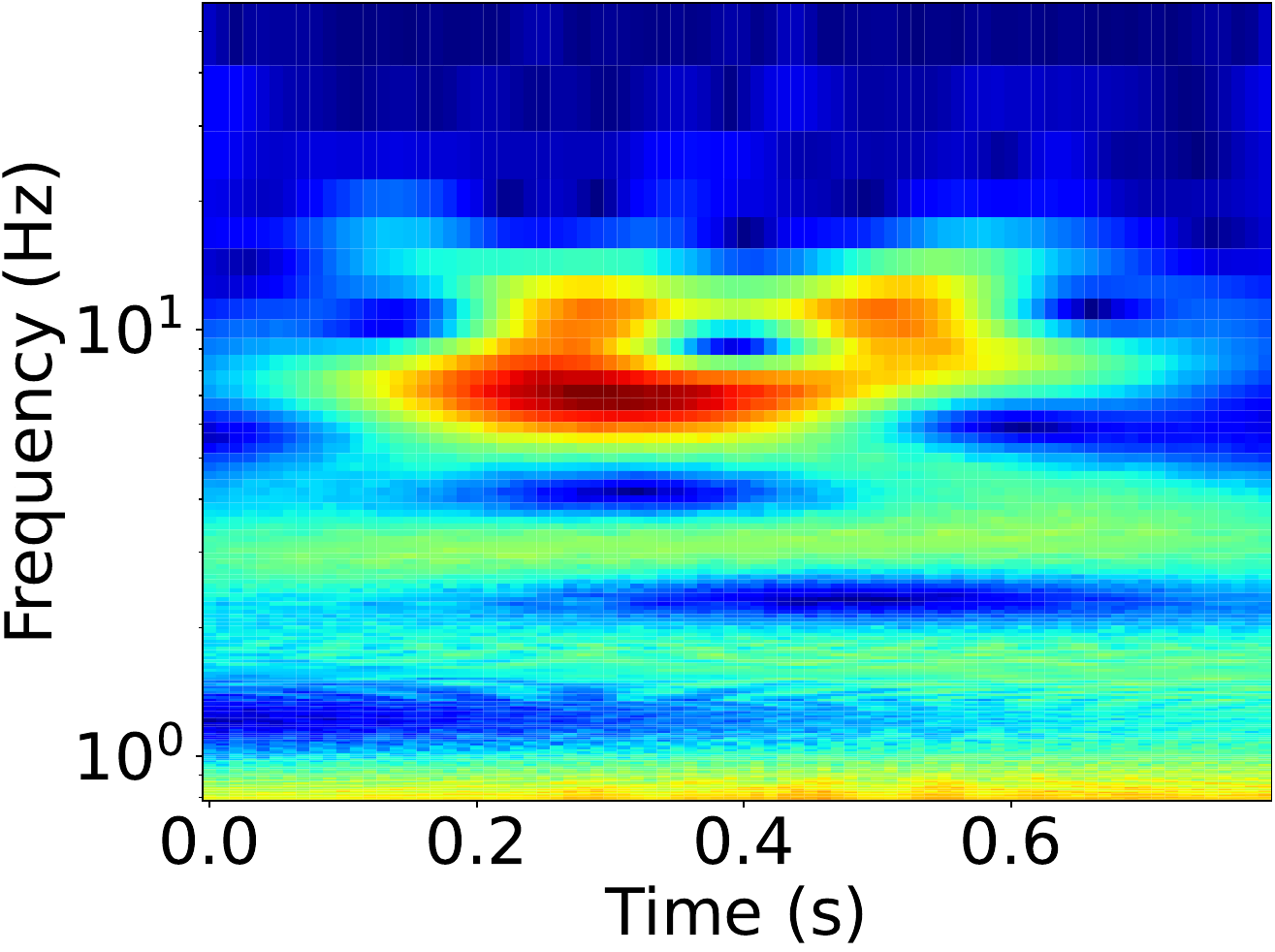}}
    \subfigure[User 2]{\label{fig:user2}
    \includegraphics[width = 0.24\textwidth]{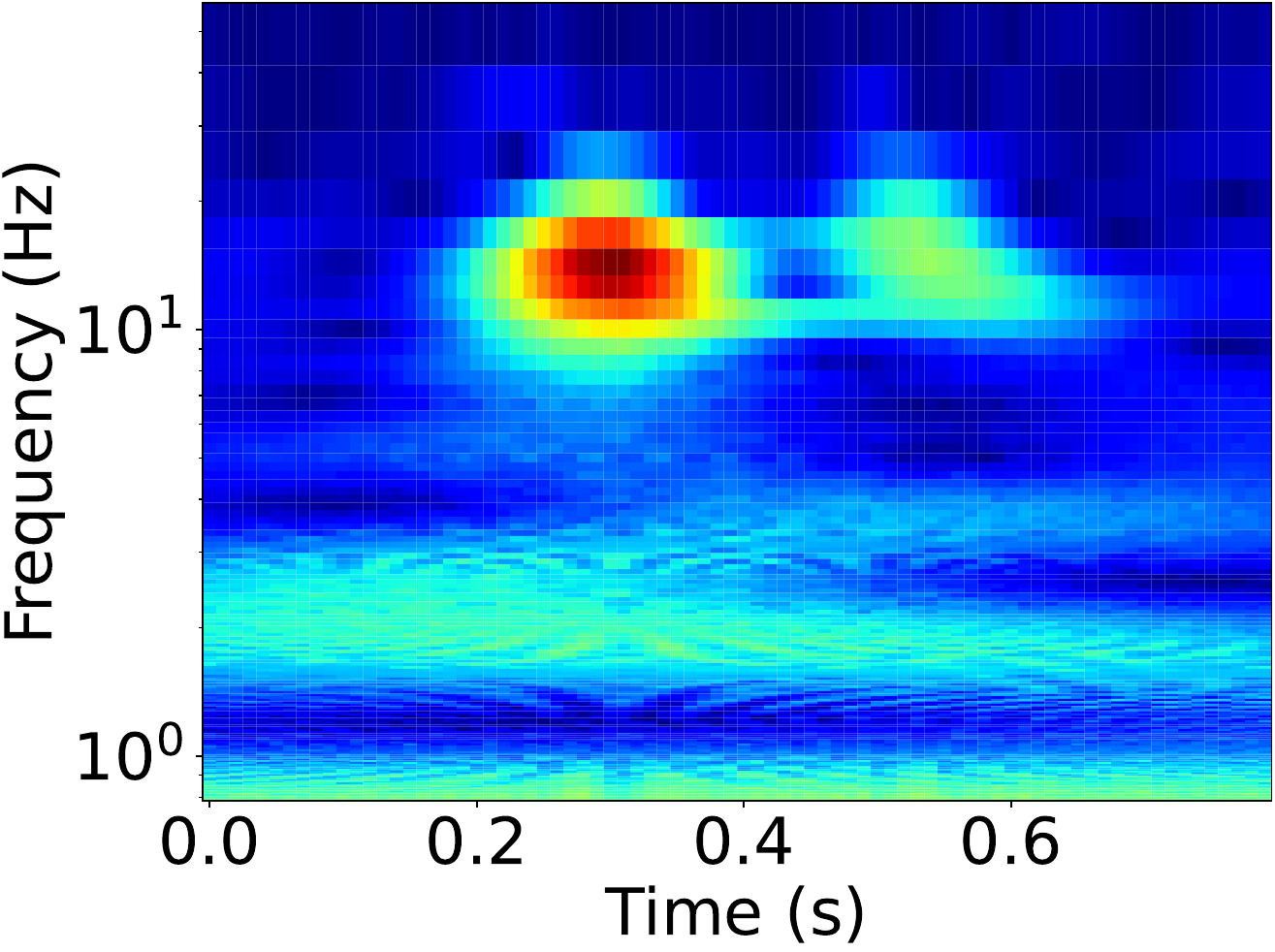}}
    \subfigure[User 3 at session 1]{\label{fig:user3_1}
    \includegraphics[width = 0.24\textwidth]{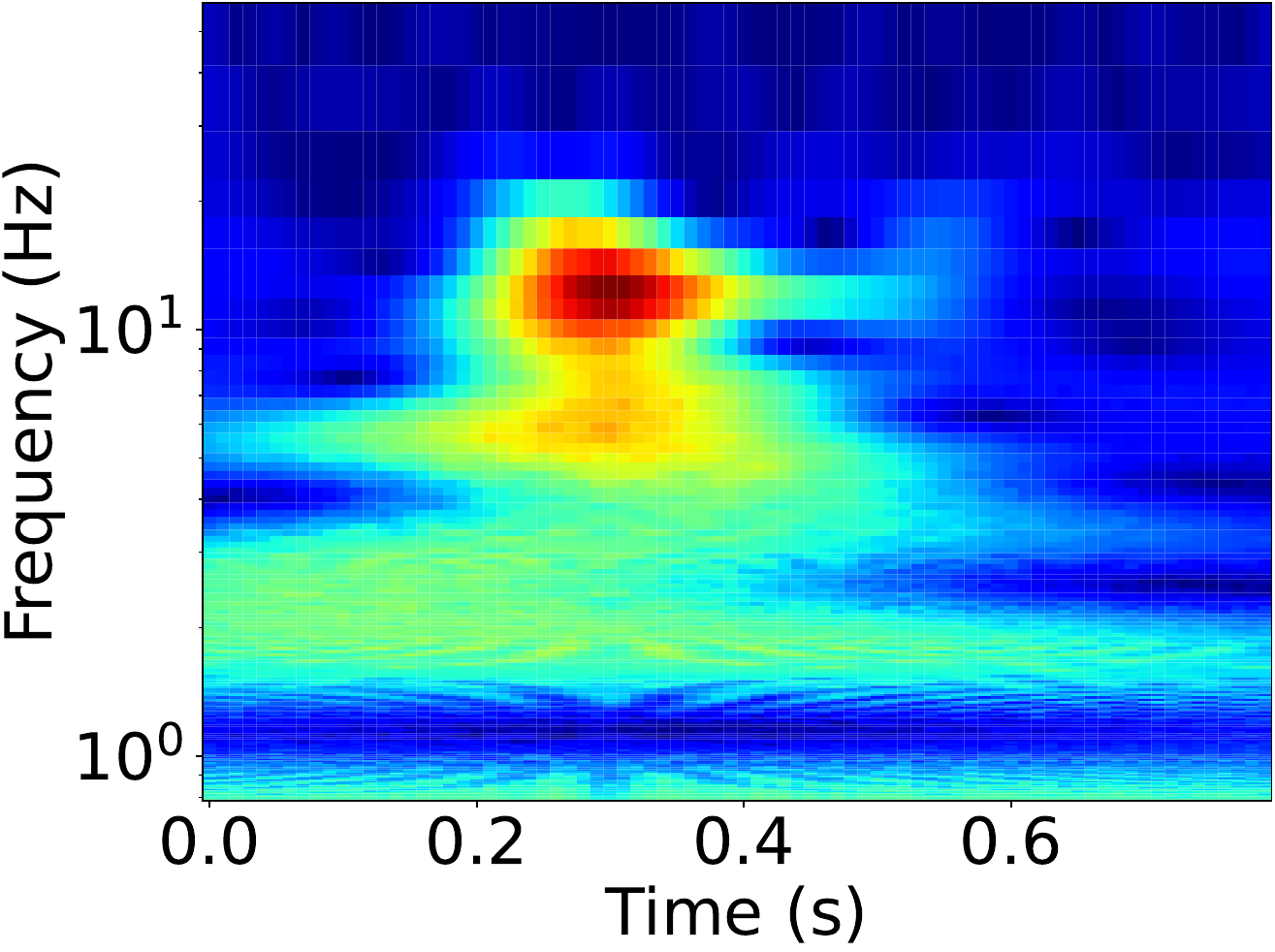}}
    \subfigure[User 3 at session 2]{\label{fig:user3_2}
    \includegraphics[width = 0.24\textwidth]{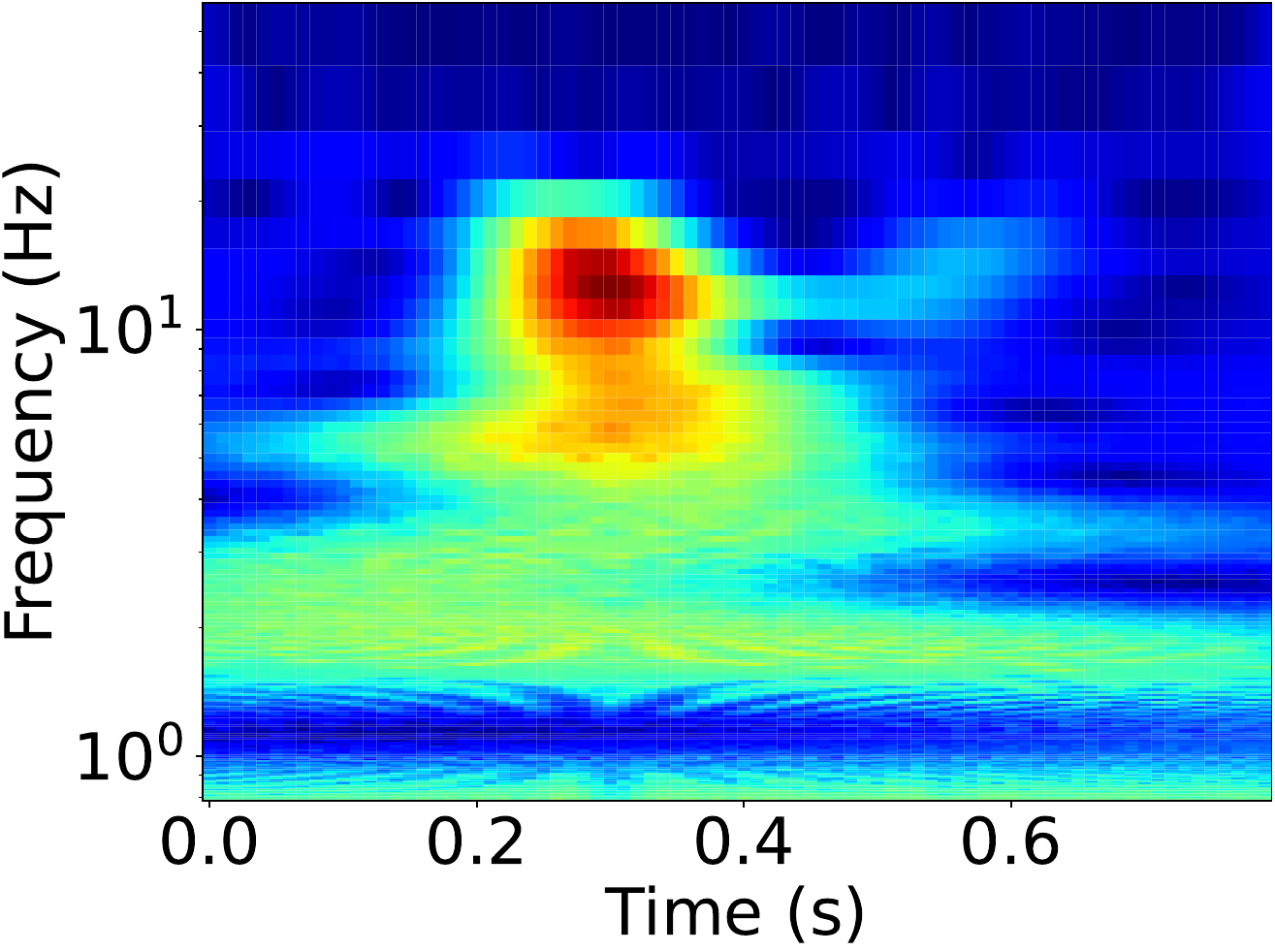}}
    \caption{CWT representations of BCG samples from different users and from the same user across sessions.}
    \label{fig:band}
\end{figure*}


\subsection{Feasibility of Utilizing the BCG Signal}

Leveraging the high sensitivity of in-ear accelerometers, \sys{} effectively captures BCG signals driven by cardiac mechanical activity.
We conduct a preliminary three-user study to explore its feasibility for biometric authentication from two aspects.

First, each participant is  asked to lie still with in-ear accelerometers, and a 1-second BCG segment is subsequently recorded.
To analyze the signal characteristics, we apply the continuous wavelet transform (CWT) \cite{aguiar2014continuous}, enabling time–frequency visualization of the extracted BCG waveforms.
As illustrated in \figref{fig:user1} to \figref{fig:user3_1}, each subplot displays the CWT spectrogram corresponding to a different user.
The results exhibit noticeable inter-user differences in both frequency composition and temporal energy distribution, indicating that in-ear BCG signals contain user-specific characteristics that can potentially support biometric authentication.

Second,  to further assess intra-user consistency, we collect BCG signals from one participant at three separate time points under consistent wearing conditions. The corresponding CWT spectrograms are presented in \figref{fig:user3_1}  and  \figref{fig:user3_2}. 
As illustrated, the time–frequency patterns remain highly consistent across sessions, demonstrating the temporal stability of BCG signals within the same individual.

\subsection{Waveform and Fiducial Variability Analysis}
\label{sec:variability_analysis}

To further understand the stability of the in-ear BCG signal under different conditions, we conduct a detailed variability analysis on both the cardiac waveform and the fiducial-point timing. 
We quantify waveform variability using Pearson correlation, and measure fiducial-point timing variability using the standard deviation of representative temporal intervals, including J--H, J--I, and J--L. 
Specifically, we consider four settings: \textit{intra-session}, \textit{inter-session with remounting}, \textit{inter-user}, and \textit{inter-device}.

For the \textit{intra-session} setting, we collect 200 four-second samples from a single participant under a fixed wearing condition. 
For the \textit{inter-session} setting, we ask five participants to repeat the earphone remounting process 100 times each, in order to explicitly evaluate the effect of remounting. 
For the \textit{inter-user} setting, we analyze the data of all 33 participants collected using the same device. 
For the \textit{inter-device} setting, we collect data from 10 participants using both our 3D-printed earphone and AirPods.

\tabref{tab:pearson} and \tabref{tab:std}  summarize the corresponding results. 
We observe that, for the same user, both the waveform similarity and fiducial-point stability remain high in the \textit{intra-session} and \textit{inter-session} settings, indicating that the in-ear BCG signal is stable over time and remains largely preserved after remounting. In contrast, the \textit{inter-user} setting shows substantially lower waveform similarity and larger timing variability, suggesting that different users exhibit sufficiently distinct cardiac patterns for authentication.

Interestingly, the \textit{inter-device} setting still shows a moderate degree of waveform similarity together with relatively small fiducial-point timing variation. 
A plausible explanation is that the dominant morphology and temporal structure of the signal are primarily determined by the same underlying cardiac mechanical activity, and are therefore preserved across devices despite hardware differences. 
This result suggests that the BCG-based representation learned by \sys{} may retain a certain degree of cross-device transferability, although more comprehensive multi-device validation is still needed in future work.

\begin{table}[t]
\centering
\caption{Cardiac waveform variability under different experimental conditions.}
\label{tab:pearson}
\begin{tabular}{ccc@{\hspace{10pt}}c@{\hspace{10pt}}c@{\hspace{10pt}}c@{\hspace{10pt}}}
\hline
Pearson correlation coefficient & \makecell{Intra-session} & \makecell{Inter-session} & \makecell{Inter-user} & \makecell{Inter-device} \\ \hline
mean ± std & 0.83 ± 0.11 & 0.80 ± 0.09  & 0.02 ± 0.31  & 0.73 ± 0.11  \\
\hline
\end{tabular}
\end{table}

\begin{table}[t]
\centering
\caption{Fiducial point timing variability under different experimental conditions.}
\label{tab:std}
\begin{tabular}{ccc@{\hspace{10pt}}c@{\hspace{10pt}}c@{\hspace{10pt}}c@{\hspace{10pt}}}
\hline
 Fiducial-point interval& \makecell{Intra-session} & \makecell{Inter-session} & \makecell{Inter-user} & \makecell{Inter-device} \\ \hline
  J-H& 6.3& 8.6  & 22.0  & 30.1  \\
   J-I& 5.0& 7.2  & 18.1  & 20.2  \\
    J-L& 4.7& 7.9  & 20.7  & 21.5  \\
\hline
\end{tabular}
\end{table}

%% file: latex/view.tex
\presec 
\section{System Overview} \label{sec:overview} 
\postsec

The architecture of \sys{} is illustrated in Fig.~\ref{fig:overview}. \sys{} is designed to authenticate users using BCG signals collected from in-ear accelerometers. The system comprises three core modules: \textit{Signal Denoising}, \textit{Feature Extraction}, and \textit{Registration \& Authentication}.
First, \sys{} performs two-stage denoising to suppress motion-induced artifacts. It applies wavelet-based denoising and hyperbolic thresholding refinement to enhance the cardiac components. To further improve robustness, a lightweight periodicity-based detector identifies sporadic motion segments, which are subsequently refined using RLS-based filtering.
Next, the refined signals are passed to the user-specific feature extraction module, \HIDNET{}, which employs adversarial training and orthogonal regularization to suppress nuisance features and preserve only user-related characteristics. 
Finally, we propose a user-agnostic authentication paradigm that enables user authentication without the need for complex model labeling or retraining, which are typically required in conventional authentication systems. 
This is achieved by first applying a Siamese-based feature embedding that maps both registration and query samples into a shared feature space. When a new user is enrolled, the system computes a feature center and determines a decision threshold based on the embedded features. During authentication, the query features are embedded using the same Siamese network, and their Euclidean distance to the user’s feature center is calculated to determine authentication validity.



%% file: latex/model.tex
\section{Signal Denoising}
\subsection{Inherent Interference Denoising}
Inherent interference refers to system noise and persistent, involuntary  movements during heartbeat monitoring, such as natural head swaying, respiration-induced torso motion.
As illustrated in \figref{fig:raw}, despite the volunteer remaining seated, the heartbeat pattern is noticeably affected by inherent interference.
In this subsection, we focus on extract cardiac components from in-ear BCG signals affected by inherent interference.

\begin{figure}[t]
\begin{center}
\includegraphics [width = 0.7\linewidth]{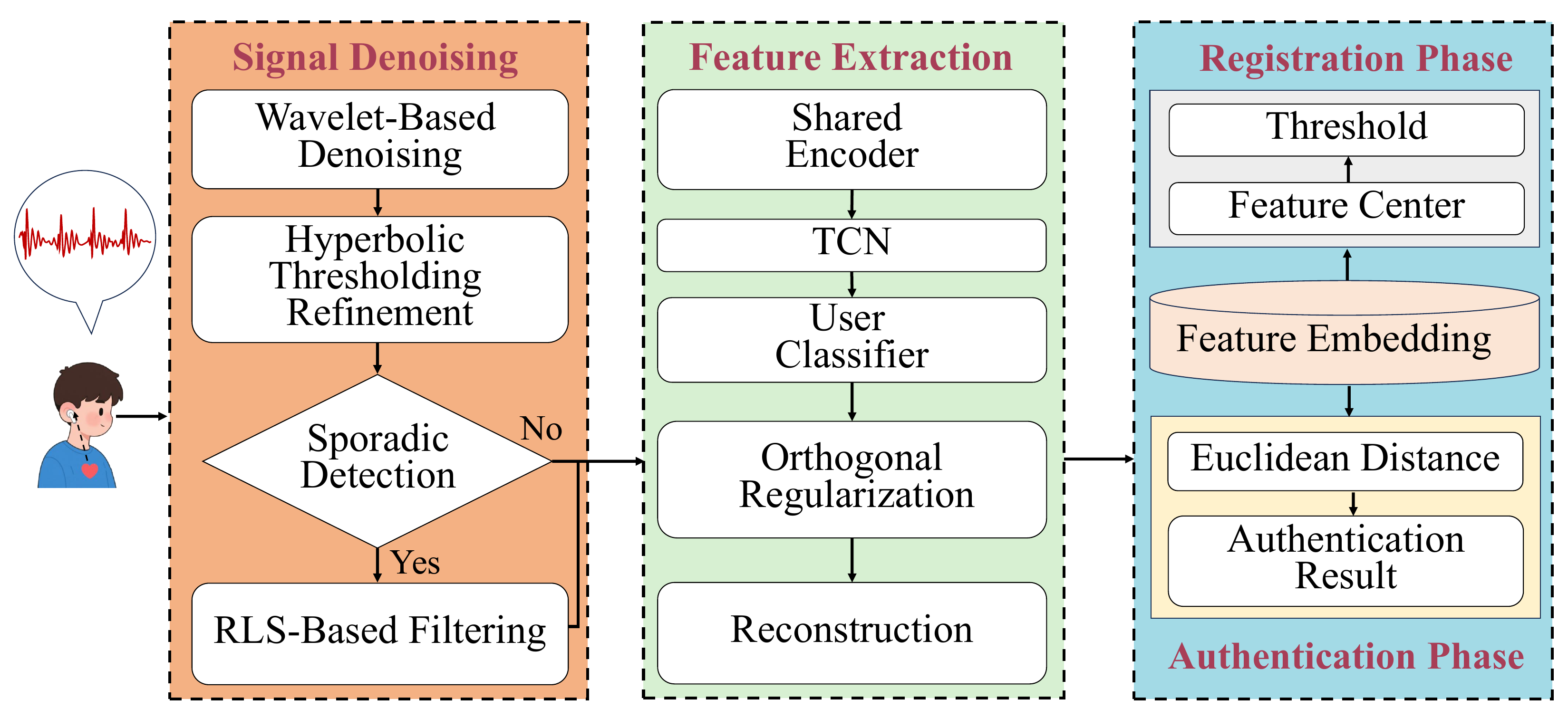}
\caption{The architecture of \sys{}.}
\label{fig:overview}
\end{center}
\end{figure}

\subsubsection{Wavelet-Based Denoising.}
Due to the ability to prevent mode mixing and maintain shift invariance in the time domain \cite{kumar2021stationary, madan2022denoising}, we select  Stationary Wavelet Transform (SWT) to extract cardiac components from contaminated signal.
We first select the Discrete Meyer wavelet as the analyzing wavelet  \cite{frikha2021source} to decompose the earphone BCG signals into six levels.
At a sampling rate of 100 Hz, the six decomposition levels correspond to the frequency range of 25–50 Hz, 12.5–25 Hz, 6.25–12.5 Hz, 3.13–6.25 Hz, 1.56–3.13 Hz, and 0.78–1.56 Hz, respectively.
Then, to suppress high-frequency noise caused by device imperfections, we discard the components at the first level.
Moreover, this process helps eliminate motion interference with frequencies below 0.78 Hz. 
For example, respiratory artifacts, which typically fall within the 0.2–0.33 Hz range \cite{bhatawadekar2013study}, can be effectively reduced.
We then reconstruct the BCG signal via inverse SWT (see \figref{fig:swt}), effectively reducing most motion artifacts and enhancing heartbeat clarity.

\subsubsection{ Hyperbolic Thresholding Refinement.}
While the wavelet-based approach effectively suppresses the majority of motion-induced artifacts, minor residual system noise may persist in the wavelet coefficient sequences across different decomposition levels.
%
%
%
To address this issue and further improve the signal quality, an additional refinement step is introduced for signal purification.
Based on the fact that the system noise is normally distributed, we use the hyperbolic thresholding algorithm \cite{he2015new,xavier2010hyperbolic} to design a new denoising scheme to further facilitate the extraction of clear vibrations of heartbeat. 
It is capable of reconstructing unknown waveforms from data characterized by independent and identically distributed Gaussian noise.
Our hyperbolic thresholding based refinement scheme undergoes the following two steps.
First, an adaptive threshold is computed for each level of the wavelet decomposition to capture the frequency-dependent properties of both the signal and the noise, which is typically defined as:
$T_j = \sigma_j \cdot \sqrt{2 \log n_j}$,
where $n_j$ is the length of wavelet coefficient sequence and $\sigma_j$ is the estimated noise standard deviation $\sigma_j = \frac{\mathrm{median}(|w_j|)}{0.6745}$, where $w_j$ denotes the detail coefficients at level $j$.
This strategy allows the threshold to adapt to the noise characteristics at each scale, improving the effectiveness of denoising while preserving essential signal features.

\begin{figure*}[t]
\centering
\begin{minipage}[t]{0.325\textwidth}
\centering
\includegraphics[width=1\textwidth]{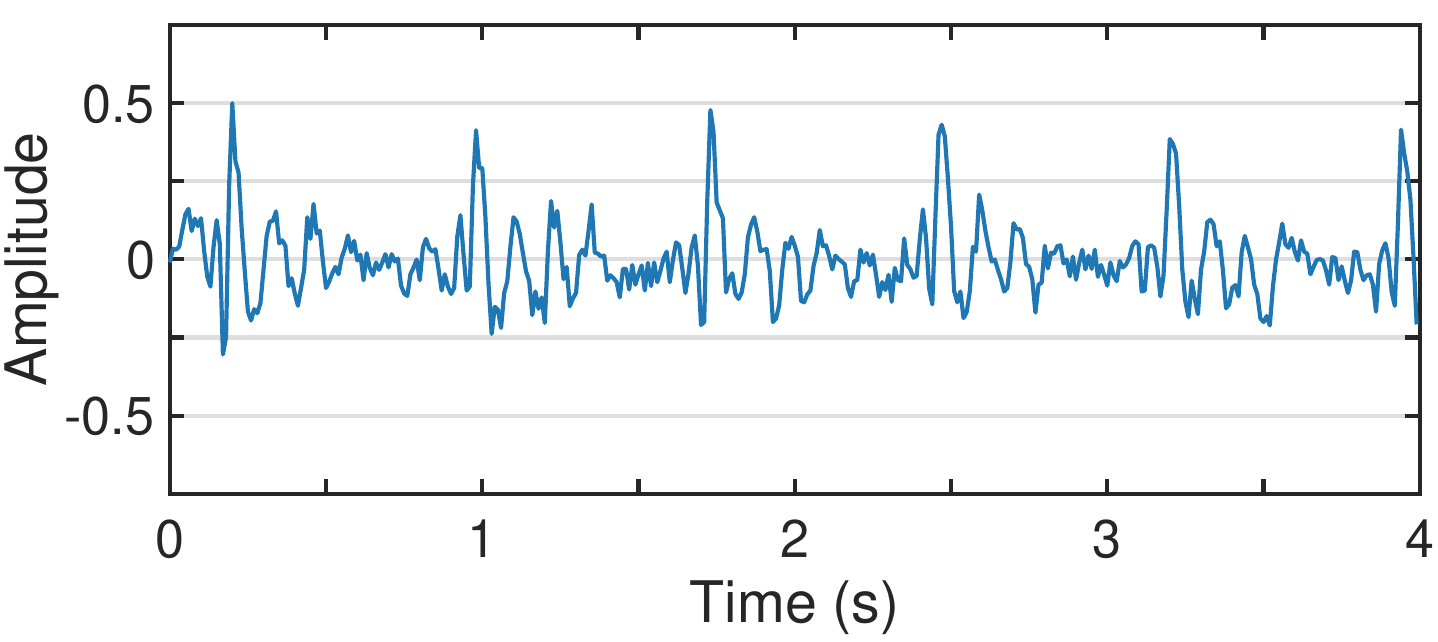}
\caption{Raw signal.}
\label{fig:raw}
\end{minipage}
\hfill
\begin{minipage}[t]{0.325\textwidth}
\centering
\includegraphics[width=1\textwidth]{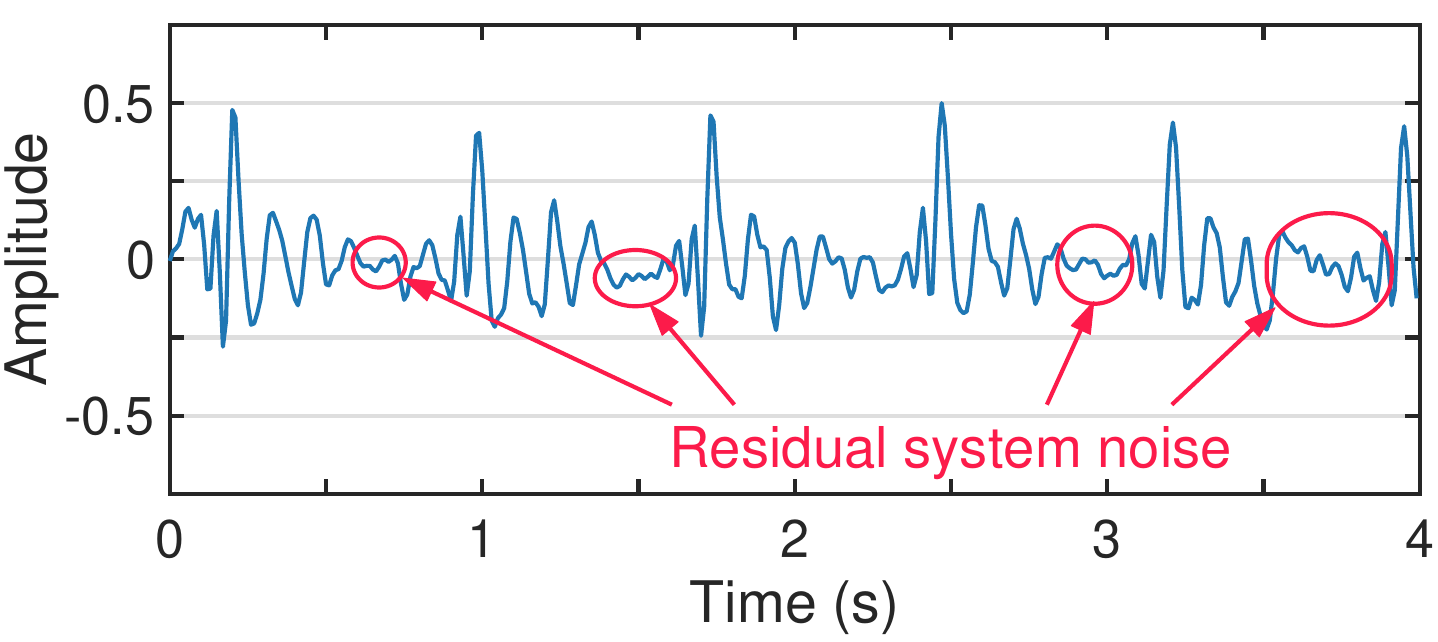}
\caption{SWT-based denoising.}
\label{fig:swt}
\end{minipage}
\hfill
\begin{minipage}[t]{0.325\textwidth}
\centering
\includegraphics[width=1\textwidth]{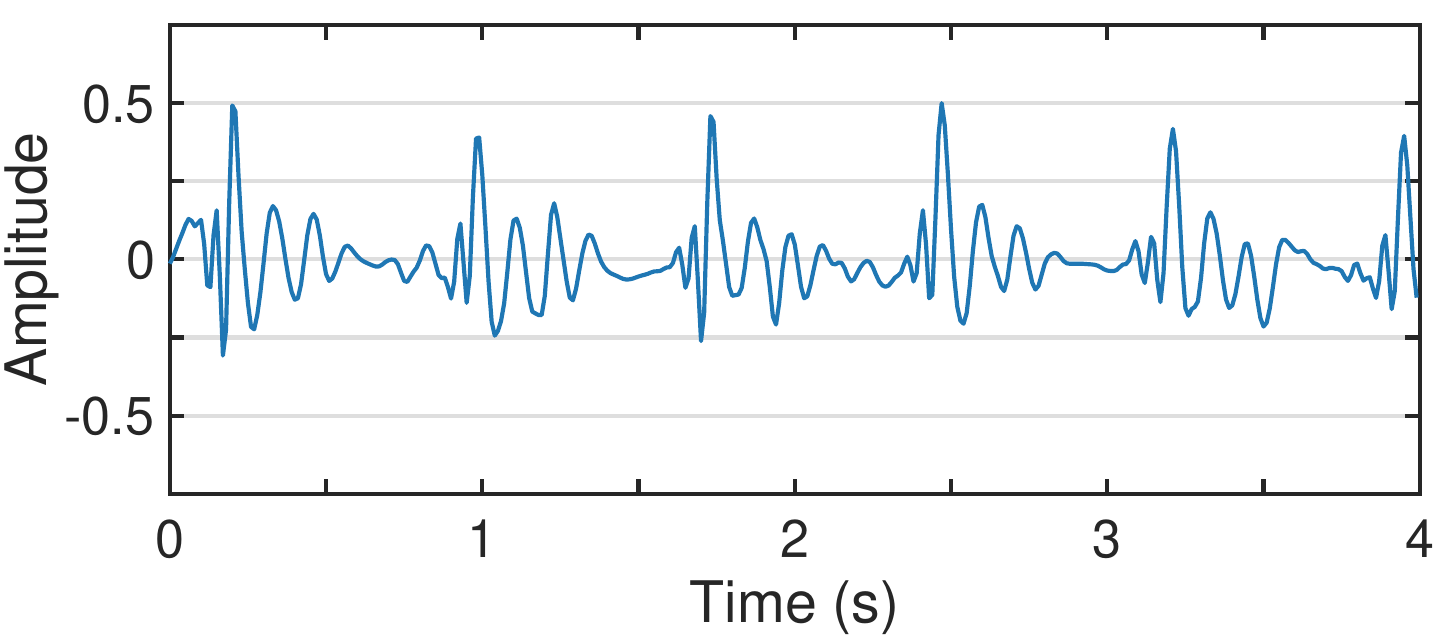}
\caption{Hyperbolic thresholding.}
\label{fig:hyper}
\end{minipage}
\end{figure*}

Second, a hyperbolic thresholding function is applied to the corresponding detail coefficients $w_j$, which is defined as:
\begin{equation}
\hat{w}_j = w_j \cdot \left(1 - e^{-\left( \frac{|w_j|}{T_j} \right)^p} \right),
\end{equation}
where $\hat{w}_j$ is the thresholded coefficient, and $p$ is a shape parameter (typically between 1 and 2) that controls the smoothness of the thresholding function. 
It provides a continuous and gradual suppression of low-amplitude coefficients, while effectively preserving significant high-amplitude signal components.
%
%
We then apply the inverse SWT to the refined data, which significantly reduces random noise and reveals a clearer heartbeat pattern, as shown in \figref{fig:hyper}.

\subsection{Sporadic Interference Denoising}
Although effective in attenuating inherent motion artifacts, the wavelet-based denoising method remains insufficient for handling sporadic disturbances, such as head shaking and blinking, as shown in \figref{fig:period}.
Therefore, it is necessary to resort to more robust noise filtering methods.
This subsection begins by identifying sporadic event.
Subsequently, we propose an RLS-based scheme to further enhance noise reduction.

\subsubsection{Sporadic Event Detection.}

We propose a lightweight approach to assess signal periodicity based on peak interval and amplitude stability, suitable for heartbeat signals with recurring patterns.
First, we normalize the signal as $x(n) \leftarrow \frac{x(n) - \mu_x}{\sigma_x}$, where $\mu_x$ and $\sigma_x$ are the mean and standard deviation of the refined signal.
Second, we detect peaks under physiological constraints. Using $HR_{\text{max}} = 120$ bpm, the minimum interval is $d_{\text{min}} = \left\lfloor \frac{f_s \cdot 60}{HR_{\text{max}}} \right\rfloor$. Peaks are retained only if their prominence exceeds 0.3 and spacing exceeds $d_{\text{min}}$ \cite{findpeaks_matlab}.
Third, we assess periodicity. If fewer than three peaks are found, the signal is labeled non-periodic. Otherwise, we compute the coefficient of variation (CV) of inter-peak intervals and amplitudes as $\text{CV}_{\text{interval}} = \frac{\text{std}(t_{i+1} - t_i)}{\text{mean}(t_{i+1} - t_i)}$ and $\text{CV}_{\text{amp}} = \frac{\text{std}(a_i)}{\text{mean}(a_i)}$, where $\{t_i\}$ and $\{a_i\}$ denote peak times and amplitudes. The signal is considered periodic if both metrics are below thresholds: $\text{CV}_{\text{interval}} < 0.1$ and $\text{CV}_{\text{amp}} < 0.2$.
%
Finally, as shown in \figref{fig:period}, we obtain $\text{CV}_{\text{interval}} = 0.08$ and $\text{CV}_{\text{amp}} = 1.02$, respectively. As the periodicity criteria are not jointly satisfied, the signal is classified as non-periodic and passed to the RLS-based denoising stage.

\subsubsection{RLS Filtering with Motion-Sensitive Forgetting Factor.}

To extract heartbeat pattern from measurements contaminated by interference, we employ an RLS filtering framework with a dynamically tuned forgetting factor. 
This adaptive mechanism improves the filter's responsiveness to abrupt motion artifacts while maintaining stability under clean conditions.
Note that the time series can serve as a temporal reference signal when periodicity is satisfied.

In traditional RLS algorithms, the forgetting factor remains fixed throughout. However, in our method, it is modulated in real time based on a fluctuation ratio between the input and a baseline reference. Let $\sigma_x(t)$ denote the local standard deviation of the input signal, and $\sigma_r(t)$ the deviation of a motion-suppressed reference (e.g., a smoothed or prior segment). The ratio $\Psi(t) = \frac{\sigma_x(t)}{\sigma_r(t)}$ is used as an indicator of motion intensity. When $\Psi(t)$ increases, it reflects higher variability in the input, likely due to motion artifacts.

We define the forgetting factor $\alpha_t$ as a monotonically decreasing function of $\Psi(t)$:
\begin{equation}
\alpha_t = \exp\left(-\kappa \cdot \Psi(t)\right),
\label{eq:forgetting_factor}
\end{equation}
where $\kappa > 0$ is a sensitivity parameter that determines how aggressively the forgetting factor responds to motion. A higher value of $\kappa$ leads to faster decay of $\alpha_t$ under noisy conditions.
This formulation ensures that when the input is stable (\ie, $\Psi(t)$ close to 1), the forgetting factor $\alpha_t$ remains close to 1, preserving long-term memory. In contrast, during motion spikes where $\Psi(t)$ increases significantly, $\alpha_t$ decreases exponentially, enabling faster adaptation to changing conditions.
Once $\alpha_t$ is determined, the RLS filter proceeds with its standard recursive updates, including gain computation, inverse correlation matrix adjustment, and weight refinement. These steps are omitted here for brevity.

\begin{figure*}[t]
\centering
\begin{minipage}[t]{0.44\textwidth}
\centering
\includegraphics[width=1\textwidth]{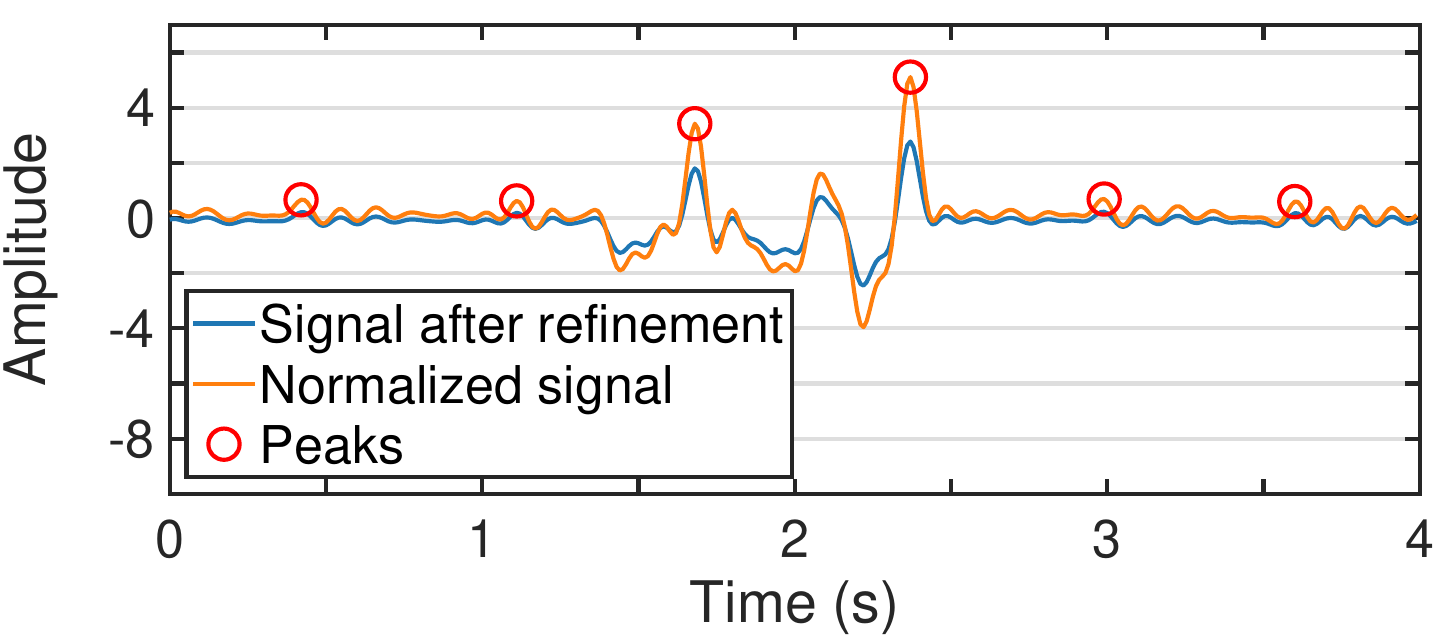}
\caption{Sporadic event detection via periodicity analysis.}
\label{fig:period}
\end{minipage}
\hspace{0.2in}
\begin{minipage}[t]{0.44\textwidth}
\centering
\includegraphics[width=1\textwidth]{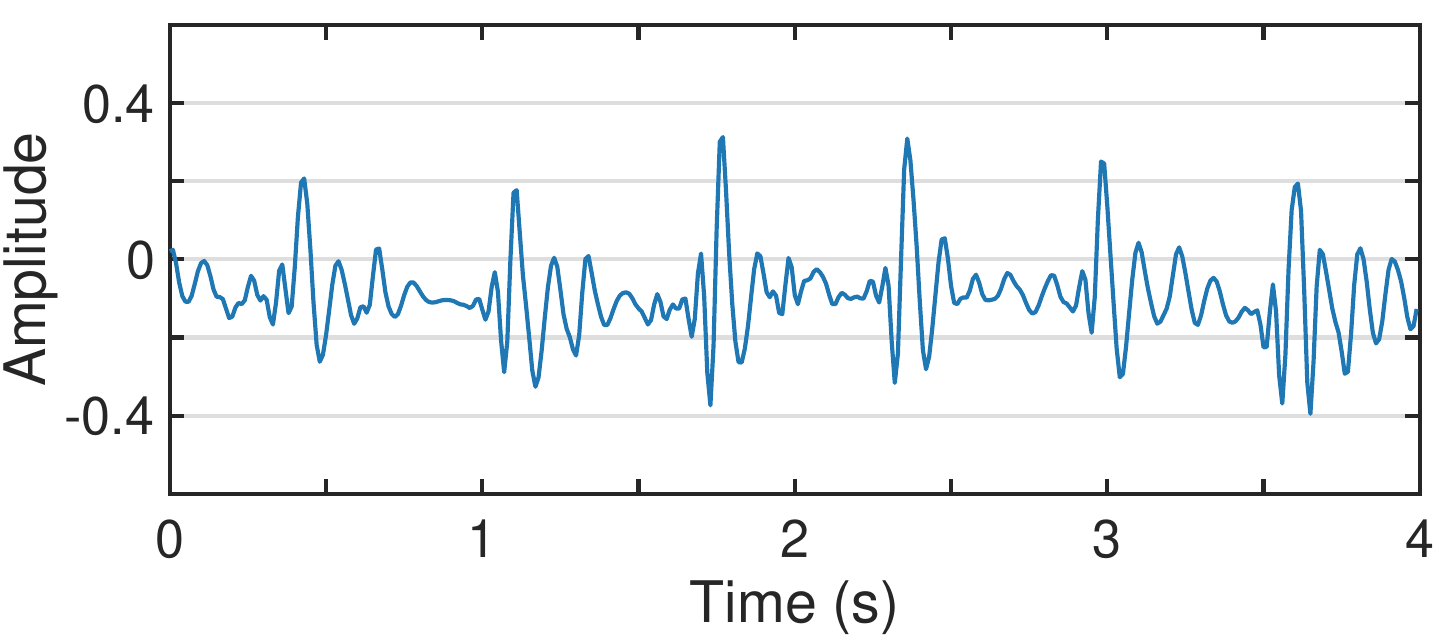}
\caption{Signal recovery using RLS filtering.}
\label{fig:rls}
\end{minipage}
\end{figure*}

This adaptive forgetting strategy makes the filter particularly effective in suppressing transient motion artifacts while preserving underlying heartbeat signals in real-world scenarios.
As shown in  \figref{fig:rls}, the heartbeat waveform becomes significantly more distinguishable after suppressing motion artifacts.



%% file: latex/hidnet.tex

\section{ User-Specific Feature Disentanglement}

\subsection{Overall architecture}

Although most noise can be mitigated through the preceding processing steps, user-related features and identity-irrelevant components remain highly entangled in BCG signals. On the one hand, user-specific features capture fine-grained, individual-dependent physiological characteristics, such as heartbeat morphology, temporal structure, and waveform dynamics. These features exhibit relative stability across different recording sessions and constitute the basis for reliable user identification. 
On the other hand, BCG signals inherently contain a substantial amount of shared physiological patterns determined by common cardiac mechanisms.
%
%
Because these shared physiological patterns often dominate BCG signals, subtle yet discriminative identity-related variations can be easily obscured, making it difficult for models to focus on truly user-specific characteristics. 
Traditional learning-based models  (\eg, CNN or Conformer) in the absence of explicit structural constraints, often struggle to effectively distinguish identity-related features from shared physiological patterns, thereby limiting their robustness under cross-session and complex real-world conditions.

\begin{figure}[t]
\begin{center}
\includegraphics [width =1\linewidth]{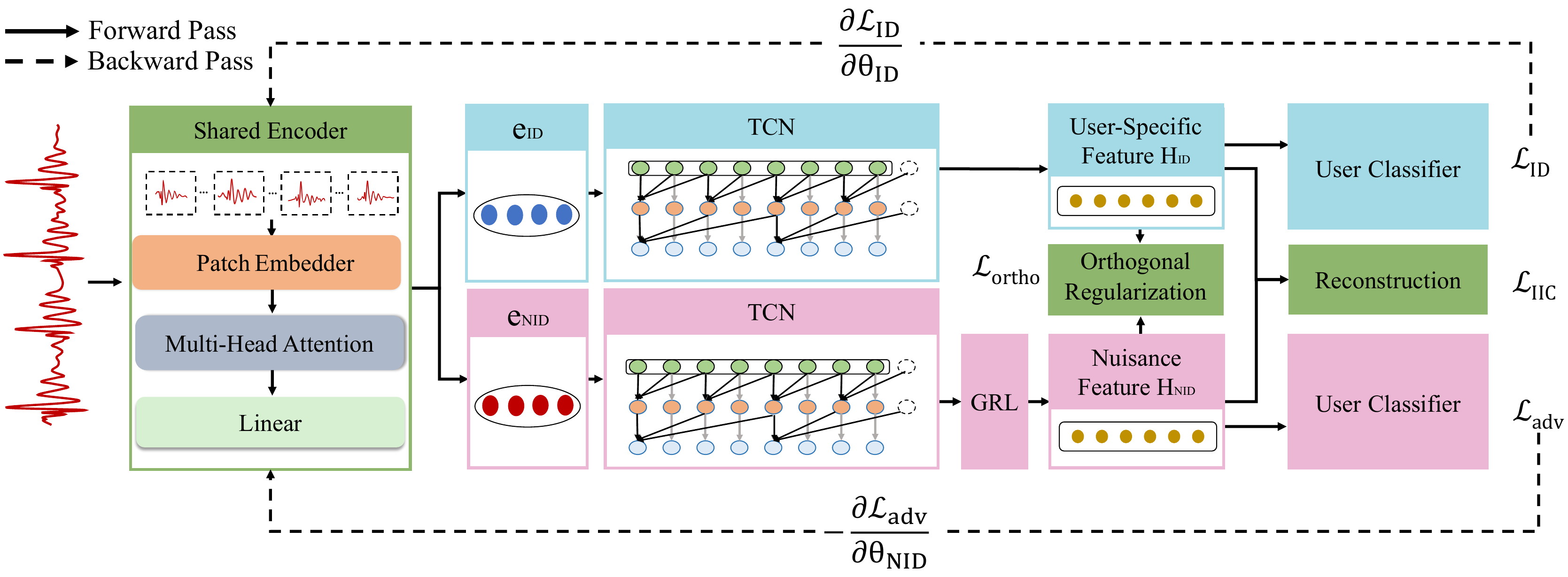}
\caption{HIDNet Architecture}
\label{fig:framework}
\end{center}
\vspace{-0.2in}
\end{figure}

To address this issue, we propose a novel user-specific feature extraction model, \HIDNET{}, which is designed to structurally disentangle identity-relevant information from identity-irrelevant factors. This disentanglement facilitates the extraction of highly discriminative and robust features for user authentication, while effectively mitigating variability introduced by nuisance factors.
As shown in \figref{fig:framework}, \HIDNET{} consists of five key components:
\begin{itemize}[leftmargin=*]
    \item \textbf{Shared encoder} encodes the input BCG segment into a hidden representation. 
    \item \textbf{Identity-relevant branch} extracts identity-related information through a subject identification task. 
    \item \textbf{Identity-irrelevant branch} models nuisance factors and suppresses their impact with adversarial training. 
    \item \textbf{Information orthogonal regularization} enforces independence between the identity-relevant and identity-irrelevant features. 
    \item \textbf{Information integrity commitment} reinforcing that the identity-relevant representation preserves all critical attributes required for robust authentication and generalization. 
\end{itemize}

\subsection{Shared Encoder for Denoised BCG Signal Embedding}
\label{subsec:shared_encoder}

The shared encoder of \HIDNET{} is designed to extract a compact representation from the denoised BCG signal and explicitly separate it into two components: one for the \textbf{identity-relevant} branch and the other for the \textbf{identity-irrelevant} branch. This disentangled design enables independent processing of subject-specific and nuisance information while maintaining a unified representation that retains essential features for both tasks.

Let the input denoised BCG signal be denoted as \( \mathbf{X} \in \mathbb{R}^T \), where \( T \) is the temporal length. The signal is first segmented into non-overlapping patches of fixed length \( L \), resulting in \( N = \left\lfloor \frac{T}{L} \right\rfloor \) patches:
\[
\mathbf{x}_p = \left\{ \mathbf{x}_1, \mathbf{x}_2, \dots, \mathbf{x}_N \right\}, \quad \mathbf{x}_i \in \mathbb{R}^L.
\]
Each patch \( \mathbf{x}_i \) is projected into a latent space via a linear patch embedder:
\[
\mathbf{z}_i = \mathbf{W}_p \mathbf{x}_i + \mathbf{b}_p,
\]
producing a sequence of embeddings \( \mathbf{Z} = [\mathbf{z}_1, \dots, \mathbf{z}_N]^\top \in \mathbb{R}^{N \times d} \). These embeddings are processed using a multi-head self-attention module to capture contextual dependencies between patches:
\[
\mathbf{Z}' = \text{MultiHeadAttn}(\mathbf{Z}).
\]

To obtain a fixed-size global representation from the sequence of contextualized patch embeddings, we apply mean pooling across the sequence dimension.
Mean pooling is chosen for its simplicity, parameter-free nature, and empirical effectiveness in various transformer-based models \cite{radford2021learning, reimers2019sentence}. 
The pooled vector \( \bar{\mathbf{z}} \) is then passed through a linear projection to produce the final shared embedding:
\[
\mathbf{e} = \bar{\mathbf{z}} \mathbf{W}_e + \mathbf{b}_e, \quad \mathbf{e} \in \mathbb{R}^{d_e}.
\]

This embedding is explicitly divided into two parts:
\begin{equation}
    \mathbf{e}_\text{ID} = \mathbf{e}[:\text{half}], \quad \mathbf{e}_\text{NID} = \mathbf{e}(\mathbf{X})[\text{half}:].
\end{equation}
The shared embedding is divided into two components, which are used to initialize the identity-relevant and identity-irrelevant branches, respectively. It is important to note that this split itself does not enforce semantic disentanglement. Instead, the two components are subsequently shaped by different learning objectives: the identity-relevant branch is optimized via subject identification, while the identity-irrelevant branch is trained adversarially to suppress identity information. As a result, the separation of identity-related and nuisance features emerges during training rather than being guaranteed by the initial partition.

%

\subsection{Identity-relevant Branch} 
\HIDNET{} extracts identity-relevant information by performing a subject identification task. Specifically, we build a Temporal Convolutional Network (TCN) to process \(\mathbf{e}_\text{ID}\) into identity-relevant features $\mathbf{H}_\text{ID}$ and predict the identity associated with the input signal \(\mathbf{X}\). This is essentially a multi-class classification problem, optimized via a cross-entropy loss:
\begin{equation} 
    \mathcal{L}_{\text{ID}} = - \sum_{i=1}^{N} y_i \log\left( \hat{y}_i \right)
\end{equation}
where \(N\) is the total number of subjects, \(y_i\) denotes the ground-truth one-hot encoded label for subject \(i\), and \(\hat{y}_i\) is the predicted probability for subject \(i\). 
By optimizing the subject identification objective, \HIDNET{} encourages the extracted features $\mathbf{H}_\text{ID}$ to be highly discriminative for identity; however, it may still retain some nuisance factors, which necessitates further mitigation.

\subsection{Identity-irrelevant Branch}  
To extract identify-irrelevant features, \HIDNET{} introduces an adversarial branch based on a Gradient Reversal Layer (GRL). Specifically, the hidden representation $\mathbf{e}_\text{NID}$ is first passed through a nuisance feature extractor, implemented using a TCN, to obtain nuisance features. These features are then input into a classifier tasked with subject identification. Crucially, a GRL is inserted between the nuisance feature extractor and the classifier.
During the forward pass, the GRL acts as an identity function:
\begin{equation}
\mathcal{R}_\lambda(\mathbf{H}_\text{NID}) = \mathbf{H}_\text{NID}.
\end{equation}
However, during the backward pass, the GRL reverses and scales the gradients:
\begin{equation}
\frac{\partial \mathcal{R}_\lambda(\mathbf{H}_\text{NID})}{\partial \mathbf{H}_\text{NID}} = -\lambda \mathbf{I},
\end{equation}
where $\lambda$ is a positive scaling coefficient controlling the strength of gradient reversal, and $\mathbf{I}$ denotes the identity matrix --- a square matrix with ones along the main diagonal and zeros elsewhere. In this context, the identity matrix ensures that each feature dimension is scaled and sign-inverted independently, without introducing cross-dimensional interactions. This design guarantees that the adversarial objective uniformly penalizes all identity-relevant information encoded in the nuisance features.
The adversarial objective is formulated as a standard cross-entropy loss:
\begin{equation}
\mathcal{L}_{\text{adv}} = - \sum_{i=1}^{N} y_i \log\left( \tilde{y}_i \right),
\end{equation}
where $N$ is the total number of subjects, $y_i$ is the ground-truth one-hot label for subject $i$, and $\tilde{y}_i$ is the predicted probability from the adversarial classifier. 

By minimizing $\mathcal{L}_{\text{adv}}$ while applying gradient reversal, the adversarial branch is explicitly encouraged to remove identity-relevant information from the nuisance features, thus achieving disentanglement between subject-relevant and subject-irrelevant components.

\subsection{Information Orthogonal Regularization}
To further improve the disentanglement of identity-relevant and identity-irrelevant features, we introduce the Information Orthogonal Regularization (IOR) component. This regularization ensures that the information captured by the identity-relevant and identity-irrelevant branches is orthogonal, meaning that they are independent of each other in the learned latent space. This orthogonality not only enforces a clearer separation of features but also promotes more efficient learning by discouraging the overlap of irrelevant information in the identity-relevant representation and vice versa.

Formally, we define the orthogonal regularization $\mathcal{R}_{\text{ortho}}$ as the squared norm of the cosine similarity between the identity-relevant and identity-irrelevant feature vectors, which should be minimized to encourage orthogonality. Specifically, the regularization term is given by:
\begin{equation}
\mathcal{R}_{\text{ortho}} = \rho \left( \frac{\mathbf{H}_\text{ID}^\intercal \mathbf{H}_\text{NID}}{\|\mathbf{H}_\text{ID}\| \|\mathbf{H}_\text{NID}\|} \right)^2,
\end{equation}
where $\mathbf{H}_\text{ID}$ and $\mathbf{H}_\text{NID}$ are the identity-relevant and identity-irrelevant feature vectors, respectively, $\|\mathbf{H}_\text{ID}\|$ and $\|\mathbf{H}_\text{NID}\|$ are the L2 norms of the corresponding feature vectors, and
$\rho$ is a regularization coefficient that controls the strength of the orthogonality constraint.

The term inside the parentheses computes the cosine similarity between the identity-relevant and identity-irrelevant features. A cosine similarity of 1 indicates perfect alignment (i.e., the features are not orthogonal), while 0 indicates perfect orthogonality. Therefore, minimizing this regularization term enforces that the two feature vectors are as orthogonal as possible, ensuring that the identity-relevant information is contained exclusively in $\mathbf{H}_\text{ID}$ and the identity-irrelevant information in $\mathbf{H}_\text{NID}$, thus improving generalization and robustness in the user authentication task.
While center loss is commonly used to enhance intra-class compactness, we adopt a Siamese-based learning (\secref{sec:siamese}), which directly optimizes pairwise relationships in the embedding space and is better suited for user-agnostic authentication scenarios involving unseen users.

\subsection{Information Integrity Commitment}
To enforce that the identity-relevant features are completely preserved, we introduce the Information Integrity Commitment (IIC) component. Specifically, the IIC aims to reconstruct the input denoised BCG signal by processing both the identity-relevant and identity-irrelevant features. This process begins by concatenating the identity-relevant and identity-irrelevant features:
$\mathbf{h}_{\text{concat}} = [\mathbf{H}_\text{ID}; \mathbf{H}_\text{NID}]$, which is then fed into an MLP-based decoder, responsible for mapping the concatenated features back to the original BCG signal space. The output of the decoder is denoted as $\hat{\mathbf{X}}$. The reconstruction is then defined as the L2 norm (mean squared error) between the original BCG signal $\mathbf{X}$ and the reconstructed signal $\hat{\mathbf{X}}$:
\begin{equation}
\mathcal{L}_{\text{IIC}} = \|\mathbf{X} - \hat{\mathbf{X}}\|_2^2.
\end{equation}
The reconstruction objective encourages the combined representation to retain sufficient information to recover the original signal, thereby reducing the risk of information loss introduced by the decomposition. We note that this constraint does not provide a strict guarantee of lossless representation; rather, it serves as a regularization mechanism that empirically promotes information preservation and stabilizes the disentanglement process.

The overall \HIDNET{} training objective is to minimize a combination of all the above terms:
\begin{equation}
\mathcal{L}_{\text{total}} = \mathcal{L}_{\text{ID}} + \lambda_{\text{adv}} \mathcal{L}_{\text{adv}} + \lambda_{\text{IIC}} \mathcal{L}_{\text{IIC}} + \lambda_{\text{ortho}} \mathcal{R}_{\text{ortho}},
\end{equation}
where \(\lambda_{\text{adv}}\), \(\lambda_{\text{IIC}}\), and \(\lambda_{\text{ortho}}\) control the relative importance of each term.
The network is optimized using a gradient with the GRL applied during backpropagation to enforce adversarial training and disentangle irrelevant features.

\begin{figure*}[t]
    \centering
    \vspace{-0.2in}
    \subfigure[Raw signals]{\label{fig:untrain}
    \includegraphics[width = 0.24\textwidth]{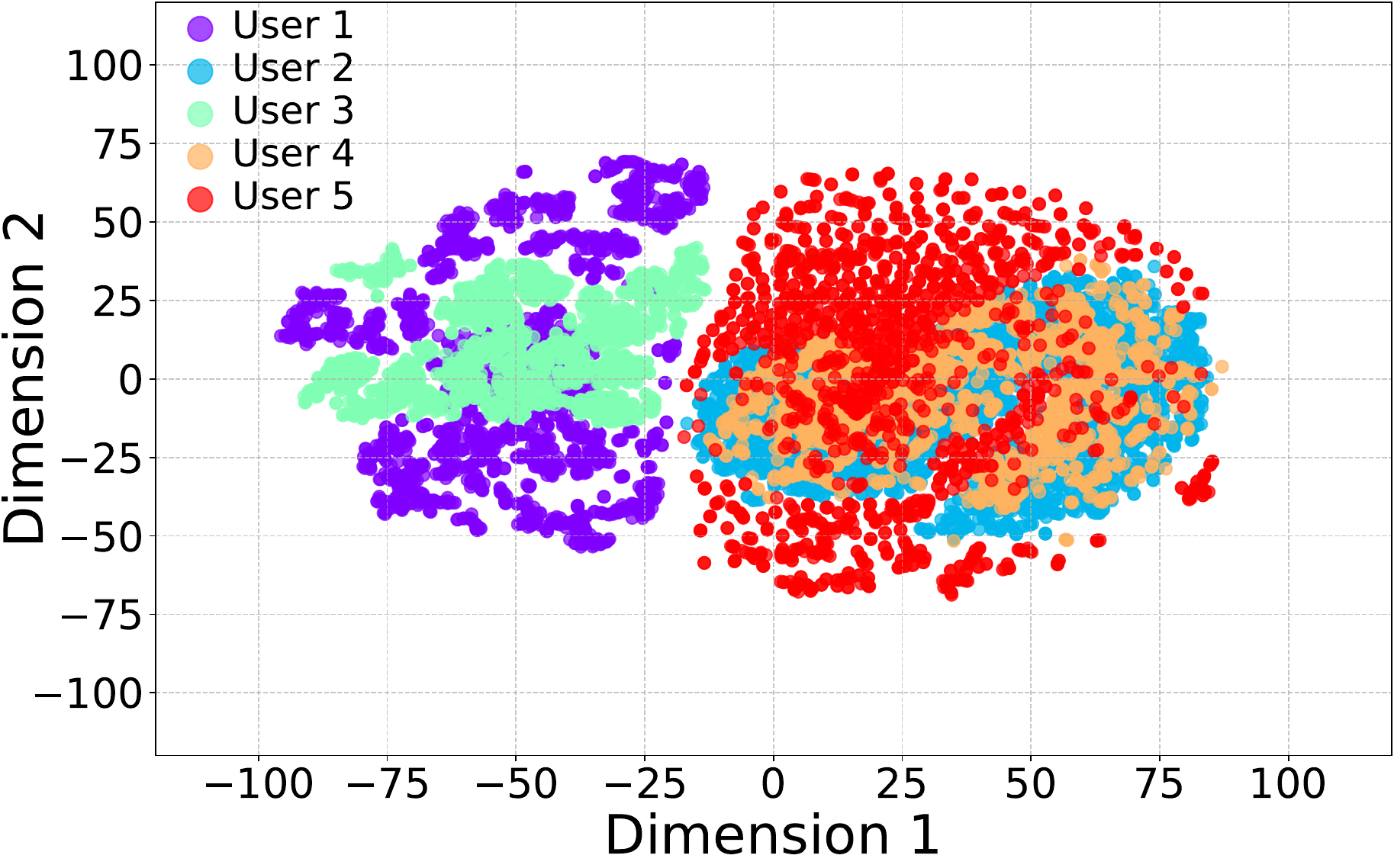}}
    \subfigure[CNN]{\label{fig:cnn}
    \includegraphics[width = 0.234\textwidth]{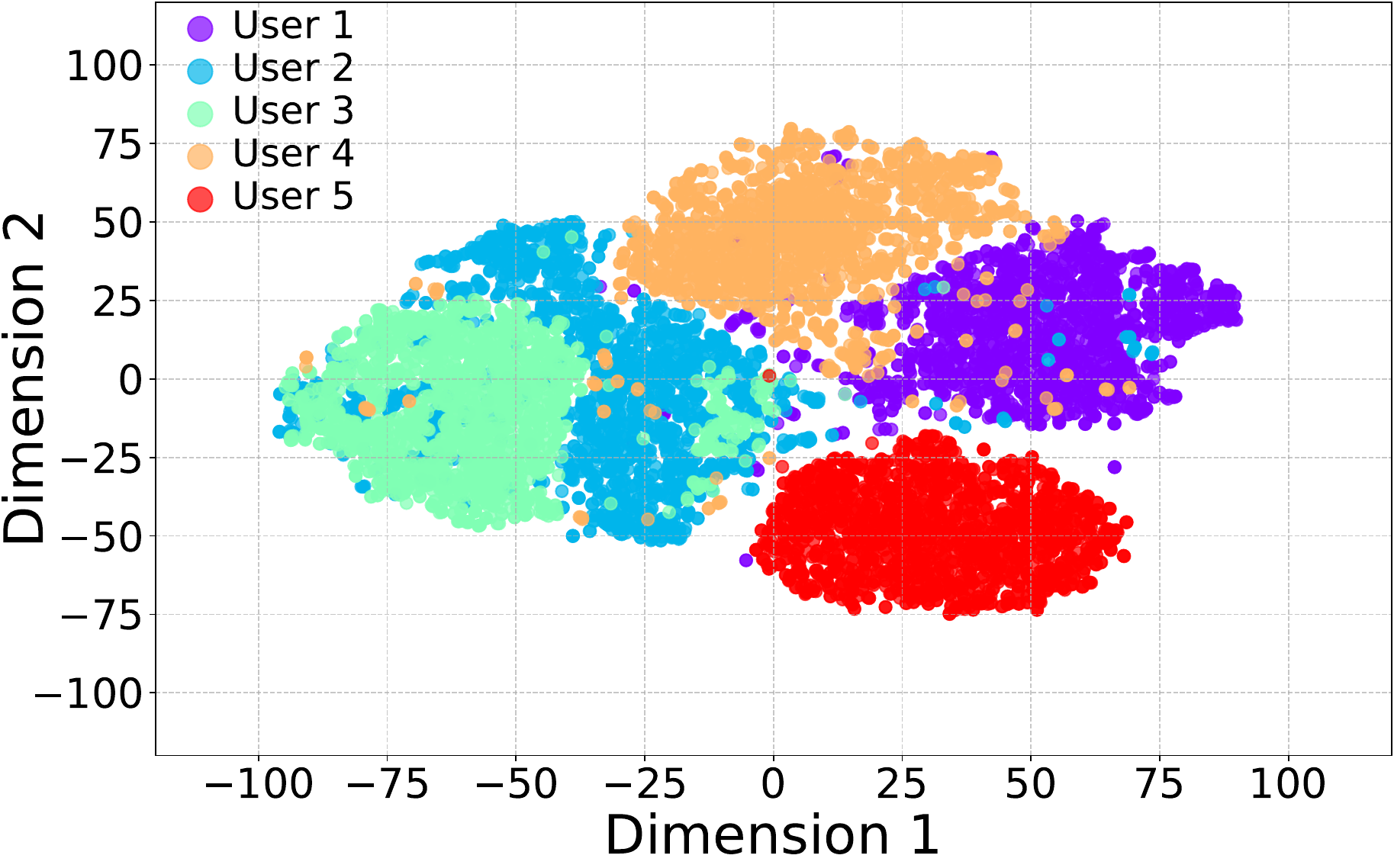}}
    \subfigure[Conformer]{\label{fig:conformer}
    \includegraphics[width = 0.24\textwidth]{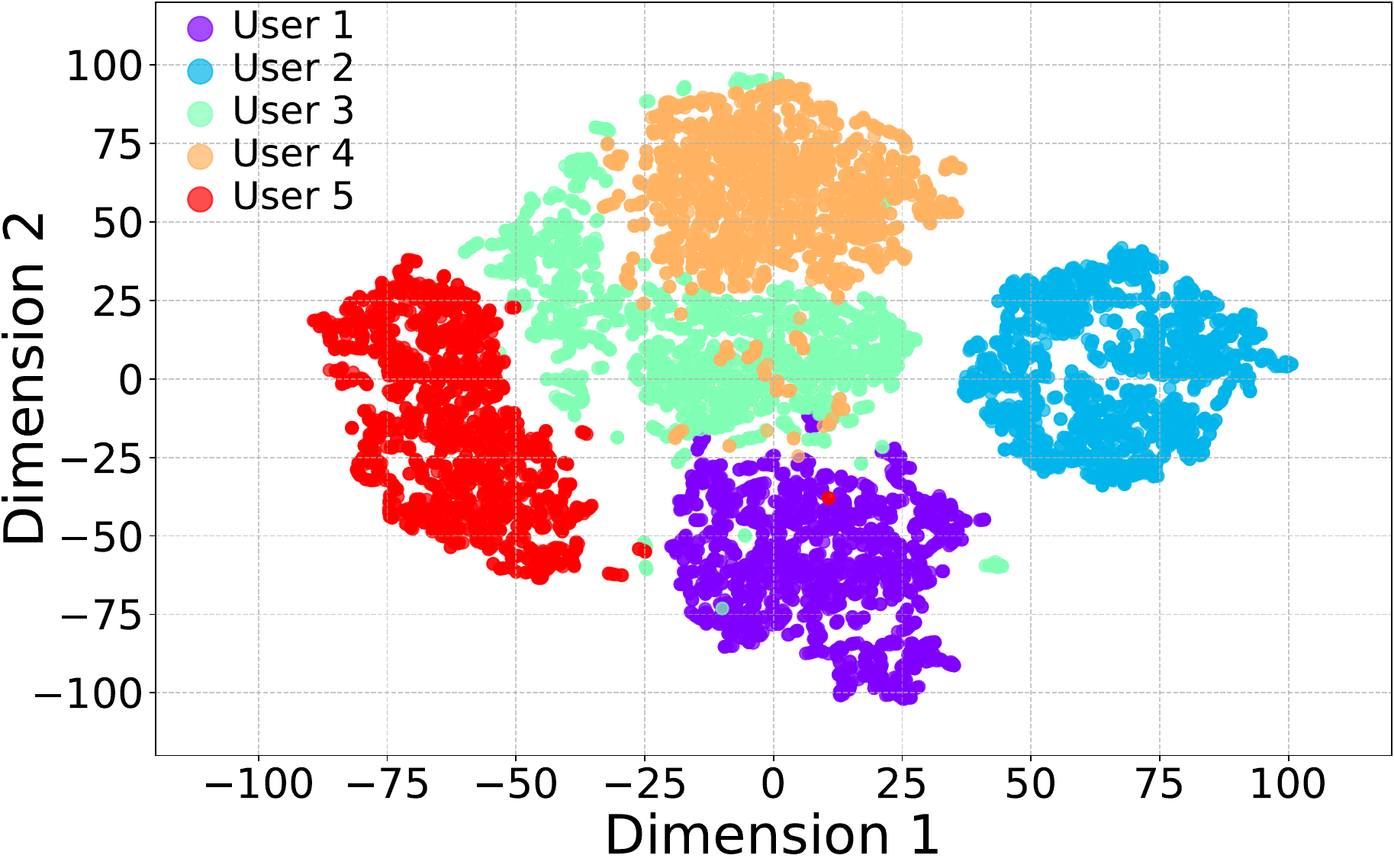}}
    \subfigure[HIDNet]{\label{fig:hidnet}
    \includegraphics[width = 0.24\textwidth]{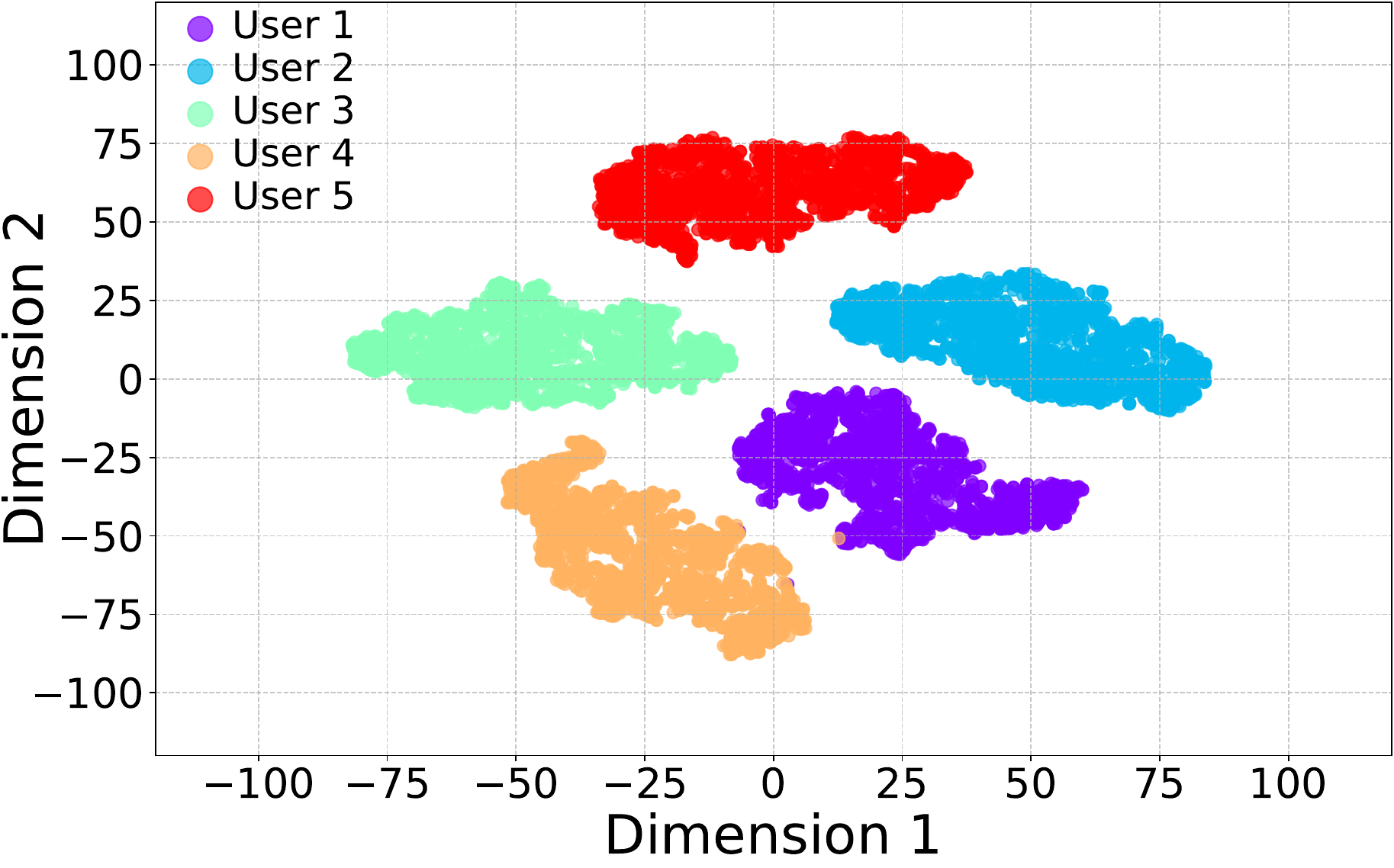}}
    \caption{t-SNE visualization of high-dimensional feature embeddings extracted by different deep learning models for five users.}
    \label{fig:tsne}
\end{figure*}

\figref{fig:tsne} presents t-SNE visualizations of high-dimensional feature embeddings extracted by different models for five users, aiming to qualitatively evaluate the discriminative capability of learned representations for user identification. Each color corresponds to one user, and better identification performance is reflected by higher intra-user compactness and lower inter-user overlap in the embedding space.
The results from the raw signals, CNN, and Conformer (\figref{fig:untrain} to \figref{fig:conformer}) show that although deep models can gradually improve the structure of the embedding space, the learned features still suffer from limited discriminability. In particular, user clusters remain partially overlapped, and intra-user distributions are relatively dispersed, indicating that user-specific characteristics are not sufficiently disentangled from shared or noisy patterns.

In contrast, the embeddings produced by HIDNet (\figref{fig:hidnet} form compact and clearly separated clusters for all users. Samples from the same user exhibit strong intra-class cohesion, while distinct users are well separated with clear margins in the embedding space. This visualization demonstrates that HIDNet is able to suppress user-irrelevant variations and enhance identity-discriminative features, resulting in a feature space that is more suitable for reliable user identification.

\presec
\section{User-Agnostic Authentication}
\postsec
\subsection{Why User-Agnostic Authentication?}
User authentication verifies the identity of a subject requesting resource access, with BCG features providing an unfalsifiable biometric basis.
Given the BCG feature represented as $\bm{X}$, the authentication problem is formulated as follows:
\begin{equation}
    \hat{Y}=f(\bm{X}), \;\text{where}\;\; \hat{Y}=
    \begin{cases}
        1 & \text{if $\bm{X}$ is from the user,} \\
        0 & \text{otherwise}.
    \end{cases}
\end{equation}
This formulation simplifies the authentication task into a binary-class classification problem. 

\begin{algorithm}[t]
\caption{User Registration and Authentication}\label{alg:authen}
\renewcommand{\algorithmicrequire}{\textbf{Input:}}
\renewcommand{\algorithmicensure}{\textbf{Output:}}
\begin{algorithmic}[1]
\REQUIRE Registration BCG: $\bm{X}_{\text{reg}}$; Attacker BCGs: $\{\bm{X}_{\text{imp}}^{(j)}\}_{j=1}^N$; Query BCG: $\bm{X}^q$
\ENSURE Authentication result $\hat{Y}$

\COMMENT{\textbf{--- Registration Phase ---}}
\STATE Segment $\bm{X}_{\text{reg}}$ into $\{\bm{X}_i\}_{i=1}^m$
\STATE $\mathbf{h}_{\text{ID}} = \frac{1}{m} \sum_{i=1}^m f(\HIDNET{}(\bm{X}_i))$
\STATE $D_{\text{gen}} = \{\|f(\HIDNET{}(\bm{X}_i)) - \mathbf{h}_{\text{ID}}\|_2\}_{i=1}^m$

\STATE $D_{\text{imp}} \leftarrow \emptyset$
\FOR{$j = 1$ to $N$}
    \FOR{$\bm{X}_k \in \texttt{Segment}(\bm{X}_{\text{imp}}^{(j)})$}
        \STATE $d = \|f(\HIDNET{}(\bm{X}_k)) - \mathbf{h}_{\text{ID}}\|_2$
        \STATE Append $d$ to $D_{\text{imp}}$
    \ENDFOR
\ENDFOR

\STATE Define $T = [\min(D_{\text{gen}} \cup D_{\text{imp}}), \max(\cdot)]$
\STATE Find $\tau = \arg\min_{t \in T} |\text{FAR}(t) - \text{FRR}(t)|$, where:
\STATE $\quad\text{FAR}(t) = \frac{|\{d \in D_{\text{imp}} : d < t\}|}{|D_{\text{imp}}|}$, 
       $\text{FRR}(t) = \frac{|\{d \in D_{\text{gen}} : d \geq t\}|}{|D_{\text{gen}}|}$

\COMMENT{\textbf{--- Authentication Phase ---}}
\STATE Segment $\bm{X}^q$ into $\{\bm{X}_k^q\}_{k=1}^{s}$ 
\STATE Extract features $\{\mathbf{h}_k^q = f(\HIDNET{}(\bm{X}_k^q))\}_{k=1}^{s}$
\STATE Compute distances $\{D_k = \|\mathbf{h}_k^q - \mathbf{h}_{\text{ID}}\|_2\}_{k=1}^{s}$
\STATE $D = \text{mean}(\{D_k\})$
\STATE $\hat{Y} = \mathbb{I}(D < \tau)$
\end{algorithmic}
\end{algorithm}

However, acquiring BCG features from all potential subjects for training is impractical. 
Moreover, a simplistic binary classifier is susceptible to overfitting by learning subject-specific features from limited training data. 
To address these challenges, we reframe the problem to seek an optimal embedding function $f^*$ that minimizes the distance between embeddings of BCG features from the same subject $\arg\min_{f} D((f(\bm{X}^a), f(\bm{X}^p)))$ while maximizes the distance between different subjects $\arg\max_f D((f(\bm{X}^a), f(\bm{X}^n)))$. 
The objective is formulated as follows:
\begin{equation}\label{eq:def}
    f^*=\arg\min_{f}[D(f(\bm{X}^a),f(\bm{X}^p))-D(f(\bm{X}^a),f(\bm{X}^n))],
\end{equation}
where $D(., .)$ denotes a distance metric, $f(.)$ represents the embedding function, $\bm{X}^a$ denotes a BCG feature from an authenticated subject, $\bm{X}^p$ denotes a BCG feature from the same subject, and $\bm{X}^n$ denotes a BCG feature from a different subject. 
For clarity of presentation, we use the term ``user'' to refer to the subject who is authorized to access a resource, and ``attacker'' to denote an unauthorized subject attempting to gain access through authentication.

\begin{figure}[t]
\begin{center}
\includegraphics [width =0.9\linewidth]{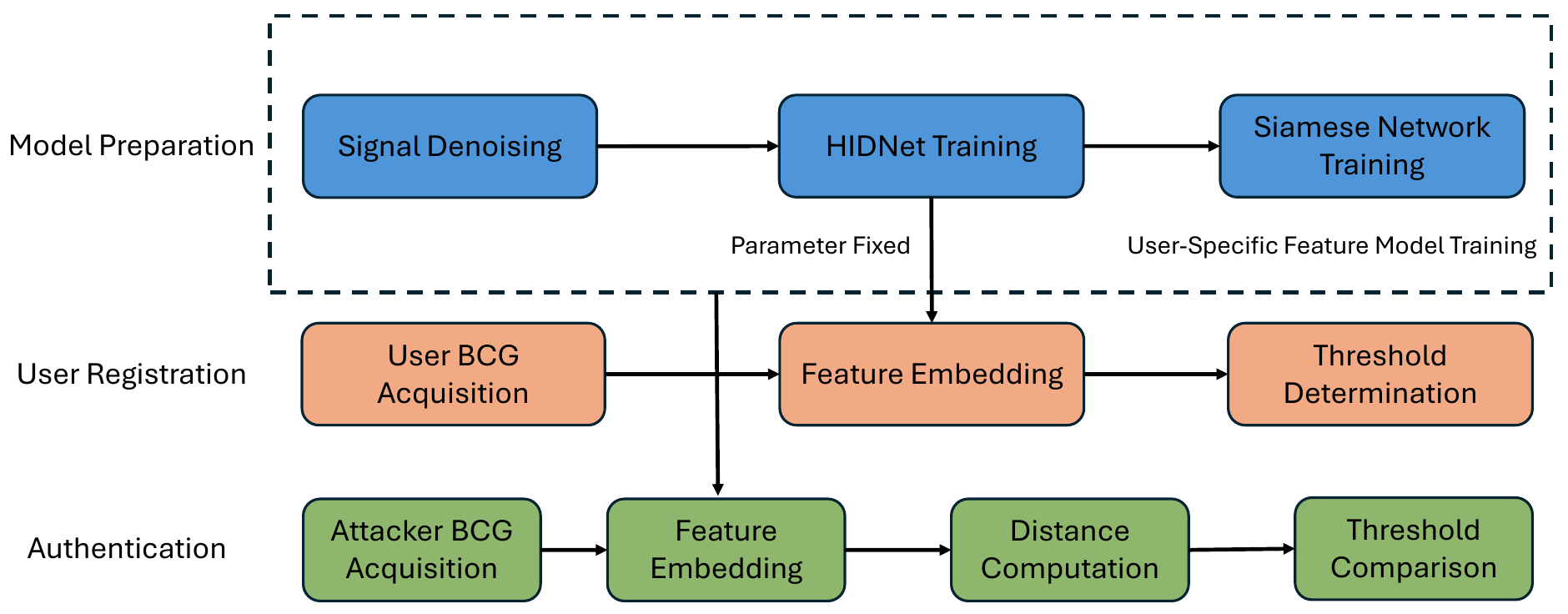}
\caption{Registration and authentication framework.}
\label{fig:fullframework}
\end{center}
\vspace{-0.2in}
\end{figure}

\subsection{Siamese-Based Feature Embedding}\label{sec:siamese}
To obtain the optimal function $f^*$ in Equation (\ref{eq:def}), we convert the extracted features from the \HIDNET{} by a Siamese network. 
Specifically, the network only adopts one MLP layer, but with the following triplet loss \cite{schroff2015facenet} for training. The optimization target is defined as:
\begin{equation}\label{eq:aim}
    \frac{1}{n}\sum^\mathcal{H}_i [cos(\mathbf{h}_\text{ID,i}^a,\mathbf{h}_\text{ID,i}^p)-cos(\mathbf{h}_\text{ID,i}^a,\mathbf{h}_\text{ID,i}^n)],
\end{equation}
where $n$ denotes the number of  triplet samples,  
$\mathbf{h}_\text{ID,i}^*=f(\mathbf{H}_\text{ID,i}^*)$ represents the embedding of the Siamese network $f(\cdot)$, and $cos(.,.)$ measures the cosine distance. 
Here, $\mathbf{h}_\text{ID,i}^a$ and $\mathbf{h}_\text{ID,i}^p$ correspond to embeddings from the user, while $\mathbf{h}_\text{ID,i}^n$ correspond to those from the authentication subject. 

Using Equation (\ref{eq:aim}) directly as a loss function for gradient descent optimization may not yield desired outcomes, as it implies equal distances for positive and negative pairs, contrary to our objective. To address this, we introduce a margin parameter $\alpha$ to enforce a larger difference for negative pairs compared to positive pairs. Formally, the modified loss function becomes:
\begin{equation}\label{eq:triloss}
    \mathcal{L}=\frac{1}{n}\sum^\mathcal{H}_i [cos(\mathbf{h}_\text{ID,i}^a,\mathbf{h}_\text{ID,i}^p)-cos(\mathbf{h}_\text{ID,i}^a,\mathbf{h}_\text{ID,i}^n)+\alpha]_+
\end{equation}

During training, we use this loss function to learn a generalized embedding space. The Siamese network is trained offline in a user-independent manner, such that samples from the same individual are mapped closer together while those from different individuals are pushed farther apart. Importantly, this training process is performed only once. After deployment, the trained network is directly applied to unseen users without any further parameter updates or fine-tuning.
We first select BCG features from the user to compute $\mathbf{h}_\text{ID,i}^a$ and $\mathbf{h}_\text{ID,i}^p$, and signals from attackers to compute $\mathbf{h}_\text{ID,i}^n$. However, only triplets satisfying $cos(\mathbf{h}_\text{ID,i}^a,\mathbf{h}_\text{ID,i}^n)-cos(\mathbf{h}_\text{ID,i}^a,\mathbf{h}_\text{ID,i}^p)<\alpha$ are used to constitute the triplet set $\mathcal{H}$ for training. Conversely, triplets where $cos(\mathbf{h}_\text{ID,i}^a,\mathbf{h}_\text{ID,i}^n)-cos(\mathbf{h}_\text{ID,i}^a,\mathbf{h}_\text{ID,i}^p)\geq\alpha$ are excluded from training, as the difference between embeddings of different subjects is already sufficiently large. The hyperparameter $\alpha$ controls the margin between the negative and positive pair differences. For each training epoch, a mini-batch of triplets is drawn from the valid set for training the Siamese network, which is repeated until convergence.

\subsection{Registration and Authentication}

Algorithm~\ref{alg:authen} outlines the complete procedure for user registration and authentication. 
When a new user is introduced, their BCG signals are first registered. 
Specifically, the system collects 6 minutes of continuous BCG data from the user. This signal is segmented into non-overlapping 4-second windows, each of which is embedded into an identity embedding $\mathbf{h}_\text{ID, i}$. All embeddings derived from the user’s data are averaged to form a representative feature center that characterizes the user’s identity. A small amount of data from other individuals (\ie, 1 minute of BCG data from non-matching users) is also processed to generate attacker embeddings. The Euclidean distances between embeddings and the user’s feature center are computed for both genuine and attacker samples. The authentication threshold $\tau$ is then determined such that FAR equals FRR. The computed center and threshold are stored as the user’s authentication profile. 
This registration process operates entirely in the learned embedding space and does not involve any model training.

In the authentication phase, new BCG test data is segmented into 4-second windows and passed through the same embedding model to obtain test embeddings. For each test embedding, its distance to the stored user center is computed. If the distance is less than the threshold $\tau$, the sample is accepted as a match to the registered user; otherwise, it is rejected.
To facilitate a clearer understanding of the overall process, we also add a table summarizing the workflow, as shown in \figref{fig:fullframework}.

\begin{figure}[t]
    \centering
        \subfigure[Earphone]{\label{fig:device}
    \includegraphics[width = 0.35\textwidth]{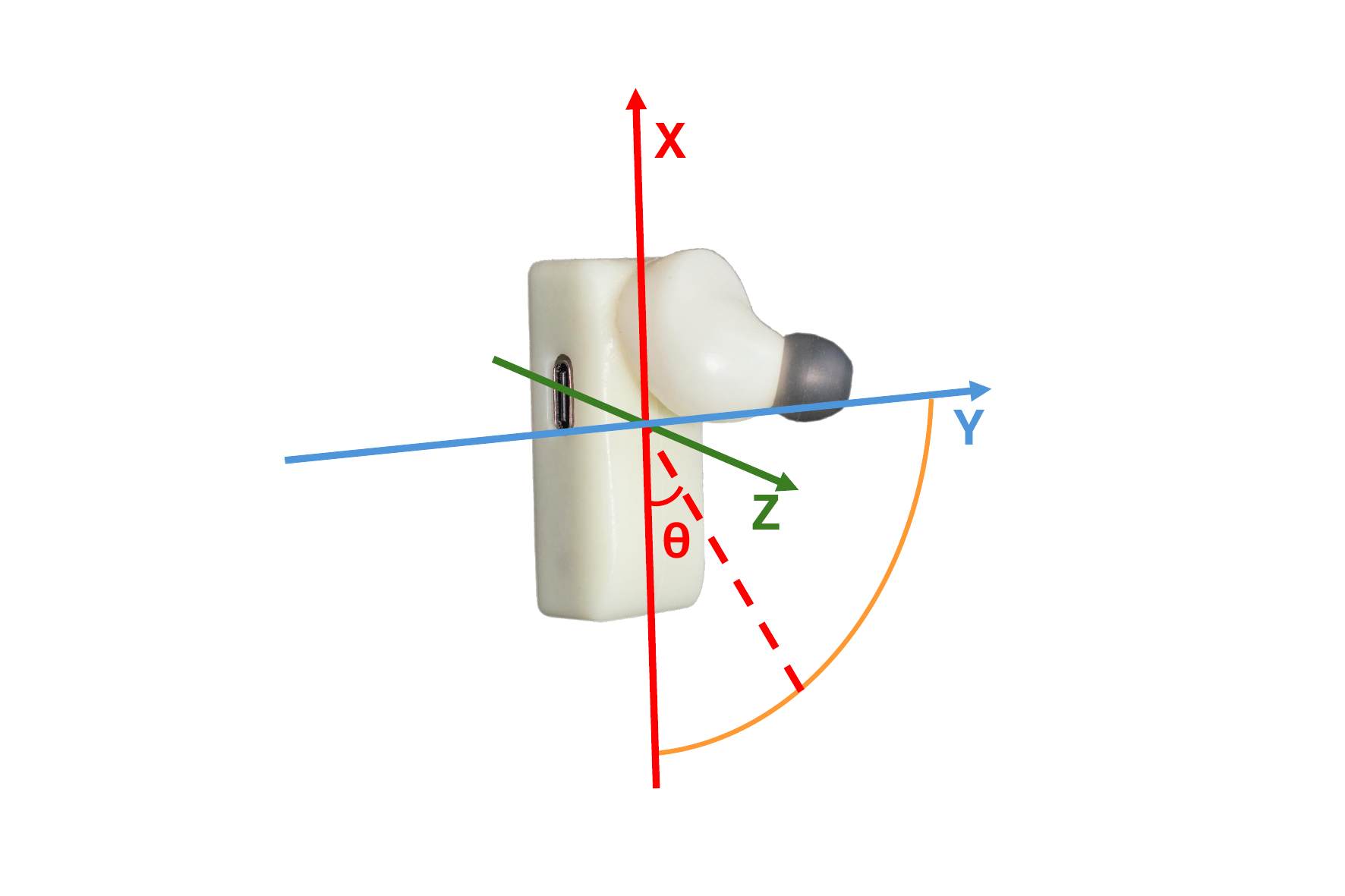}}
        \hspace{0.1 in} 
    \subfigure[Scenario]{\label{fig:experiment}
    \includegraphics[width = 0.35\textwidth]{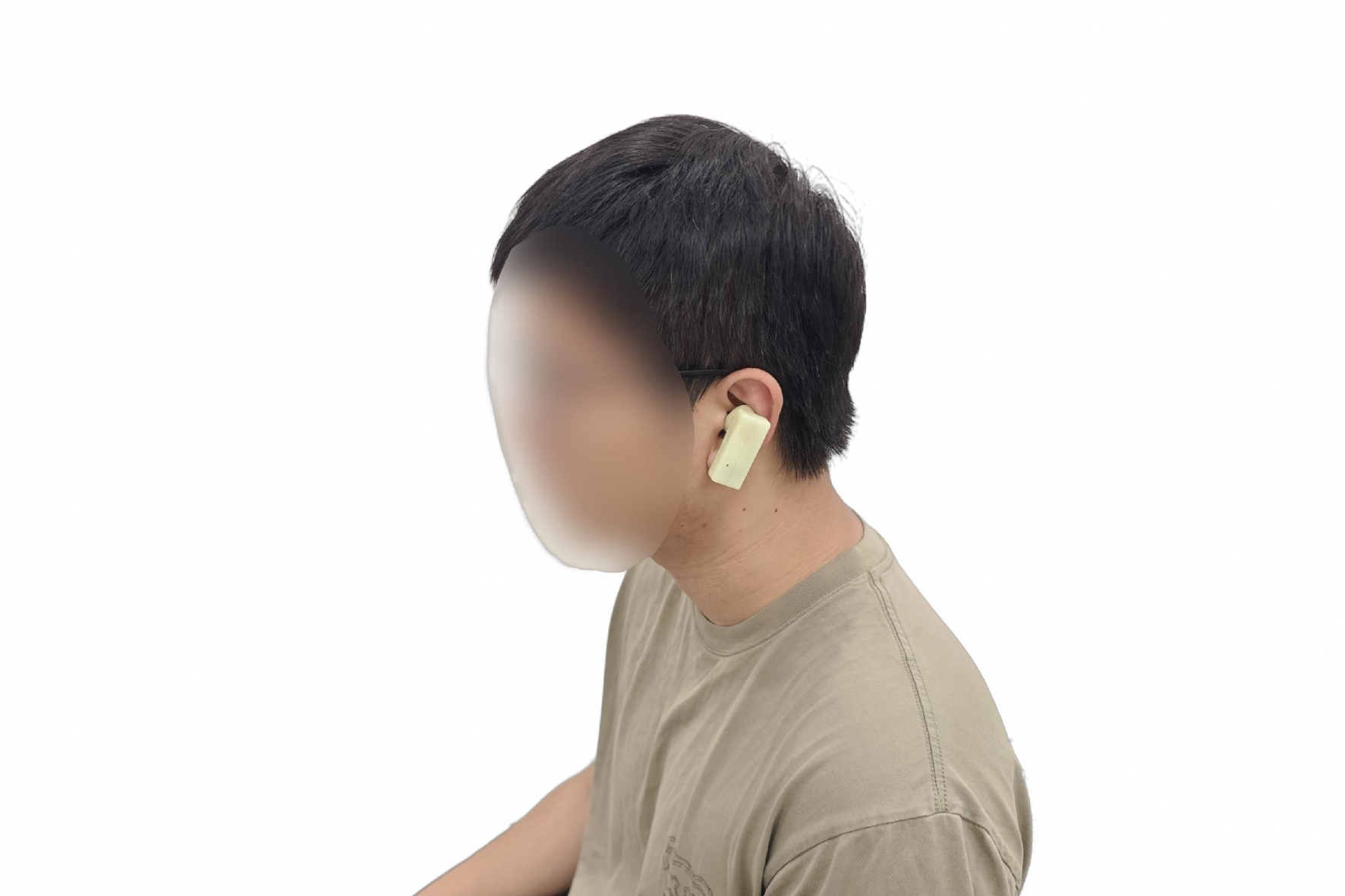}}
        \caption{Experimental setup.}
    \label{fig:eva}
    \vspace{-0.2in}
\end{figure}

\section{Evaluation}\label{sec:evaluation}
\subsection{Experimental Setup}
\subsubsection{Implementation.}\label{sec:implementation}
%
%
Although commercial earphones such as AirPods are equipped with inertial sensors, their raw inertial measurement unit (IMU) signals are not accessible to developers. 
Only heavily processed and downsampled motion cues are exposed at limited rates (\eg, 25 Hz for AirPods \cite{mollyn2023imuposer}), which are insufficient for fine-grained physiological or biometric analysis. 
Therefore, our system employs a commercial IMU sensor (ICM-42688-P \cite{ICM-42688-P}) embedded in a 3D-printed earphone, as shown in \figref{fig:device}, to acquire in-ear accelerometer readings at a sampling rate of 100 Hz.
The collected data is transmitted in real time via Bluetooth to a server for processing.
%
We recruit 33 participants (17 males and 16 females), aged between 18 and 75, for system evaluation.
All experimental procedures are reviewed and approved by our institutional review board (IRB) before implementation. 
%
The entire data collection process spans six months, during which participants are instructed to wear the earphones in their usual manner in the classroom, as shown in \figref{fig:experiment}.
Each participant contributes 16 minutes of data, resulting in 240 samples, each 4 seconds long.
Before the experiment begins, a brief five-minute instruction session is provided to explain the proper use of the earphones and the standardized data collection procedure.
%
%
We train neural network models on an NVIDIA 4090 GPU using PyTorch. The \HIDNET{} was trained for 100 epochs with an initial learning rate of $1\times10^{-4}$ using the Adam optimizer~\cite{kingma2014adam}, decayed by a factor of 0.5 every 20 epochs. For the Siamese network, we train for 80 epochs with an initial learning rate of $5\times10^{-4}$, keeping the same decay schedule.
A three‑fold cross‑validation protocol is adopted: in each fold, two groups of subjects are used for \HIDNET{} and Siamese network training, and the remaining group serves exclusively for testing. This design guarantees subject independence between training and evaluation.

\subsubsection{Evaluation Metrics.}
We evaluate \sys{} using two standard metrics. 
The  FAR measures the proportion of unauthorized users who are mistakenly granted access, while the  FRR  quantifies the proportion of legitimate users who are incorrectly denied access.
Lower values of both FAR and FRR indicate superior authentication performance.

\begin{figure*}[t]
\centering
\begin{minipage}[t]{0.24\textwidth}
\centering
\includegraphics[width=1\textwidth]{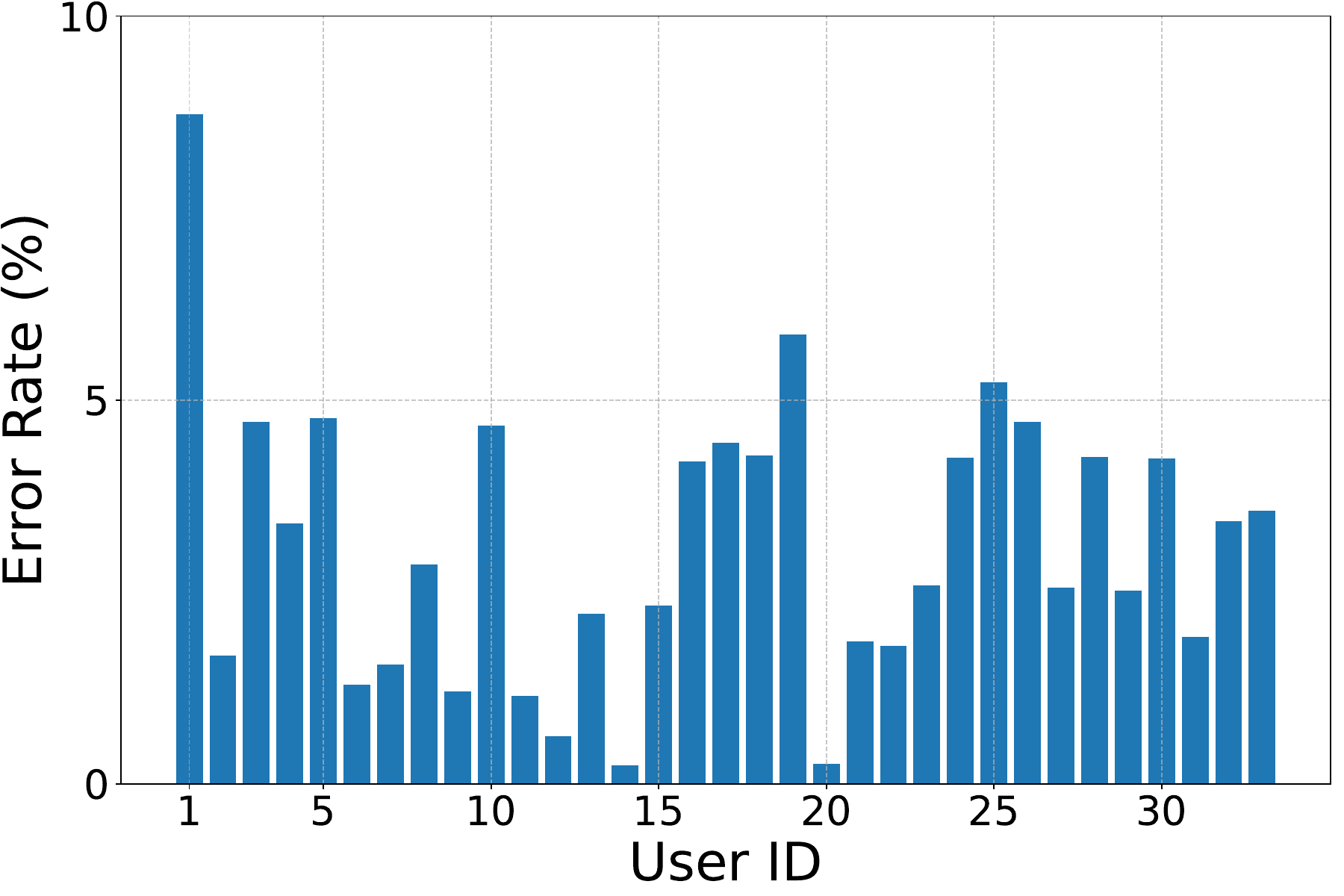}
\caption{FAR vs. user.}
\label{fig:userfar}
\end{minipage}
\hfill
\begin{minipage}[t]{0.24\textwidth}
\centering
\includegraphics[width=1\textwidth]{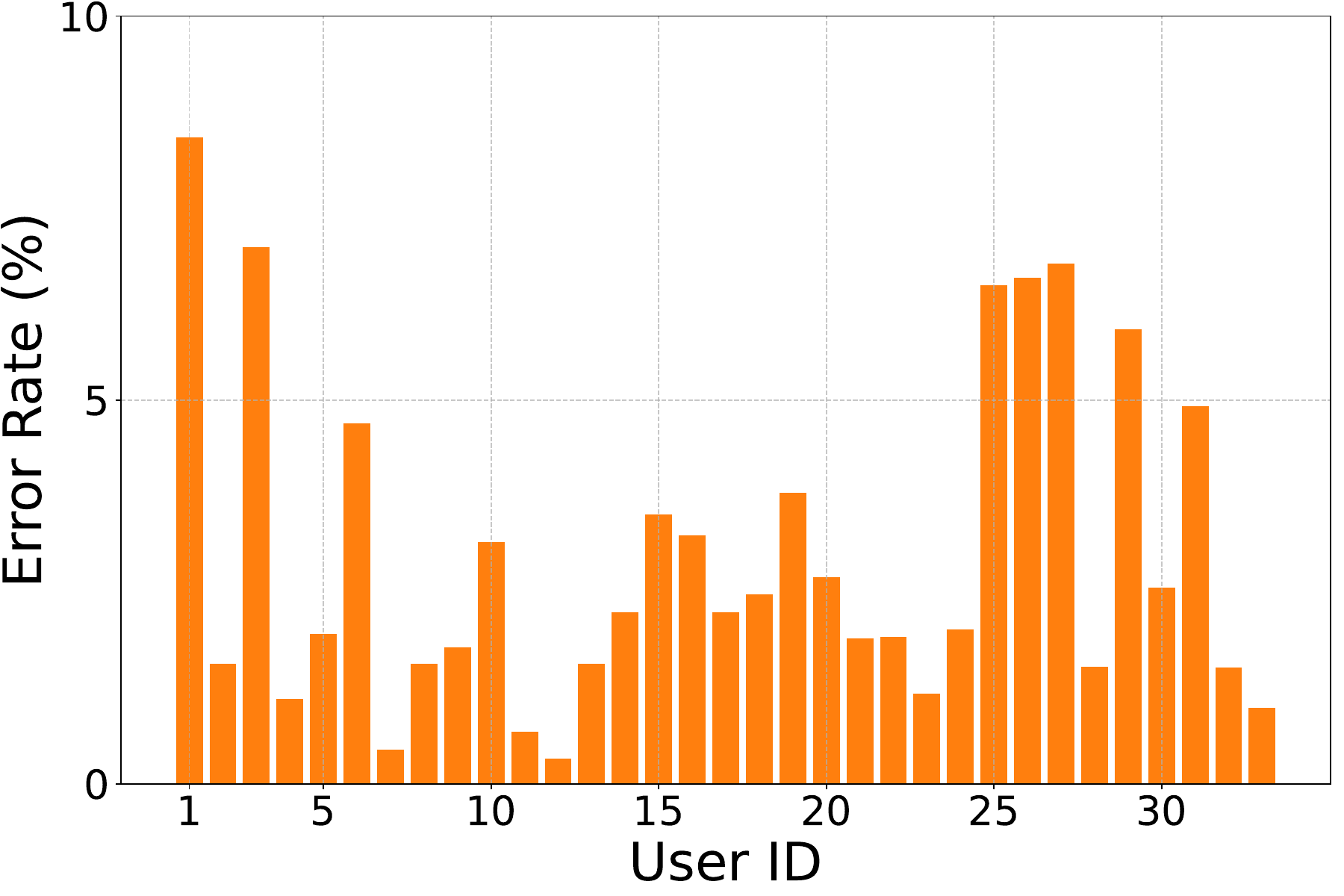}
\caption{FRR vs. user.}
\label{fig:userfrr}
\end{minipage}
\hfill
\begin{minipage}[t]{0.24\textwidth}
\centering
\includegraphics[width=1\textwidth]{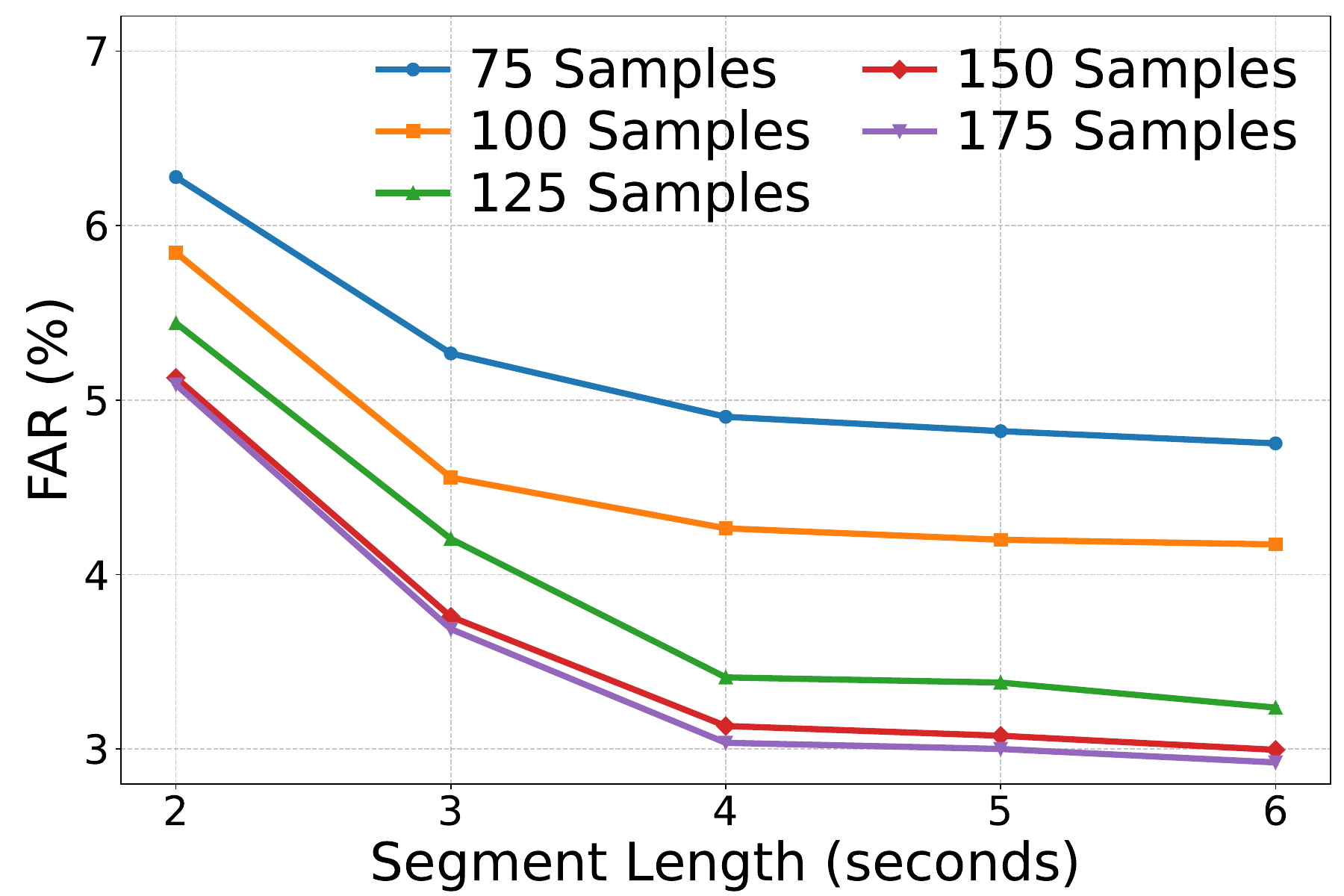}
\caption{FAR vs. training size and length.}
\label{fig:sizefar}
\end{minipage}
\hfill
\begin{minipage}[t]{0.24\textwidth}
\centering
\includegraphics[width=1\textwidth]{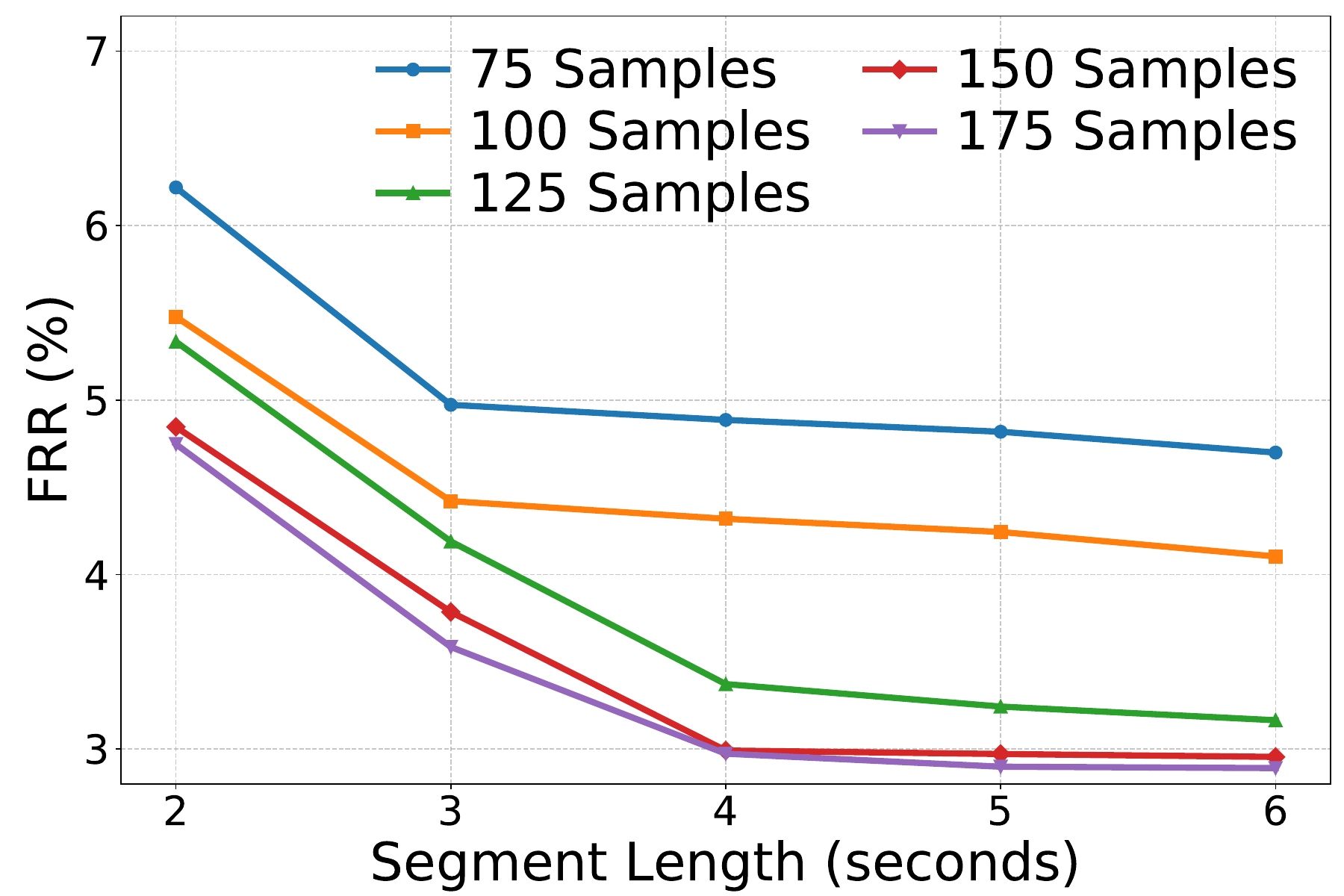}
\caption{FRR vs. training size and length.}
\label{fig:sizefrr}
\end{minipage}
\vspace{-0.2in}
\end{figure*}

\subsection{Overall Performance}
\subsubsection{Performance vs. Users.} 
We evaluate the system using three-fold cross-validation across all users in the static scenario, as detailed in Section~\ref{sec:implementation}.
In each fold, 11 participants are selected for testing, with one designated as the authenticated user and the remaining 10 serving as attackers.
To ensure fairness, the role of the authorized user is rotated such that every participant is evaluated as the legitimate user exactly once.
To enable robust evaluation, the 240 BCG samples collected from each user are randomly partitioned into three equal groups of 80 samples. 
In each iteration, one group is used for registration and the remaining for testing the authorized user. 
This process is repeated three times, rotating the registration group in each round, and the average FAR and FRR are reported for the corresponding user.
Note that, unless otherwise specified, all FAR and FRR reported in Section~\ref{sec:evaluation} are averaged over 33 experiments, with each participant serving once as the authorized user under the described settings.
As shown in \figref{fig:userfar} and \figref{fig:userfrr}, the average FAR and FRR across 33 users are 3.13\% and 2.99\%, respectively. Even the worst-case FAR and FRR remain below 8.72\% and 8.42\%. Overall, \sys{} demonstrates strong performance, sufficient to meet the requirements of daily authentication scenarios.

\subsubsection{ Performance vs. Training Size and Segment Length.}
To examine the impact of training sample size and segment length in the \HIDNET{} network on the feature extraction capability, we systematically assess authentication performance under various training configurations.
We vary the total size of static BCG samples from 75 to 175 and adjust the segment length from 2 to 6 seconds per participant.
As shown in \figref{fig:sizefar} and \figref{fig:sizefrr}, both FAR and FRR decrease consistently with larger sample sizes and longer segments, with performance plateauing at higher values.
When the training size reaches 150  with 4-second segment, the system achieves a FAR of 3.13\% and an FRR of 2.99\%. 
Beyond this point, further increases in training size or segment length yield only marginal improvements.
These findings indicate that 150 training samples, each 4 seconds long, are sufficient to capture representative features for robust and reliable authentication.
Note that although the initial authentication is performed using a 4-second segment, this design does not limit the system’s responsiveness in practice. 
As introduced in Section \ref{sec:intro}, the earphone-based authentication system is primarily intended for device pairing and access control, where preventing post-authentication takeover is critical. 
Therefore, authentication is conducted in a continuous manner rather than as a one-time decision.
Specifically, we can adopt a sliding-window strategy with a short step size (\eg, 0.5 s), enabling overlapping authentication decisions over time. 
As a result, although each authentication model operates on a 4-second segment, the system effectively updates its authentication status every 0.5 s. This design reduces the effective latency to the sliding step, rather than the full segment length, thereby supporting real-time and continuous authentication without compromising security.

\subsubsection{Performance vs. Registration Samples.}

\begin{figure}[t]
\begin{center}
\includegraphics [width = 0.5\linewidth]{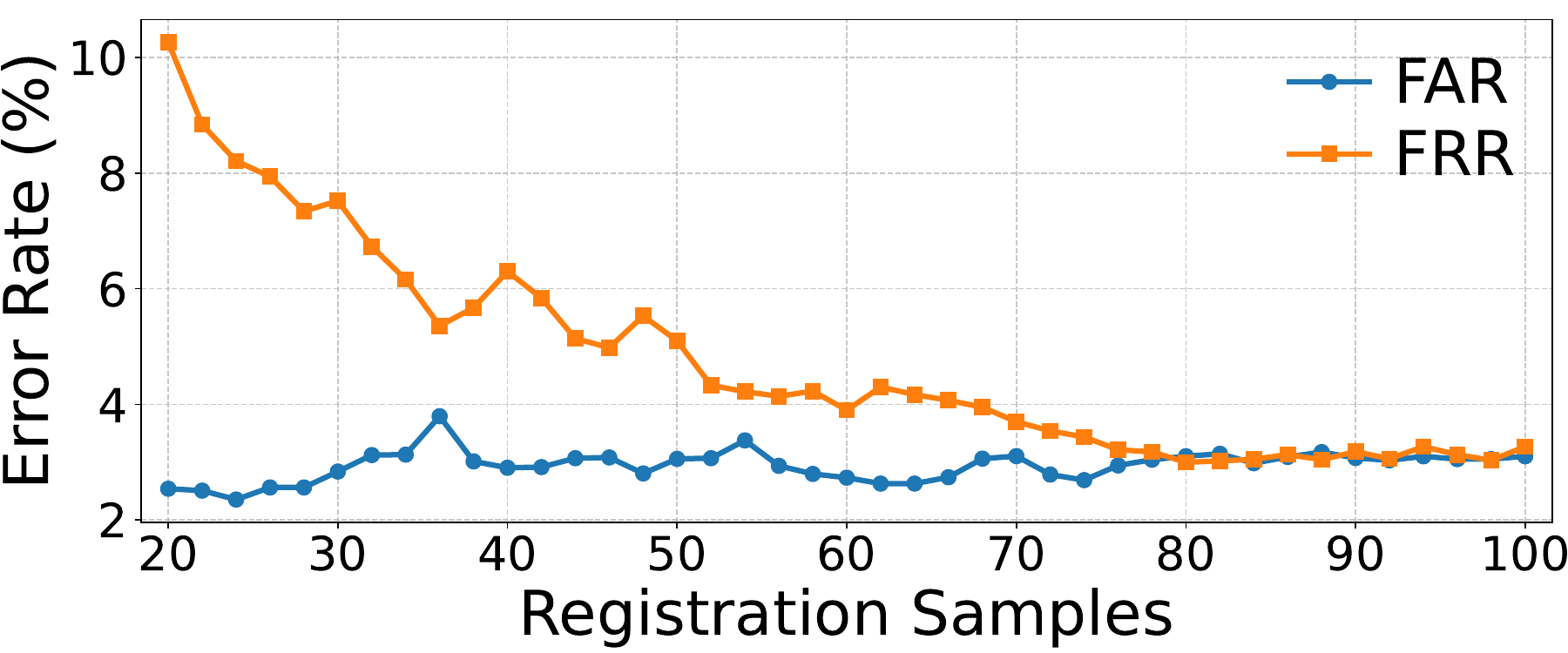}
\caption{Performance vs. registration samples}
\label{fig:registration}
\end{center}
\vspace{-0.1in}
\end{figure}

We evaluate how the number of registration samples affects authentication performance. As shown in \figref{fig:registration}, both FAR and FRR remain below 10\% when using more than 22 samples, and gradually converge to approximately 3\% with 80 samples. 
We observe that when the number of samples is small, the estimated threshold tends to be tight, leading to high FRR and low FAR. 
As more samples are used, the threshold becomes more relaxed, reducing FRR while slightly increasing FAR. 
Eventually, the two rates balance, indicating that the registration data better represents the true distribution. 
Overall, our system achieves satisfactory accuracy even with a small number of registration samples.
For better user experience, we can keep the initial registration lightweight, requiring only approximately 2 minutes of data (30 samples). 
Under this setting, the system achieves an average FAR of 2.83\% and an average FRR of 7.52\%, which is sufficient for everyday authentication scenarios.
Then, during regular usage, the system can unobtrusively collect additional data for incremental registration, further improving the authentication performance.

\subsubsection{Comparison with Baseline Models.}
To evaluate the effectiveness of our proposed \HIDNET{} model, we compare its authentication performance against four popular deep learning baselines: CNN, ResNet50 (R50), Conformer (CF) \cite{gulati2020conformer}, and LSTM. 
In this comparison, we replace \HIDNET{} with each baseline model while keeping the Siamese-based feature embedding and authentication pipeline unchanged.
As shown in \figref{fig:baseline}, \HIDNET{} achieves the lowest FAR and FRR, significantly outperforming all competitors.
This advantage primarily stems from \HIDNET{}'s disentanglement-based design, which effectively separates identity-specific features from shared nuisance components, enabling more robust and discriminative feature learning for reliable authentication.
We further conduct paired statistical tests between the proposed method and baseline models across users. The results indicate that the improvements in both FAR and FRR are statistically significant (p $<$ 0.05), suggesting that the observed gains are unlikely to be caused by random variation.

\subsection{Ablation Study}
We perform an ablation study by systematically removing individual modules from our model to evaluate their contributions to authentication performance.
The tested variants are as follows:
\textbf{V1} (w/o Orthogonal Regularization): removes the orthogonality constraint between identity-relevant and identity-irrelevant features.
\textbf{V2} (w/o Reconstruction): removes the reconstruction module,
\textbf{V3} (w/o Orthogonal Regularization and Reconstruction): removes both components in \textbf{V1} and \textbf{V2},
\textbf{V4} (w/o \HIDNET{}): disables the disentangled feature extraction network, and
\textbf{V5} (w/o Siamese-Based Feature Embedding): omits the Siamese embedding mechanism.

\begin{figure*}[t]
\centering
\begin{minipage}[t]{0.3\textwidth}
\centering
\includegraphics[width=1\textwidth]{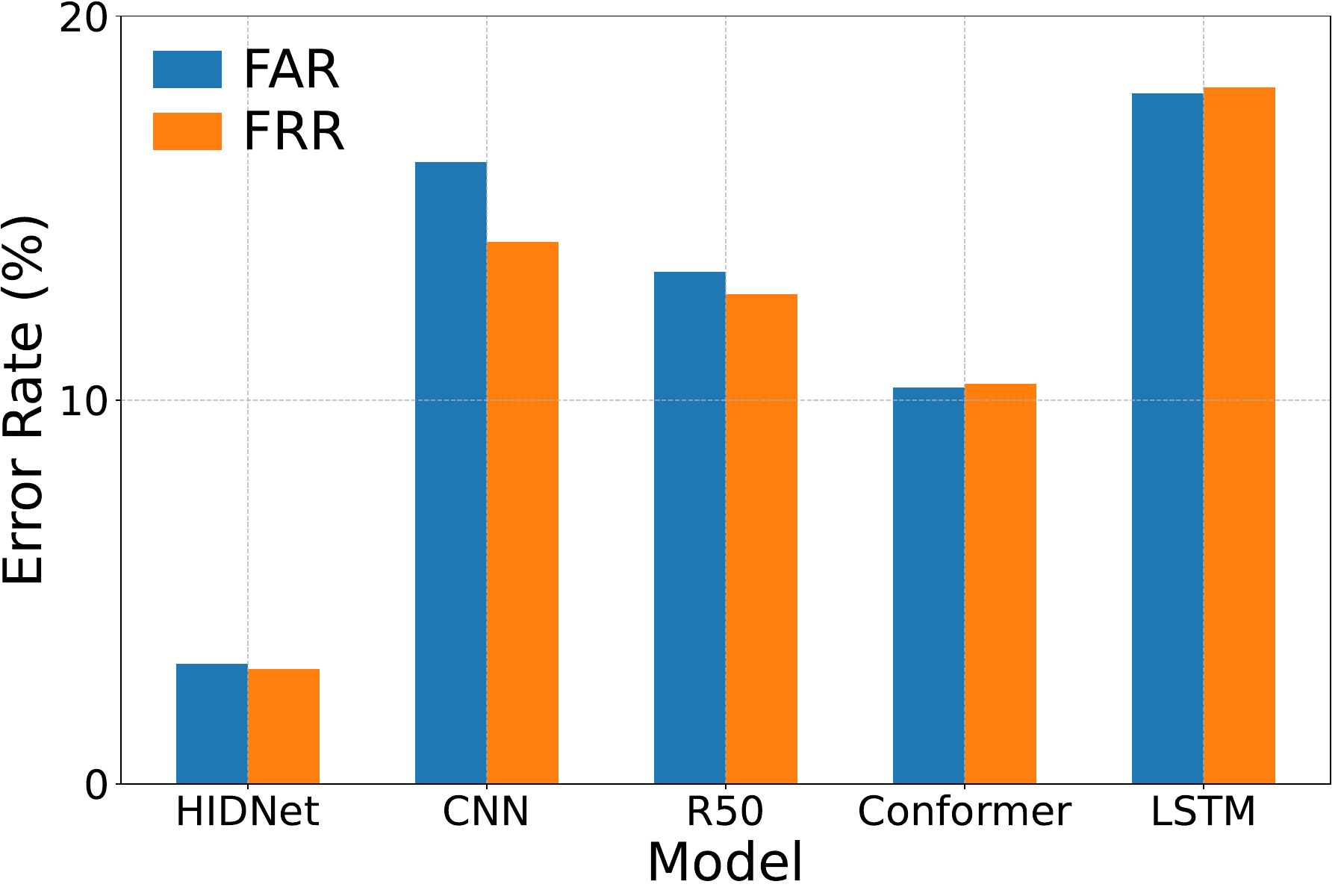}
\caption{Comparison with baseline models.}
\label{fig:baseline}
\end{minipage}
\hfill
\begin{minipage}[t]{0.3\textwidth}
\centering
\includegraphics[width=1\textwidth]{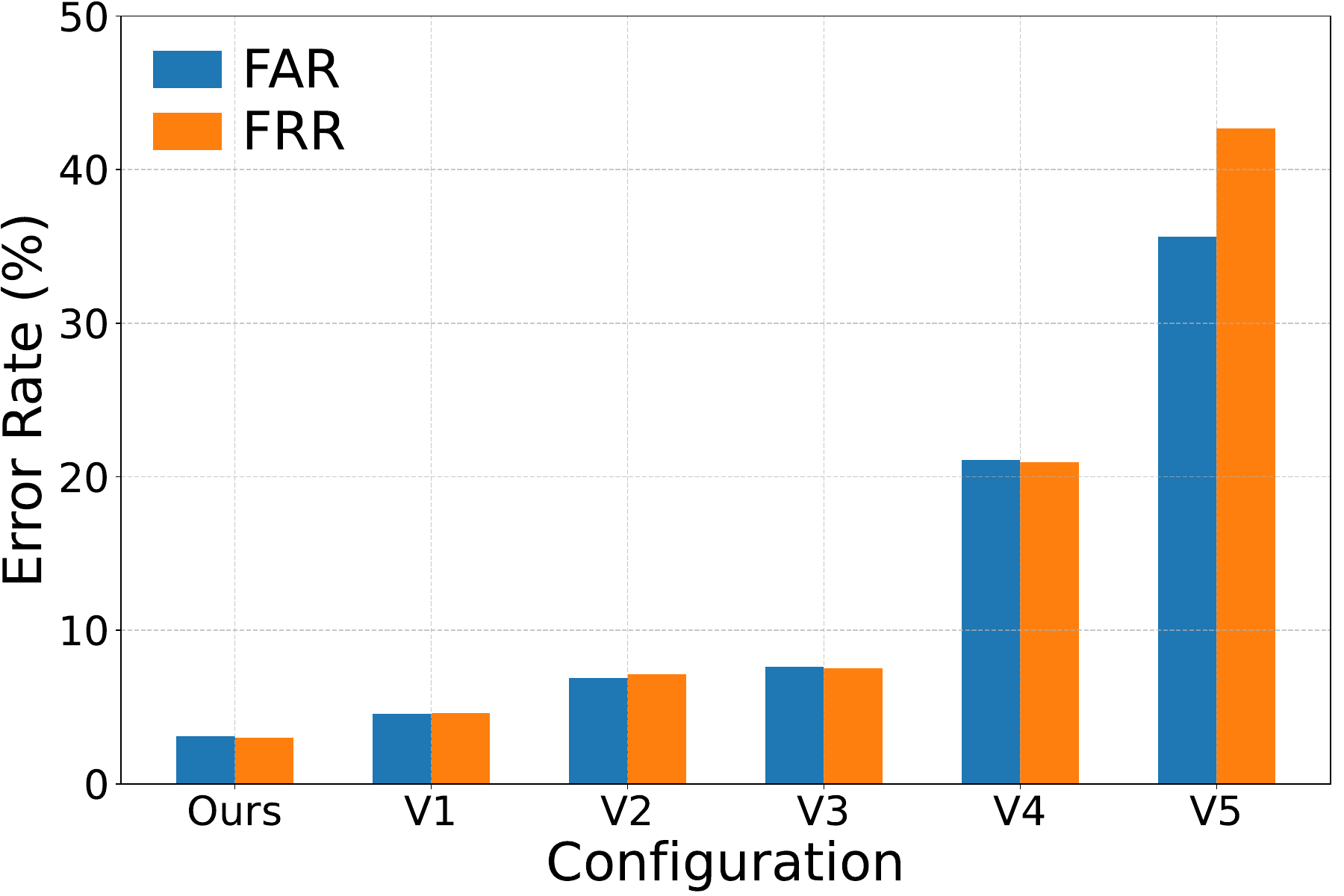}
\caption{Ablation study.}
\label{fig:ablation}
\end{minipage}
\hfill
\begin{minipage}[t]{0.3\textwidth}
\centering
\includegraphics[width=1\textwidth]{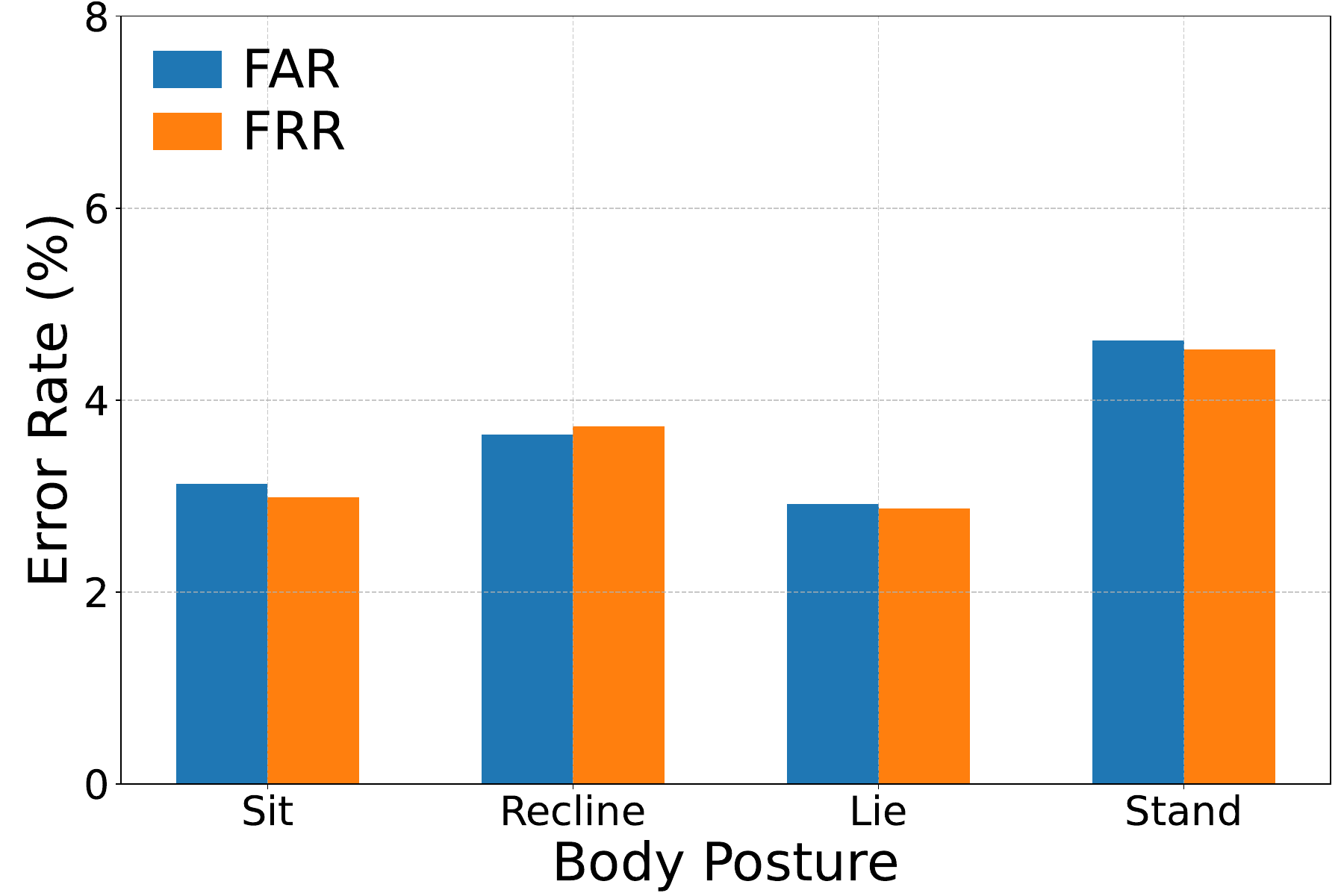}
\caption{Impact of body postures.}
\label{fig:posture}
\end{minipage}
\vspace{-0.2in}
\end{figure*}

As shown in \figref{fig:ablation}, the full model (``Ours’’) consistently achieves the best performance, yielding the lowest FAR and FRR across all settings. Removing orthogonal regularization (\textbf{V1}) leads to noticeable performance degradation, as identity-relevant features become entangled with nuisance factors, reducing inter-class separability and increasing both false acceptance and rejection rates. Similarly, eliminating the reconstruction module (\textbf{V2}) weakens the model’s ability to preserve identity-discriminative information, since the latent representations are no longer constrained to retain sufficient input semantics, resulting in less stable and less informative features.
When both orthogonal regularization and reconstruction are removed (\textbf{V3}), performance deteriorates further, indicating that these two components play complementary roles: orthogonal regularization promotes effective feature disentanglement, while reconstruction enforces representation completeness. Their joint absence significantly compromises the quality and robustness of identity-relevant embeddings.
The most severe degradation is observed in \textbf{V4}, where \HIDNET{} is removed. Without disentangled feature learning, the model collapses to learning entangled representations that are highly sensitive to environmental and behavioral variations, severely impairing authentication reliability. This confirms that \HIDNET{} is a core component for isolating stable identity characteristics from irrelevant factors.
Finally, removing the Siamese-based feature embedding (\textbf{V5}) also results in clear performance drops. Without metric learning supervision, the model cannot explicitly enforce intra-user compactness and inter-user separation in from the required measurement perspective, reducing its generalization capability in user-agnostic authentication scenarios.
Quantitatively, the full model achieves the lowest average FAR (3.13\%) and FRR (2.99\%), while \textbf{V4} and \textbf{V5} exhibit the highest error rates, further validating the necessity of each proposed module and the effectiveness of their integration.
We also perform statistical significance testing between the full model and each ablated variant. The results show that the performance differences are statistically significant (p $<$ 0.05), demonstrating the effectiveness of each component in the proposed framework.


\subsection{Impact of Different Factors}

\subsubsection{Impact of Movements.}
\begin{figure*}[t]
\centering
\begin{minipage}[t]{0.3\textwidth}
\centering
\includegraphics[width=1\textwidth]{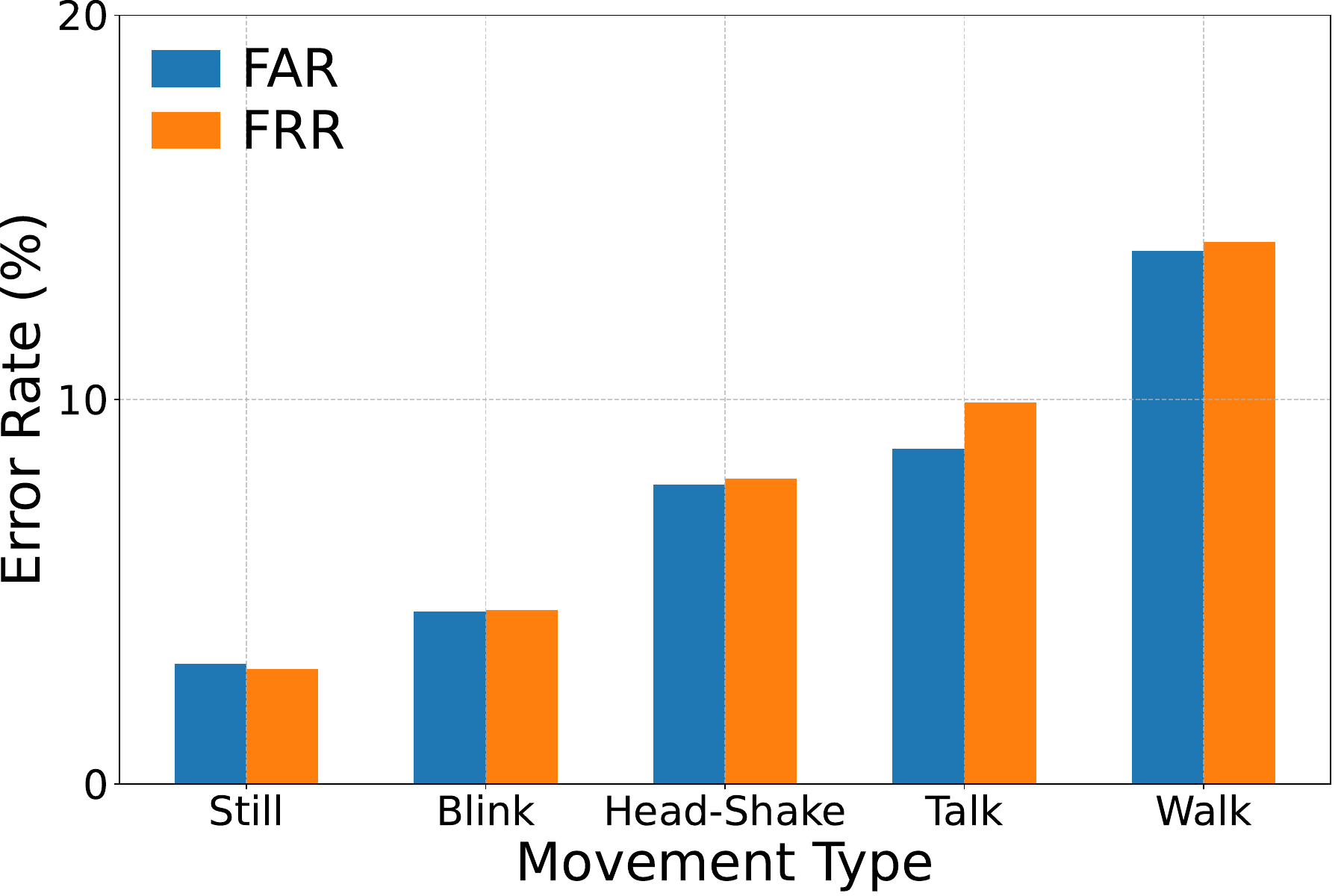}
\caption{Impact of movements.}
\label{fig:movement}
\end{minipage}
\hfill
\begin{minipage}[t]{0.3\textwidth}
\centering
\includegraphics [width=1\textwidth]{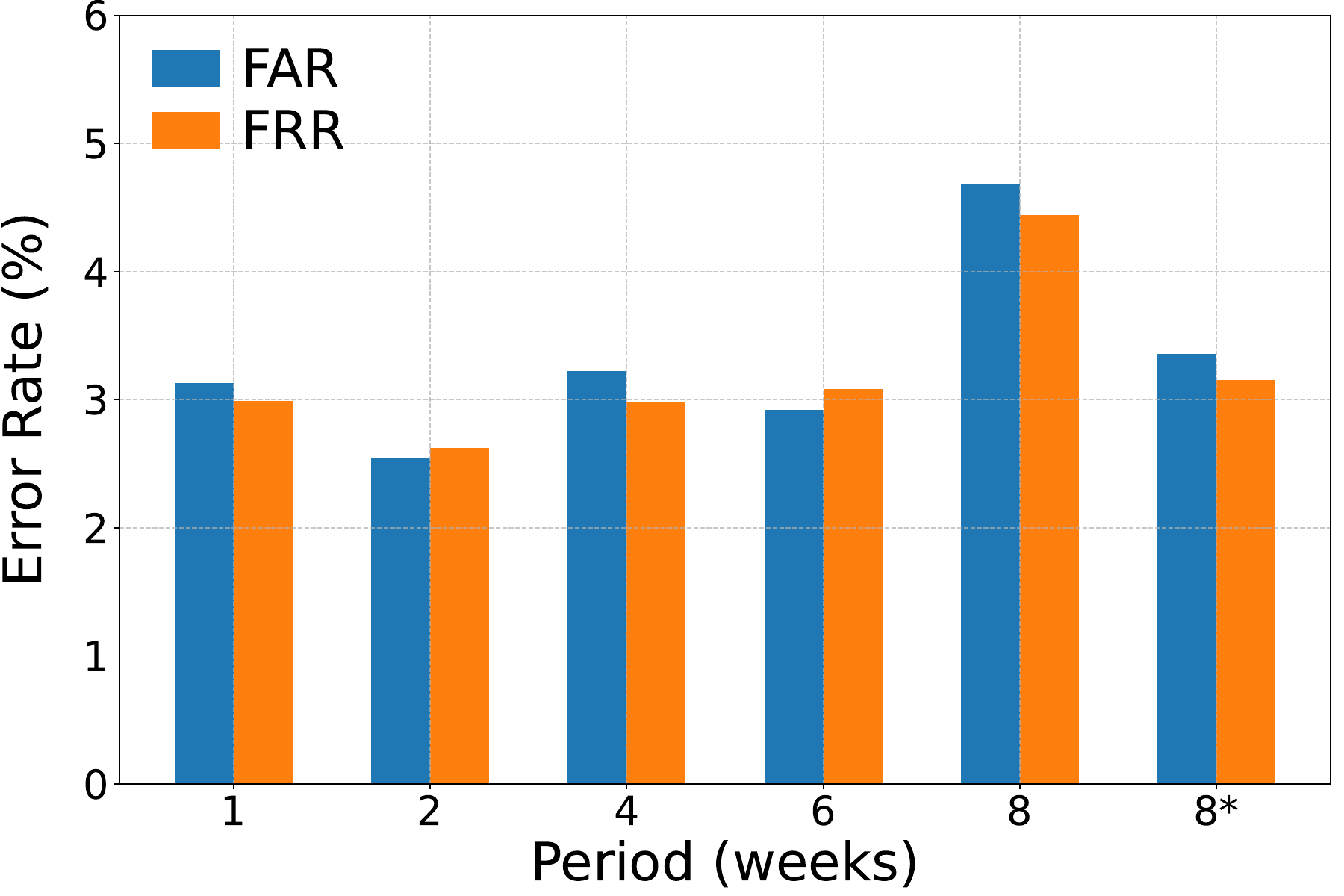}
\caption{Impact of long-term use.}
\label{fig:longterm}
\end{minipage}
\hfill
\begin{minipage}[t]{0.3\textwidth}
\centering
\includegraphics[width=1\textwidth]{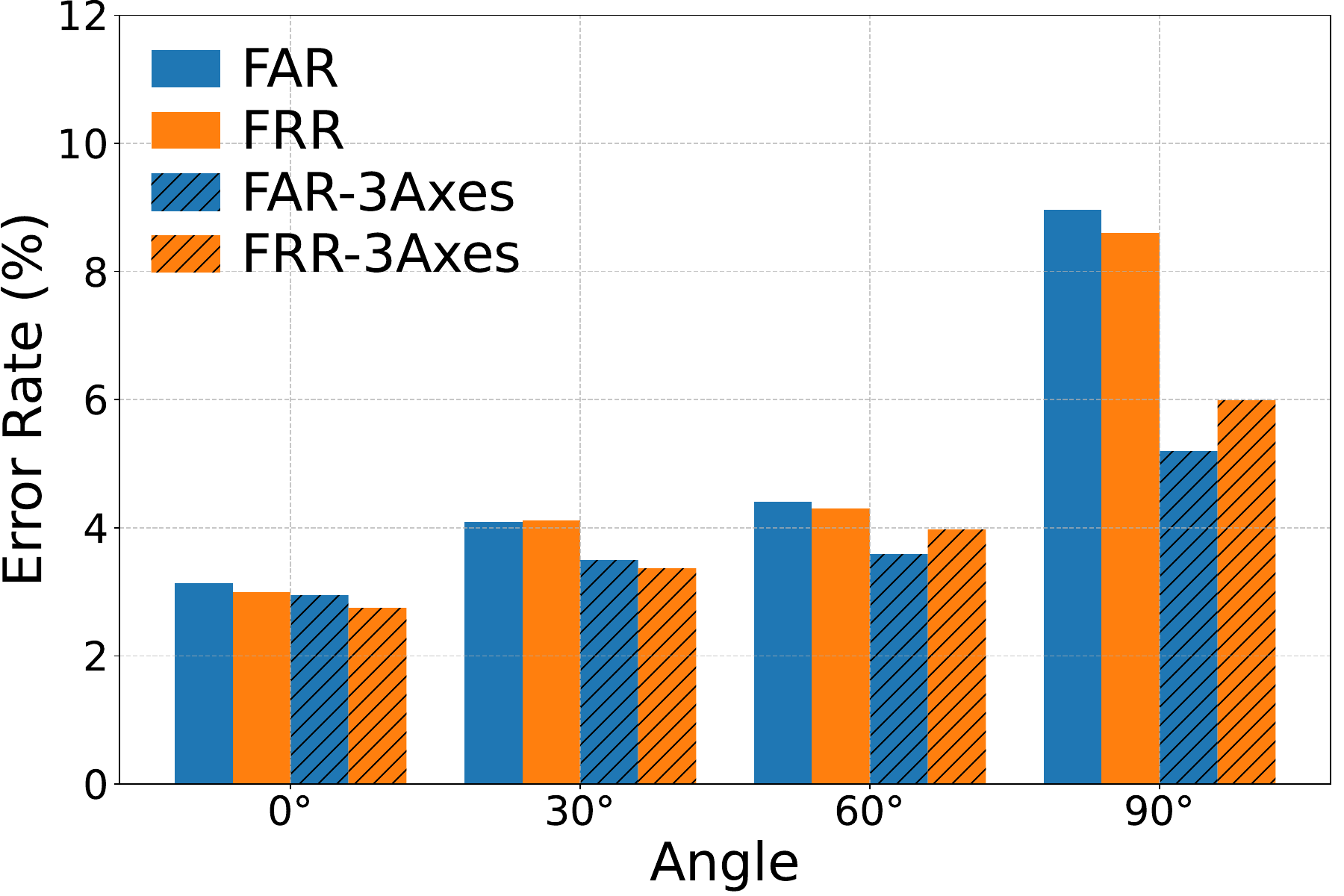}
\caption{Impact of earphone angles.}
\label{fig:angle}
\end{minipage}
\vspace{-0.2in}
\end{figure*}

To evaluate motion robustness, we test \sys{} under five daily movement scenarios: still, blink, head-shake, talk, and walk. 
As shown in \figref{fig:movement}, our system maintains low error rates under mild motions. Even with head-shaking, the FAR and FRR remain at 7.79\% and 7.95\%, respectively, demonstrating the effectiveness of our two-step denoising scheme in suppressing sporadic artifacts. 
No that we observe a noticeable performance drop in this condition, which we attribute mainly to speech-related jaw motion and contact variation. The resulting interference contains frequency components that substantially overlap with the heartbeat-induced BCG signal, limiting the effectiveness of the current denoising pipeline.
To examine whether this issue can be mitigated in practice, we augment the registration stage with 10 additional 4-second talking samples for each user, and then evaluate the system under the talking condition using different speech content. 
The results show an average FAR of 7.21\% and an average FRR of 8.72\%, compared with 8.71\% and 9.91\%, respectively, in the original setting.
Although the performance is worse than that under the static condition, it suggests that the degradation can be alleviated by incorporating representative talking patterns during registration.
Additionally,  walking introduces complex whole-body motion that cannot be fully mitigated by the RLS filter, resulting in degraded authentication performance (FAR: 13.86\%, FRR: 14.10\%). 
Despite this limitation, the error rates remain bounded below 14.1\%, suggesting that the system retains a practical level of usability in real-world authentication settings.

\subsubsection{Impact of Body Posture.}
We evaluate the authentication performance under four body postures: sitting, reclining, lying, and standing. 
As shown in \figref{fig:posture}, the FAR/FRR under sitting, reclining, and lying conditions remain low and comparable (e.g., 3.13\%/2.99\%, 3.64\%/3.73\%, and 2.92\%/2.87\%, respectively). 
Standing results in slightly higher error rates (4.62\%/4.53\%) due to inevitable body sway. 
Nonetheless, our two-step denoising scheme effectively suppresses such motion-induced noise, maintaining reliable performance across all postures.

\subsubsection{Impact of Long-Term Use.}
To evaluate the temporal stability of the proposed system, we conduct experiments over an 8-week period. 
As shown in \figref{fig:longterm}, the system achieves consistent results across different time points, with average FAR and FRR remaining below 3.2\% in the first 6 weeks and increasing slightly to 4.68\% and 4.44\% in week 8. 
This demonstrates the robustness of heartbeat patterns over time and suggests that user-specific features remain relatively stable for at least two months. 
In practice, to address minor temporal drifts, the system can be incrementally updated with fresh samples to maintain performance.

Note that the there is a noticeable performance degradation after eight weeks. We believe this trend mainly reflects distribution drift during prolonged use. In practice, subtle variations in wearing position, device state, and personal behavior patterns can gradually shift the feature distribution over time. Under such conditions, a static registration template may no longer fully capture the user’s current signal characteristics, leading to degraded authentication performance.
To mitigate this issue, we design a lightweight incremental template update mechanism based on high-confidence samples. 
Specifically, at the end of each week, the system selects 10 4-second samples from successfully authenticated data whose authentication distance is smaller than half of the current threshold \(\tau\). These high-confidence samples are then used to incrementally update the template center and the threshold \(\tau\). The update process is performed entirely in the background, without requiring explicit user re-registration or changing the normal authentication workflow. Restricting the update to only high-confidence samples helps the system adapt to long-term drift while reducing the risk of template contamination.
The updated result at Week 8, denoted by \(8^*\), exhibits a clear reduction in error compared with the original Week-8 result, bringing the performance back to a level close to that of the first six weeks. This result suggests that a lightweight background update mechanism is effective in maintaining authentication reliability under long-term drift.

\subsubsection{Impact of Earphone  Angles.}
We examine the system robustness under different earphone rotation angles (\ie, $0^\circ$, $30^\circ$, $60^\circ$, and $90^\circ$), which simulate variations in real-world wearing conditions.
As shown in \figref{fig:angle}, the system maintains a stable performance when the angle is below $60^\circ$, with both FAR and FRR remaining under 4.5\%. 
However, at $90^\circ$, the error rates rise significantly (FAR: 8.96\%, FRR: 8.60\%), likely due to reduced sensitivity of the accelerometer's X-axis to cardiac vibrations. 
%
To improve robustness, we extend the system to use all three accelerometer axes and introduce a channel attention module based on multi-head attention to adaptively fuse the tri-axis signals. This design allows the model to assign different weights to the three axes according to their reliability under different wearing orientations, rather than relying on a single fixed axis.
The updated results show consistent improvements across all tested rotation angles from $0^\circ$ to $90^\circ$. The gain is especially notable at $90^\circ$, where the performance improves from an FAR of 8.96\% and an FRR of 8.60\% in the original version to an FAR of 3.58\% and an FRR of 5.19\% after tri-axis fusion. These results indicate that axis sensitivity is a real factor under rotation, and that adaptive multi-axis fusion is effective in mitigating the resulting performance degradation.

\subsubsection{Impact of Playing Music.}
We evaluate the impact of speaker-induced vibrations on our system by playing music through the earphone at different volume levels, namely``Low'', ``Medium'', and ``High'', where ``High'' corresponds to the maximum adjustable volume of the earphone. 
As shown in \figref{fig:music}, system performance exhibits a slight degradation with increasing volume.
This is because, although most noise components can be filtered out, speaker-induced vibrations partially overlap with the dominant heartbeat frequency band, making complete suppression difficult. Nevertheless, even under the ``High''-volume setting, the system achieves average FAR and FRR of 4.14\% and 4.65\%, respectively, which are only marginally worse than those under the Mute condition. These results indicate that our system is not impacted by the speaker during normal use.

\subsubsection{Impact of Environments}
We evaluate the robustness of our system under different real-world environments. As shown in \figref{fig:environment}, compared to the reference setting (\ie, ``Classroom''), we observe a slight increase in error rates in the ``Station'' and ``Playground'' environments. 
This is because vibrations induced by train movements in the station and basketball impacts on the ground in the playground can propagate to the volunteers’ bodies, introducing mild interference to the sensed signals. Nevertheless, even in the ``Playground'' scenario, the system achieves average FAR and FRR of 3.71\% and 4.19\%, respectively. These results demonstrate that our system remains robust across diverse environments and can be widely deployed in everyday settings.

\subsubsection{Impact of Cardiac Diseases.}
\begin{figure*}[t]
\centering
\begin{minipage}[t]{0.3\textwidth}
\centering
\includegraphics[width=1\textwidth]{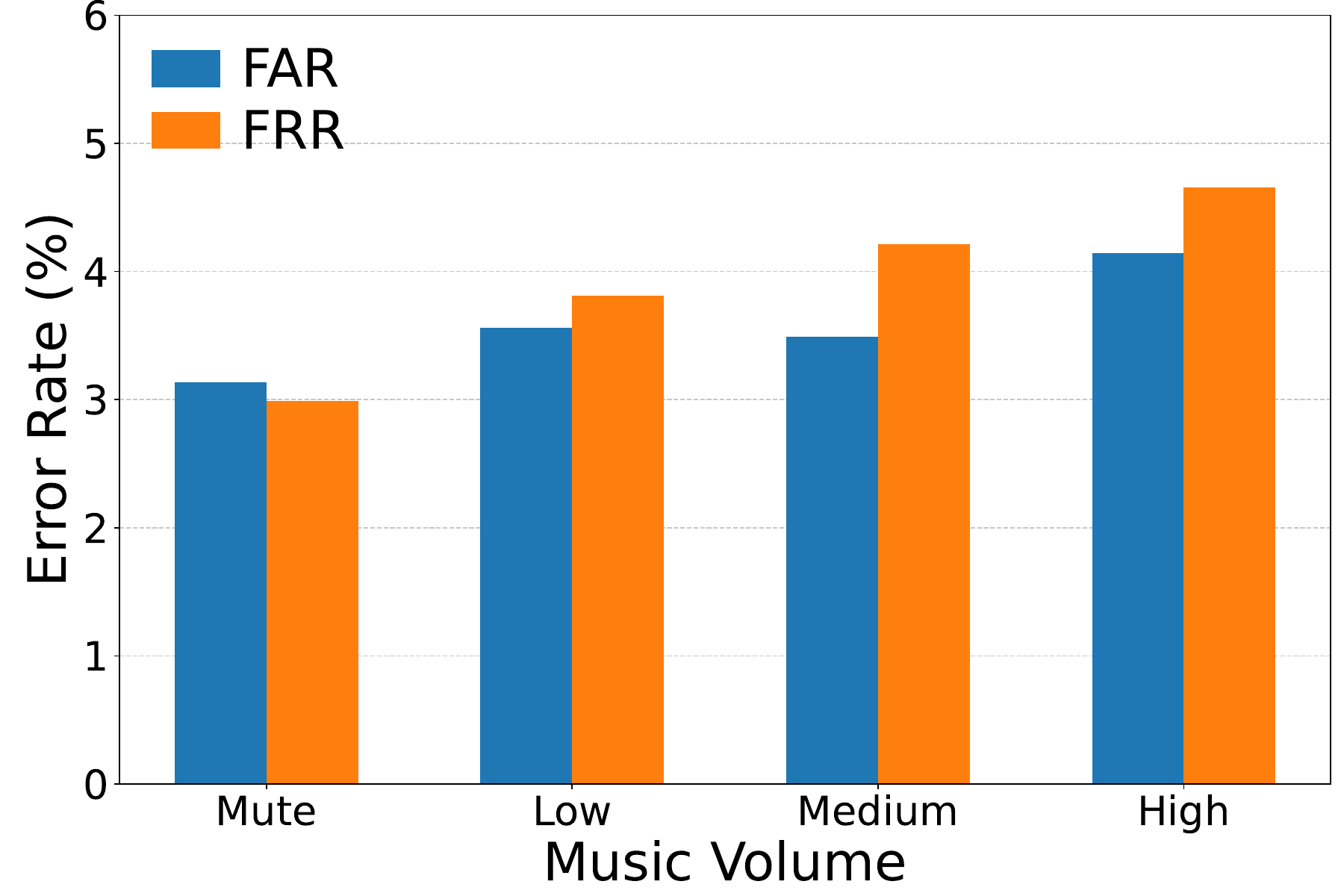}
\caption{Impact of playing music through earphones.}
\label{fig:music}
\end{minipage}
\hfill
\begin{minipage}[t]{0.3\textwidth}
\centering
\includegraphics[width=1\textwidth]{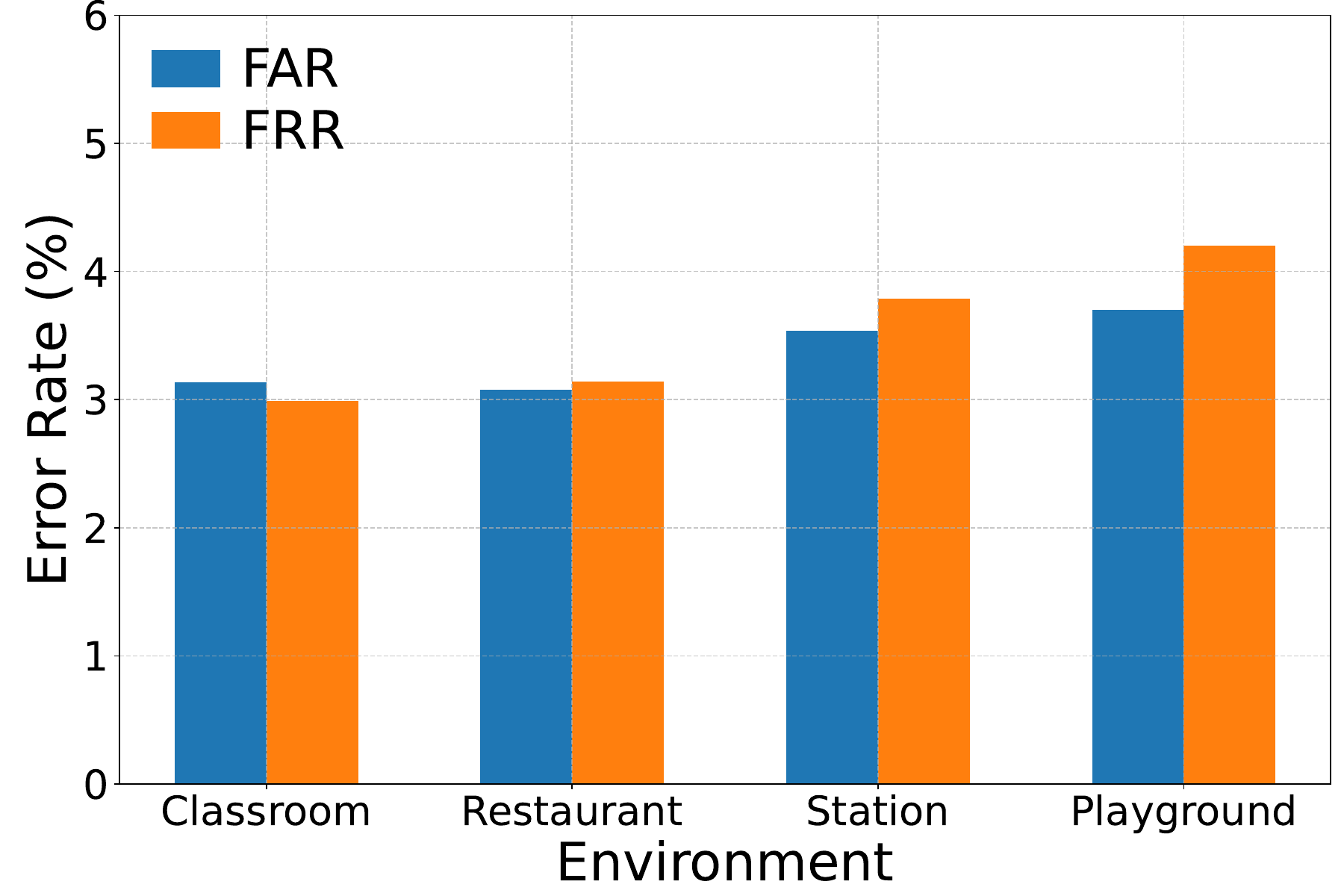}
\caption{Impact of environments.}
\label{fig:environment}
\end{minipage}
\hfill
\begin{minipage}[t]{0.3\textwidth}
\centering
\includegraphics[width=1\textwidth]{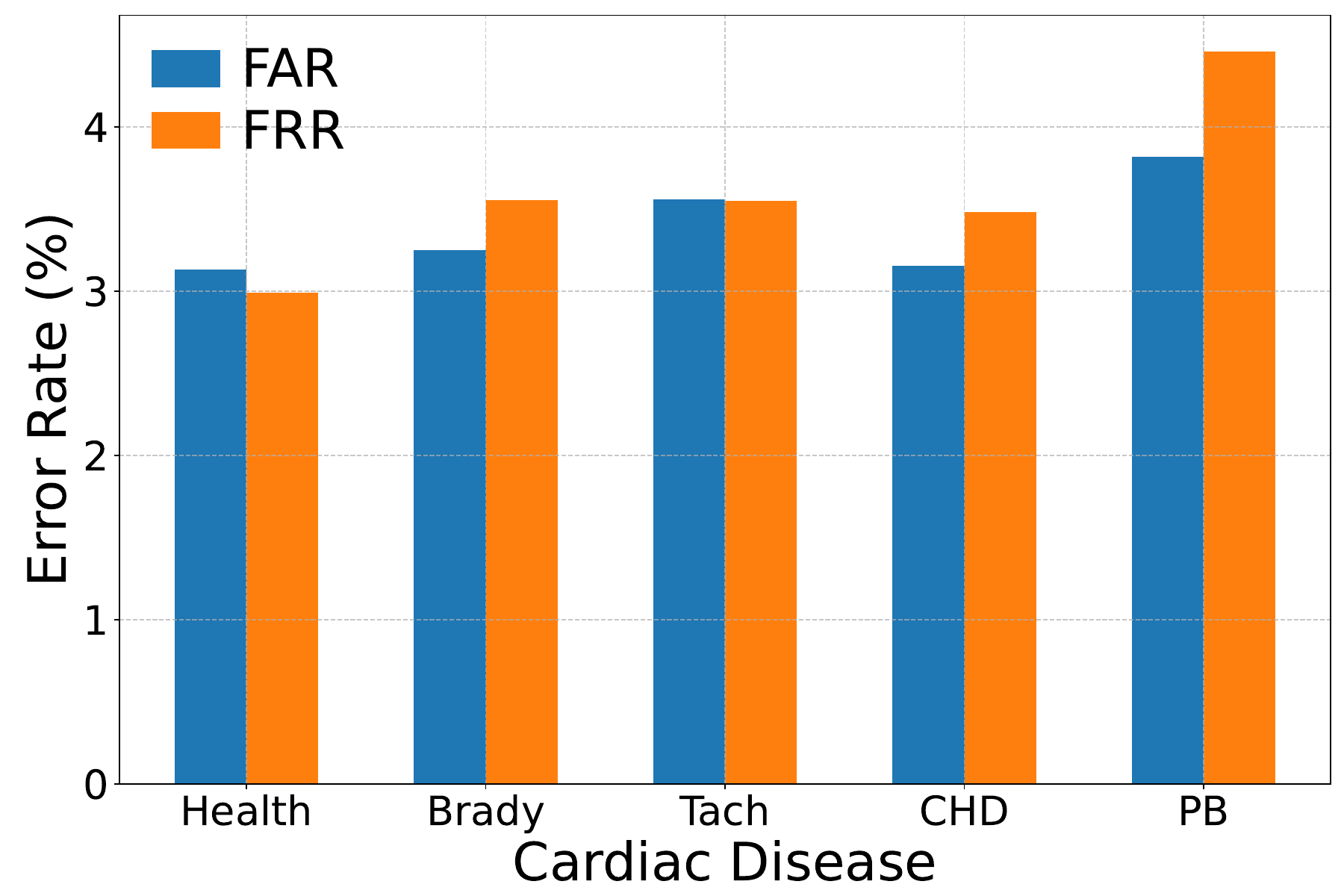}
\caption{Impact of cardiac diseases.}
\label{fig:disease}
\end{minipage}
\vspace{-0.1in}
\end{figure*}

We further evaluate the robustness of our system across subjects with different cardiac diseases. 
We consider four common heart diseases, including Bradycardia (Brady), Tachycardia (Tach), Coronary Heart Disease (CHD), and Premature Beat (PB). 
For each condition, we recruit five subjects, whose data are used only for registration and testing and are excluded from the training of both \HIDNET{} and the Siamese network. 
As shown in \figref{fig:disease}, compared to healthy subjects, the system consistently achieves low authentication errors across all disease groups. 
We observe slightly higher authentication errors for subjects with PB, which is expected, as premature beats introduce intermittent rhythm irregularities that disrupt the temporal consistency and morphological stability of heartbeat signals. Unlike sustained rhythm disorders, PB occurs sporadically, leading to increased intra-subject variability and making reliable identity verification more challenging. Overall, these results demonstrate that \HIDNET{}  can extract discriminative heartbeat features even from subjects with cardiac diseases, further validating the robustness and broad applicability of \sys{} in real-world scenarios.

\subsubsection{Impact of Sampling Rates}
\label{sec:sampling_rate}
\begin{figure*}[t]
\centering
\begin{minipage}[t]{0.3\textwidth}
\centering
\includegraphics[width=1\textwidth]{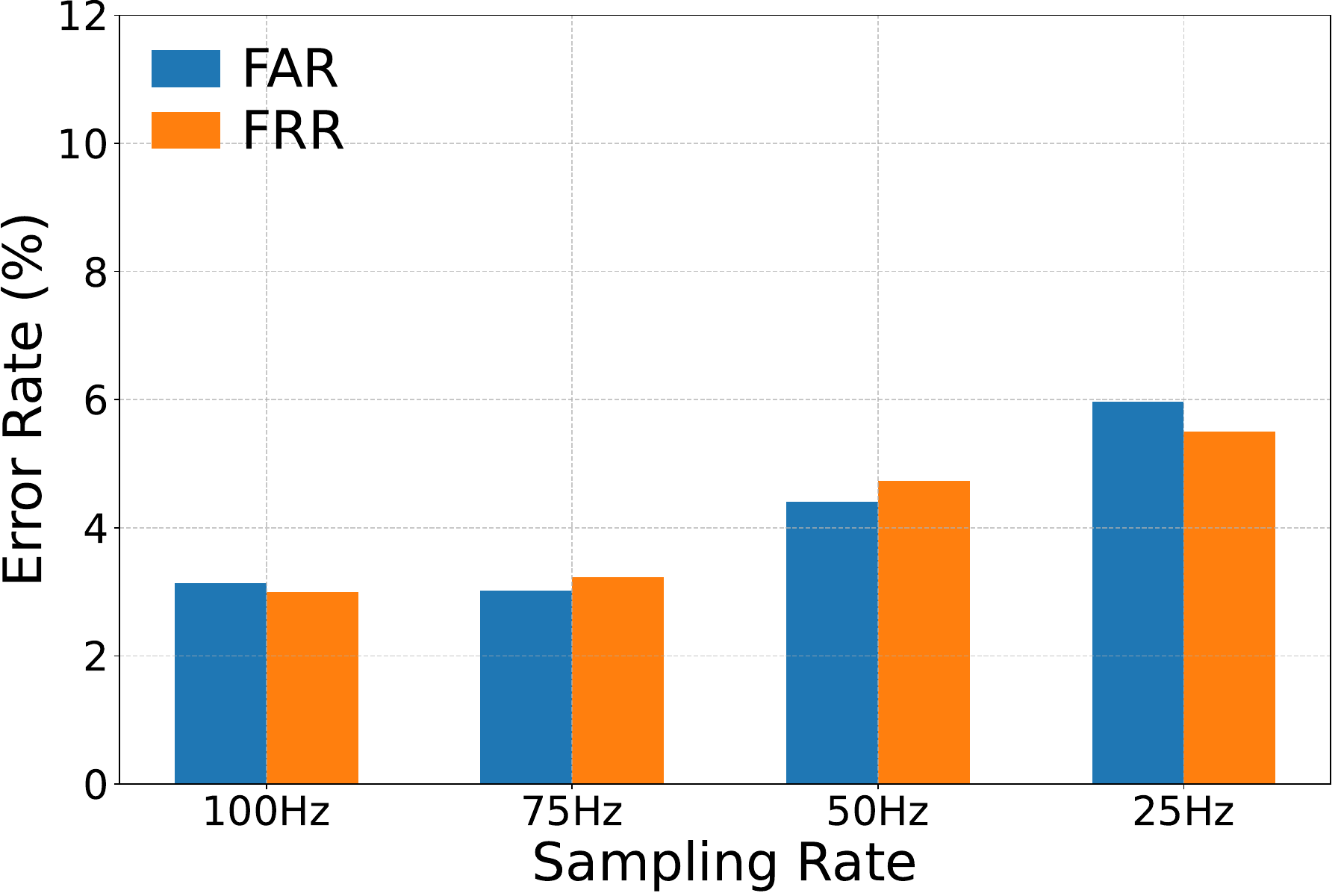}
\caption{Impact of sampling rates.}
\label{fig:sampling}
\end{minipage}
\hfill
\begin{minipage}[t]{0.3\textwidth}
\centering
\includegraphics[width=1\textwidth]{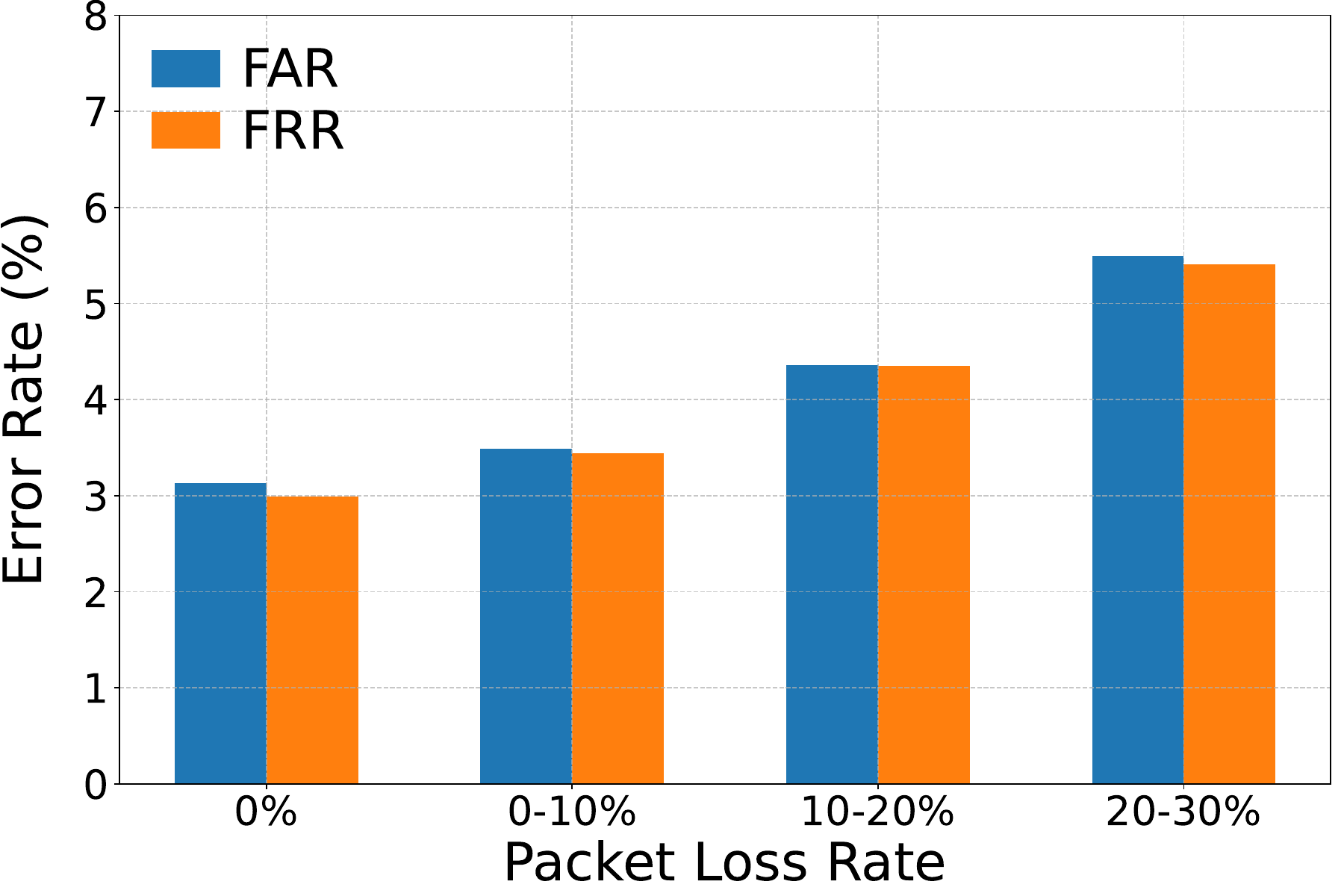}
\caption{Impact of BLE packet loss.}
\label{fig:packetloss}
\end{minipage}
\hfill
\begin{minipage}[t]{0.3\textwidth}
\centering
\includegraphics[width=1\textwidth]{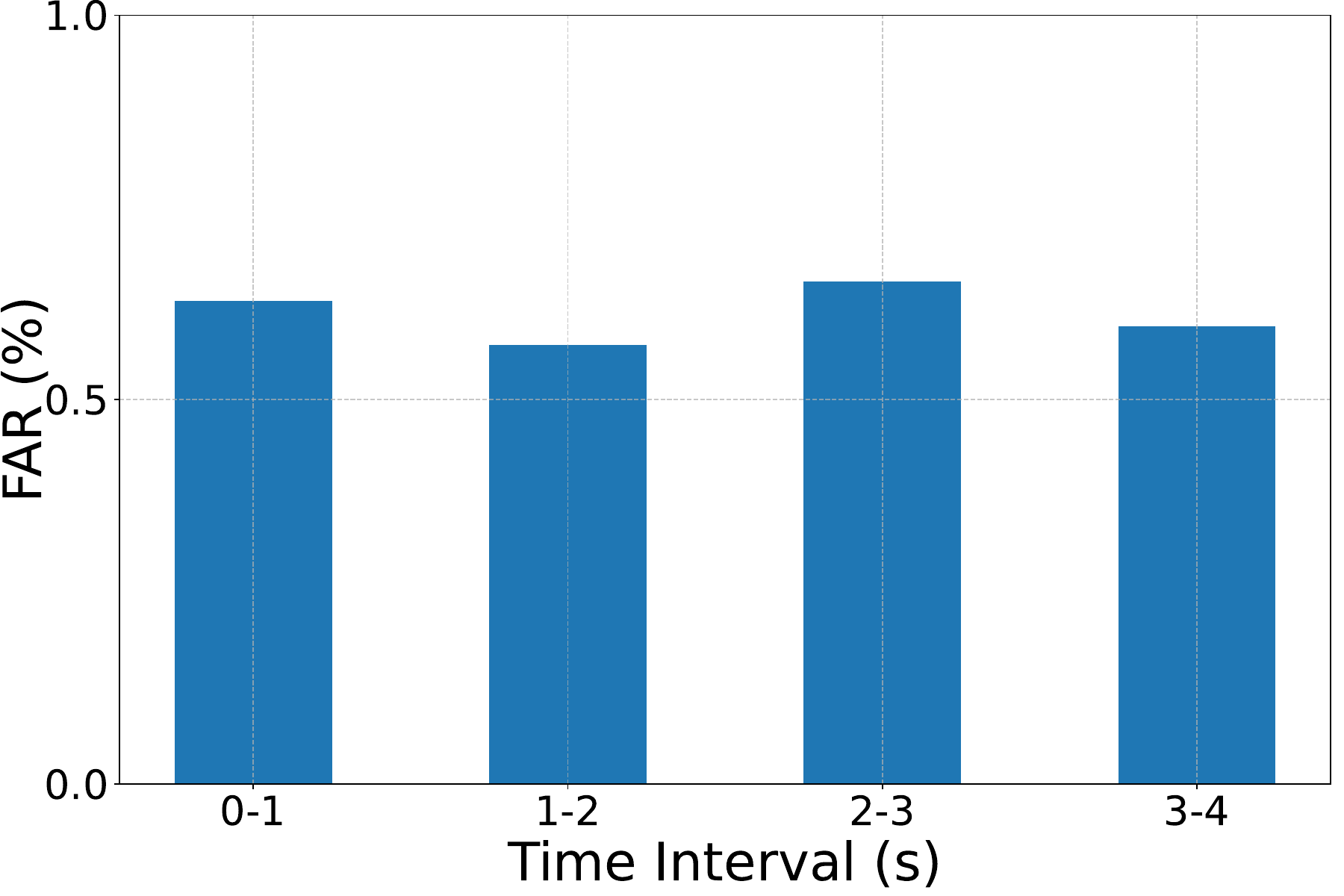}
\caption{Impact of earphone removal.}
\label{fig:remove}
\end{minipage}
\vspace{-0.2in}
\end{figure*}

To understand the effect of IMU sampling rate on authentication performance, we evaluate \sys{} under four settings: 100 Hz, 75 Hz, 50 Hz, and 25 Hz. 
We generate the 75 Hz signals from the original 100 Hz data using a polyphase-filtering-based resampling method (\texttt{scipy.signal.resample\_poly}), and obtain the 50 Hz and 25 Hz signals through direct downsampling. For each sampling rate, we independently repeat the full training, registration, and testing pipeline, while keeping the data split, model architecture, and training strategy unchanged. This setup allows a fair comparison of the complete system under different sampling-rate settings.
\figref{fig:sampling} shows that authentication performance degrades as the sampling rate decreases. This trend is expected, since lower sampling rates reduce signal fidelity and thus weaken the discriminative information available for authentication. Nevertheless, even at 25 Hz, \sys{} still achieves an average FAR of 5.97\% and an average FRR of 5.51\%, which remains acceptable for practical daily use.

\subsubsection{Impact of BLE Packet Loss}
\label{sec:packet_loss}
Bluetooth Low Energy (BLE) transmission may introduce packet loss during IMU streaming, especially when the communication channel is shared with other earphone functions. To study this practical issue, we evaluate the robustness of \sys{} under different packet-loss conditions and introduce a packet-loss-aware augmentation strategy during training.
Specifically, we simulate packet loss by randomly zero-masking samples in the IMU time series. In each epoch, 25\% of the samples in a batch remain clean, while the remaining 75\% are evenly assigned packet-loss rates of 0--10\%, 10--20\%, and 20--30\%. For each corrupted sample, the dropped positions are randomly selected and set to 0. The packet-loss pattern is resampled in every epoch, so that the model learns to handle diverse missing patterns rather than memorizing fixed corruption structures. We apply the same random packet-loss simulation strategy in the training, registration, and testing stages. Except for this augmentation, the model architecture, loss function, and evaluation protocol remain the same as those in the baseline system.
\figref{fig:packetloss} shows that the authentication performance degrades as the packet-loss rate increases. This trend is expected, since packet loss reduces the completeness of the physiological signal and weakens the discriminative information available for authentication. Nevertheless, even at a packet-loss rate of 30\%, \sys{} still achieves an average FAR of 5.49\% and an average FRR of 5.41\%, which remains acceptable for practical daily use. These results indicate that \sys{} preserves useful robustness under substantial BLE packet loss and remains feasible for deployment in realistic wireless settings.

\subsubsection{Impact of Earphone Removal}
\label{sec:earphone_removal}
A practical threat scenario arises when the initial authentication has already succeeded, but the legitimate user later removes the earphone and an attacker obtains it to exploit the still-trusted session.
In this case, the role of continuous authentication is not to protect first access, but to promptly revoke trust once the current wearer is no longer the legitimate user.
To evaluate whether \sys{} can effectively handle this scenario, we conduct an additional experiment under earphone removal. 
Specifically, we evaluate whether the system can correctly reject the user when the earphone-removal action occurs within four time intervals: 0--1 s, 1--2 s, 2--3 s, and 3--4 s, \ie, whether it can output \texttt{False} in time to terminate the trusted session.
\figref{fig:remove} shows that the average FAR remains below 0.65\% across all intervals. These results indicate that AccLock can quickly reject the removed-earphone state after removal and thus effectively prevent subsequent misuse by an attacker who obtains the earphone after the initial authentication has already succeeded. This result highlights the practical value of continuous authentication beyond one-time access control.

\subsection{Deployment on Commercial Earphone.}\label{sec:airpods} 
To further validate the practical deployability of \sys{}, we deploy the system on a commercial earphone platform, \ie, AirPods. 
Since AirPods expose IMU signals at a much lower sampling rate than our prototype, we adopt a two-stage training strategy to adapt \sys{} to this realistic hardware setting.
In the first stage, we pretrain the model on the dataset under the same 25 Hz setting used in the sampling-rate study, as illustrated in \secref{sec:sampling_rate}. 
We keep the same preprocessing pipeline and model architecture as those used in the main system, so that the resulting model learns a robust representation for low-sampling-rate IMU signals. 
In the second stage, we use real AirPods data collected from 10 users for lightweight adaptation. 
Specifically, we freeze the backbone network and update only the LoRA parameters inserted into the convolutional layers of the feature encoder and the Query/Value projections of the Transformer attention layers. We optimize these lightweight adaptation parameters with triplet loss for 20 epochs.
After model adaptation, we recruit an additional 11 users and collect 16 minutes of AirPods data from each user for evaluation. Following a protocol similar to that used in our main experiments, we evaluate the authentication performance on the commercial device. 
The results show an average FAR of 7.54\% and an average FRR of 7.1\%, demonstrating that AccLock can be successfully adapted to AirPods despite the low IMU sampling rate and practical hardware constraints.
These results provide direct evidence that AccLock is not restricted to our prototype platform and can be deployed on commercial earphones in practice. We believe that larger-scale AirPods data collection and full retraining under the same training strategy as AccLock can further improve the deployment performance in future work.

\subsection{System Performance}

\subsubsection{System Latency.} To evaluate the feasibility of deploying \sys{} on earphone–host device pairs, we transmit accelerometer signals from the earphone to multiple host devices, including three smartphones (Xiaomi 13, OPPO Find X8, and Samsung Galaxy S23) and a laptop (Lenovo IdeaPad 5), via Bluetooth, and examine the system’s runtime performance on the host devices.
Our implementation consists of two parallel threads: a display thread and a data processing thread. To handle the inertial data, the system maintains a 4-second buffer segment, updated every 100 ms to ensure real-time authentication. The system comprises three main blocks: Signal Denoising, Feature Disentanglement, and Authentication. On each device, we measure the processing time for each block 100 times and calculate the average. \tabref{tab:latency} summarizes the results. Specifically, \sys{} achieves average total runtimes of 29.11 ms, 20.80 ms, 26.97 ms, and 9.13 ms on the Xiaomi 13, OPPO Find X8,  Samsung Galaxy S23, and Lenovo IdeaPad 5, respectively. 
While processing delays vary across host devices due to hardware differences, the total latency consistently remains well below the 100 ms threshold, fully meeting the requirements for real-time authentication.

Beyond the cross-device latency comparison, we further examine the effect of sampling rate on runtime efficiency by evaluating AccLock on Xiaomi 13 under different IMU sampling rates.
As shown in \tabref{tab:latency}, the average latency is 21.43 ms, 23.14 ms, and 23.58 ms at 25 Hz, 50 Hz, and 75 Hz, respectively, all of which are slightly lower than that at 100 Hz. 
This result is expected, since a lower sampling rate reduces the input size and thus decreases the processing cost. 
However, the latency reduction remains modest, while the authentication performance degrades much more noticeably at lower sampling rates. 
As shown in \figref{fig:sampling}, the performance at 75 Hz and 100 Hz remains substantially better than that at 25 Hz and 50 Hz. 
Therefore, considering both runtime latency and authentication performance, 75 Hz and 100 Hz provide a more favorable trade-off and are more suitable for practical deployment.
\begin{table}[t]
\centering
\caption{System Latency}
\label{tab:latency}
\begin{tabular}{ccc@{\hspace{10pt}}c@{\hspace{10pt}}c@{\hspace{10pt}}c@{\hspace{10pt}}}
\hline
Devices & \makecell{Sampling\\Rate (Hz)} & \makecell{Signal\\Denoising (ms)} & \makecell{Feature\\Disentanglement (ms)} & \makecell{Authentication (ms)} & \makecell{Sum (ms)}\\ \hline
Xiaomi 13 & 100& 0.70  & 28.34  & 0.07 & 29.11  \\
Find X8 &100& 0.47 & 20.29 & 0.04 & 20.80 \\
S 23 &100&  0.68  & 26.25 & 0.04 & 26.97  \\
IdeaPad 5 &100 & 0.47  & 8.63 & 0.03 & 9.13  \\
Xiaomi 13 &25& 0.23  & 21.17  & 0.03& 21.43  \\
Xiaomi 13 &50& 0.41  & 22.70  & 0.03& 23.14 \\
Xiaomi 13&75& 0.60  & 24.92  & 0.06& 25.58  \\
\hline
\end{tabular}
\end{table}

\subsubsection{Energy Efficiency.}  We measure the energy consumption on three smartphones (\ie, Xiaomi 13, OPPO Find X8, and Samsung Galaxy S23) using Android Debug Bridge (ADB).
Specifically, for each device, we 10 consecutive authentication sessions, each lasting 5 minutes, and report the average energy consumption.
The results show that the average energy consumption of the Xiaomi 13, OPPO Find X8, and Samsung Galaxy S23 is 27.30 mAh, 29.01 mAh, and 23.77 mAh, respectively.
Given that a typical smartphone battery capacity is approximately 3,000 mAh, the energy overhead is minimal, enabling sustained daily use without noticeable battery impact.

We further evaluate the runtime overhead of AccLock on the phone side, as continuous authentication should run as a lightweight background service without affecting normal user interactions.
On Xiaomi 13, each inference takes less than 30 ms, leading to a low CPU duty cycle during continuous authentication.
The average CPU usage remains around 2.5\%, and the memory footprint stays stable at 139 MB.
Given the computing capability of modern mobile SoCs, such overhead is lightweight for background execution.
We do not observe noticeable interference with foreground applications during our experiments, suggesting that AccLock can operate continuously on the phone side with limited runtime overhead.

We also examine the battery impact on the earphone side, which is critical for continuous authentication in daily use.
Since the 3-axis accelerometer is already kept ``always-on'' to support default earphone functions, such as gesture recognition, AccLock does not introduce additional sensing power for inertial signal acquisition.
The main incremental overhead comes from the BLE transmission of 100 Hz inertial data from the earphone to the phone.
To quantify this overhead, we measure the total operating time of the earphone from a full charge to automatic shutdown under two modes.
In the \textit{Idle Connection Mode}, the earphone remains connected to the phone but does not transmit sensing data, and the operating time is $T_{idle}=4.15$ h.
In the \textit{AccLock Continuous Mode}, the earphone continuously streams inertial data for authentication, and the operating time is $T_{active}=3.25$ h.
Given the battery capacity $C=25$ mAh, we estimate the additional energy drain introduced by AccLock as $(C/T_{active})-(C/T_{idle})$.
The result shows that AccLock introduces only 1.67 mAh/h additional energy consumption on the earphone side.
This overhead is small compared with the battery budget of commodity earbuds, which typically integrate batteries at the scale of tens of mAh.

%% file: latex/related.tex
\presec
\section{Related Work}\label{sec:related}
\postsec

\nosection{Earable-Based Authentication:}
The rapid advancement of earable technologies has opened new avenues for designing user authentication systems. 
For instance, Nakamura et al. develop an in-ear EEG-based user authentication system \cite{nakamura2017ear}.
However, such an approach depends on dedicated sensors that are typically not integrated into commercial products, limiting its scalability in practical deployments. 
In contrast, our approach utilizes  in-ear accelerometers that are widely integrated into commercial earphones, providing a more practical and scalable solution for large-scale adoption.
EarEcho \cite{gao2019earecho} and EarDynamic  \cite{wang2021eardynamic} employ ultrasonic signals to probe the unique anatomical structure of the ear canal for user identification.
However, these techniques occupy the speaker component of the earphones, compromising their primary audio functionality and negatively affecting user experience. 
Several prior studies leverage microphones or IMUs to capture behavioral biometrics based on user activities such as tooth clicking \cite{xie2022teethpass, wang2022toothsonic}, mandible vibrations \cite{liu2021mandipass}, or walking patterns \cite{ferlini2021eargate}. However, these approaches require explicit user participation, which may compromise user convenience.
In contrast, our method passively and unobtrusively monitors in-ear bone-conducted BCG signals for authentication, preserving the standard usage experience of the earphones.
Recent work \cite{cao2023heartprint} enables passive user authentication via fine-grained audio signals, requiring no user interaction. However, high ANC earphone costs limit scalability, and denoising  techniques can only achieve limited effectiveness in suppressing interference.
In contrast, our system leverages distinctive features extracted from an alternative  modality, \ie, BCG signals,  captured by widely deployed accelerometers in standard earphones, offering strong resilience to environmental noise.
A closely related recent work  TWSCardio explores BCG-based authentication on earables using IMU signals \cite{fu2025enabling}. 
While TWSCardio demonstrates the feasibility of earable cardiac biometrics, our work differs from it in two important aspects. 
First, TWSCardio introduces an intermediate cardiogram reconstruction stage that converts in-ear IMU signals into SCG-like signals before downstream authentication, whereas AccLock directly performs authentication from single-modality in-ear accelerometer signals. 
This design avoids an additional reconstruction pipeline and leads to a simpler deployment path. 
Moreover, our framework can be adapted to commodity earphones such as AirPods through the fine-tuning-based deployment scheme described in \secref{sec:airpods}.
Second, TWSCardio requires model adaptation for new users, including fine-tuning of the reconstruction model and retraining of the authentication stage. In contrast, once AccLock is trained, a new user only needs a one-time registration process without model fine-tuning or retraining. Therefore, compared with TWSCardio, the contribution of AccLock lies not in the general feasibility of earable BCG authentication, but in a more lightweight and deployment-friendly design for practical use.

\nosection{Biometric-Based Authentication:}
Biometrics-based authentication has been widely studied due to its inherent uniqueness.
Typical biometric features include fingerprints \cite{shabrina2016fingerprint,clancy2003secure,ratha2000robust}, face \cite{fathy2015face,bicego2006use,jonsson2002support}, voice \cite{feng2017continuous,zhang2016voicelive,zhang2017hearing}, and iris \cite{thavalengal2015iris,revenkar2010secure}. 
However, they are relatively vulnerable to spoofing, making them susceptible to replay attacks. 
In contrast, heartbeat is inherently difficult to forge and have thus emerged as a recent focus of research.
Existing approaches are primarily based on physiological signals such as ECG \cite{arteaga2015ecg,sufi2010ecg,safie2011electrocardiogram}, PPG \cite{zhao2020trueheart,yadav2018evaluation,karimian2017non},  PCG \cite{abo2014biometric,el2021pcg}, and SCG \cite{wang2018unlock,hsu2021motion}.
Among these, ECG-based authentication systems  require additional hardware to be integrated into mobile or wearable devices, which leads to higher costs and consequently limits large-scale deployment.
PPG has been extensively adopted in consumer devices such as smartwatches and fitness trackers due to its convenience. 
%
%
However, variations in ambient lighting can distort the PPG signal, compromising the quality of subsequent signal acquisition \cite{frey2022blood,chen2021movi}.
%
Furthermore, the LED illumination during monitoring can significantly increase user discomfort, particularly in dark environments.
Both PCG and SCG signals exhibit distinctive individual characteristics that can be utilized as physiological traits in authentication systems \cite{khan2020biometric,beritelli2007biometric,wang2018unlock}.
However, they only work in specific collecting conditions such as holding the sensor against the chest, which involve cumbersome efforts for frequent use. 
In contrast, our work introduces a passive, user-friendly, and environmentally resilient authentication system that leverages accelerometers commonly integrated into earphones to capture heartbeat-induced vibrations within the ear canal.
Note that recent work shows that earables can also capture gyrocardiography (GCG) signals using built-in gyroscopes \cite{zhang2026lubdubdecoder}. 
In this work, we do not include GCG mainly for practical reasons: gyroscopes are less widely available on commodity earphones (e.g., Apple AirPods (2nd generation) and Samsung Galaxy Buds FE) and incur higher power consumption than accelerometers \cite{bmi270}, making them less suitable for always-on continuous authentication.
Moreover, our accelerometer-only design already achieves strong average performance, with an FAR of 3.13\% and an FRR of 2.99\%, which is sufficient for practical daily use.

\section{Limitations and Discussion}
 Currently, our system faces two key issues:

\nosection{Large-scale movements:}
Although the system is robust to inherent and sporadic movements (e.g., respiration, head shaking, and talking), large-scale movement remains challenging.
The substantial performance drop observed during running (FAR/FRR $> 40\%$) underscores the requirement for more sophisticated denoising schemes. 
In future work, we plan to explore  generative AI models to reconstruct corrupted heartbeat waveforms. 
By integrating the  advanced scheme, we aim to enhance the system's robustness in highly dynamic real-world environments


\nosection{Commercial earphones:} 
Although our system achieves high authentication accuracy using in-ear accelerometer signals, the current implementation relies on a custom hardware prototype to collect accelerometer data at a sampling rate of 100 Hz. In contrast, while many commercial earphones (e.g., AirPods) are equipped with inertial sensors capable of high-rate sampling, access to  raw accelerometer measurements is not exposed to third-party developers. Instead, only processed and downsampled motion signals are available, which limits their applicability for fine-grained physiological or biometric analysis.
In future work, we will pursue two directions.
First, we will explore opportunities for closer technical collaboration with earphone vendors to examine the feasibility of accessing higher-fidelity raw sensor data. 
Second, we will examine commercial sensing platforms with greater openness and higher sampling rates to evaluate the system’s generalizability on off-the-shelf hardware.

\nosection{User Variability}
The per-user FAR/FRR distribution indicates that a small subset of users shows higher error rates than the majority of the cohort. We believe this trend mainly reflects user-level variability in in-ear BCG quality and separability, which are influenced by factors such as anatomy, wearing fit, contact stability, and subtle earphone shifts during use. As \sys{} aims to support practical deployment without per-user retraining or fine-tuning after onboarding, the current system adopts a unified framework for users with diverse signal characteristics. While this design improves usability and deployment simplicity, it also makes a small number of users more challenging to authenticate reliably. Improving robustness under such user variability remains an important direction for future work.

%% file: latex/conclusion.tex

\presec
\section{Conclusion}  
\postsec

In this paper, we make the following three contributions.
First, we present \sys{}, the first passive earphone-based authentication system using in-ear BCG signals. \sys{} requires no user interaction or active speaker output, and is robust to environmental noise.
Second, we design a two-stage denoising scheme and propose a disentanglement-based model, \HIDNET{}, to extract user-specific features while suppressing shared physiological patterns.
Third, we develop a scalable Siamese-based framework that removes the need for per-user model training. 
Experiments with 33 participants show that \sys{} achieves high authentication performance, confirming its practical feasibility.
